\documentclass{elsart}

\usepackage{amsmath}

\usepackage{epsfig} \graphicspath{ {./figs/}  }

\usepackage{natbib} \newcommand{\newblock}[0]{\newline}
\def\aap{A\&A}

\def\aj{Astron. J.}  
\def\apj{Ap. J.}

\def\apjl{Ap. J. Lett.} 
\def\mnras{MNRAS}
\def\nat{Nature} \def\baas{BAAS}
\def\apss{Astrophys.and Space Sc.}
\def\prl{Phys. Rev. Lett.}  \def\prd{Phys. Rev. D}

\def\AbergelEtal94{1994ApJ...423L..59A}
\def\AghanimEtalq96{1996A&A...311....1A}
\def\AghanimEtala97{1997AA...325....9A}
\def\BahcallCen93{1993ApJ...407L..49B}
\def\BartlettSilk94{1994ApJ...423...12B}
\def\BennettEtal94{1994ApJ...436..423B}
\def\BirkinshawGull83{1983Natur.302..315B}
\def\BlainLongair93{1993MNRAS.264..509B}
\def\BondEfstaTegmark97{1997MNRAS.291L..33B}
\def\BondEfstathiou87{1987MNRAS.226..655B}
\def\BouchetEtalb95{1995proc...frb...1}
\def\BouchetEtalc95{1995proc...frb...2}
\def\BouchetEtald96{1996...fb...thisbook}
\def\BouchetGispert99{BouchetGisperta98}
\def\BouchetBennettStebbins88{1988Natur.335..410B}
\def\BoulangerEtal96{1996A&A...312..256B}
\def\BunnEtal94{1994ApJ...432L..75B}
\def\BurtonHartmann94{1994Ap&SS.217..189B}
\def\ChengEtal94{1994ApJ...422L..37C}
\def\ClemensEtal98{1997AAS...191.6305C}
\def\ColafrancescoEtal94{1994ApJ...433..454C}
\def\CondonEtal89{1989AJ.....97.1064C}
\def\DaviesEtalb95{1996MNRAS.278..883D}
\def\DaviesEtala96{1996MNRAS.278..925D}
\def\DebernardisEtal94{1994ApJ...422L..33D}
\def\DebernardisEtala92{1992issa.proc..315D}
\def\Delabrouille98{1998A&AS..127..555D}
\def\DelucaEtal95{1995A&A...300..335D}
\def\DesertEtal90{1990A&A...237..215D}
\def\DodelsonStebbins94{1994ApJ...433..440D}
\def\DraineLazarian97{1998ApJ...494L..19D}
\def\DwekEtal96{1997ApJ...475..565D}
\def\EfstathiouBonda86{1986MNRAS.218..103E}
\def\EfstathiouBond87{1987MNRAS.227P..33E}
\def\FischerEtala92{1992ApJ...388..242F}
\def\FischerEtal95{1995ApJ...444..226F}
\def\GaierEtal92{1992ApJ...398L...1G}
\def\GautierEtal92{1992AJ....103.1313G}
\def\GangaEtal93{1993ApJ...410L..57G}
\def\Gorski94{1994ApJ...430L..85G}
\def\GorskiEtal94{1994ApJ...430L..89G}
\def\GuiderdoniEtal98{1998MNRAS.295..877G}
\def\HaeneltEtal96{1996MNRAS.279..545H}
\def\HaslamEtal82{1982A&AS...47....1H}
\def\HancockEtala97{1997MNRAS.289..505H}
\def\HancockEtal94{1994Nature367..333}
\def\HauserEtal98{1998ApJ...508...25H}
\def\HobsonEtal98{1998MNRAS.300....1H}
\def\HenryArnaud91{1991ApJ...372..410H}
\def\HildebrandDragovan95{1995ApJ...450..663H}
\def\HuWhite97{1997NewA....2..323H}
\def\JanssenGulkis92{1992issa.proc..391J}
\def\JonesEtal96{1998MNRAS.294..582J}
\def\Knox95{1995PRD...52..4307}
\def\KogutEtala95{1995BAAS..18712002K}
\def\KogutEtal96{1996ApJ...464L...5K}
\def\LagacheEtal96{1996A&A...TBP1}
\def\LaureijsEtal93{1991ApJ...372..185L}
\def\LawsonEtal87{1987MNRAS.225..307L}
\def\LineweaverBarbosa98{1998A&A...329..799L}
\def\LeitchEtal97{1997ApJ...486L..23L}
\def\LockmanEtal86{1986ApJ...301..380L}
\def\LonsdaleEtal90{1990ApJ...358...60L}
\def\MeinholdEtal93{1993ApJ...409L...1M}
\def\MyersBond93{1993BAAS..182.7707M}
\def\NetterfieldEtal97{1997ApJ...474...47N}
\def\OliveiraEtal97{1997ApJ...482L..17D}
\def\OliverEtal97{1997MNRAS.289..471O}
\def\OstrikerVishniac86{1986ApJ...306L..51O}
\def\Peebles87{1987ApJ...315L..73P}
\def\PlattEtal97{1997ApJ...475L...1P}
\def\PointecouteauEtal97{1998A&A...336...44P}
\def\PugetEtal96{1996A&A...308L...5P}
\def\ReachEtal95{1995ApJ...451..188R}
\def\ReichReich88{1988A&AS...74....7R}
\def\Rephaeli95{1995ApJ...445...33R}
\def\Rephaelib95{1995ARA&A..33..541R}
\def\Reynolds89{1989ApJ...339L..29R}
\def\ReynoldsEtal95{1995ApJ...448..715R}
\def\RushEtal93{1993ApJS...89....1R}
\def\SavageEtal77{1977ApJ...216..291S}
\def\SchlegelEtal97{1998ApJ...500..525S}
\def\SchusterEtal93{1993ApJ...412L..47S}
\def\Seljaka96{1996ApJ...463....1S}
\def\Seljak97{1997ApJ...482....6S}
\def\TanakaEtal96{1996ApJ...468L..81T}
\def\Tegmarka96{1996ApJ...464L..35T}
\def\Tegmark97{1997ApJ...480L..87T}
\def\TegmarkTaylorHeavens97{1997ApJ...480...22T}
\def\TegmarkEfstathiou95{1996MNRAS.281.1297}
\def\ToffolattiEtal97{1998MNRAS.297..117T}
\def\TuckerEtal97{1997ApJ...475L..73T}
\def\VianaLiddle96{1996MNRAS.281..323V}
\def\WollackEtal93{1993ApJ...419L..49W}
\def\WilnerWright97{1997ApJ...488L..67W}
\def\WhiteFrenk91{1991ApJ...379...52W}
\def\WrightEtal91{1991ApJ...381..200W}
\def\Wright97{1998ApJ...496....1W}
\def\ZaroubiEtal95{1995ApJ...449..446Z}
\def\ZaldarriagaSpergelSeljak97{1997ApJ...488....1Z}

\def\BouchetGispert96{1996...frb...inprep}
\def\GispertBouchet96{1996...rg...thisbook}

\newcommand{\bce}[0]{\begin{center}} \newcommand{\ece}[0]{\end{center}}
\newcommand{\ben}[0]{\begin{enumerate}} \newcommand{\een}[0]{\end{enumerate}}
\newcommand{\bit}[0]{\begin{itemize}} \newcommand{\eit}[0]{\end{itemize}}

\renewcommand{\bar}[1]{\overline{#1}} 
\renewcommand{\hat}{\widehat} \renewcommand{\tilde}{\widetilde}

\newcommand{\simlt}[0]{{\lower.5ex\hbox{$\; \buildrel < \over \sim \;$}}}
\newcommand{\simgt}[0]{{\lower.5ex\hbox{$\; \buildrel > \over \sim \;$}}}
\newcommand{\VEV}[1]{\left\langle #1\right\rangle}
\newcommand{\Cov}[1]{\mathrm{Cov}\left[#1\right]}
\newcommand{\prim}[0]{^\prime} \newcommand{\dprim}[0]{^{\prime\prime}}

 \newcommand{\microK}{~$\mu\mathrm{K}$}
\newcommand{\mic}{~$\mu\mathrm{m}$} 
\newcommand{\HI}{{H\,I}} 

\newcommand{\dg}{\nobreak^\circ}\newcommand{\rms}{{\em rms } }
\newcommand{\cf}{{\it cf. }} \newcommand{\eg}{{\it e.g. }}
\newcommand{\ie}{{\it i.e. }}

\newcommand{\planck}{{\sc Planck}} \newcommand{\plancks}{{\sc Planck} }
\newcommand{\hfi}{{\sc HFI}} \newcommand{\hfis}{{\sc HFI} }
\newcommand{\lfi}{{\sc LFI}} \newcommand{\lfis}{{\sc LFI} }
\newcommand{\cobe}{{\sc COBE}} \newcommand{\cobes}{{\sc COBE} }
\newcommand{\dirbe}{{\sc DIRBE}} \newcommand{\dirbes}{{\sc DIRBE} }
\newcommand{\map}{{\sc MAP}} \newcommand{\maps}{{\sc MAP} }
 \newcommand{\dmrs}{{\sc DMR} }
\newcommand{\firas}{{\sc FIRAS}} \newcommand{\firass}{{\sc FIRAS} }
\newcommand{\iras}{{\sc IRAS}} \newcommand{\irass}{{\sc IRAS} }

 \newcommand{\bb}[0]{\mathbf{b}}
 
\newcommand{\be}[0]{\mathbf{e}}

 \newcommand{\bl}[0]{\boldsymbol{\ell}}
 \newcommand{\bn}[0]{\mathbf{n}}

\newcommand{\bs}[0]{\mathbf{s}} \newcommand{\bt}[0]{\mathbf{t}}
 
 \newcommand{\bx}[0]{\mathbf{x}}
\newcommand{\by}[0]{\mathbf{y}} 

\newcommand{\bA}[0]{\mathbf{A}} \newcommand{\bB}[0]{\mathbf{B}}
\newcommand{\bC}[0]{\mathbf{C}} 
\newcommand{\bE}[0]{\mathbf{E}} 
\newcommand{\bI}[0]{\mathbf{I}} 
 \newcommand{\bL}[0]{\mathbf{L}}
 \newcommand{\bN}[0]{\mathbf{N}}
 
\newcommand{\bR}[0]{\mathbf{R}} \newcommand{\bS}[0]{\mathbf{S}}
\newcommand{\bW}[0]{\mathbf{W}}

\newcommand{\beps}[0]{\boldsymbol{\varepsilon}}

\newcommand{\calB}[0]{\mathcal{B}} \newcommand{\calC}[0]{\mathcal{C}}
\newcommand{\calF}[0]{\mathcal{F}} 
 
 \newcommand{\calT}[0]{\mathcal{T}}

\newcommand{\bxh}[0]{\hat{\bx}}
\begin{document}
%******************************************************************************

\begin{frontmatter}

\title{FOREGROUNDS AND CMB EXPERIMENTS}

\subtitle{I. Semi-analytical estimates of contamination}

\author{Fran\c{c}ois R. Bouchet}

\address{Institut d'Astrophysique de Paris, CNRS, Paris, France}

\author{Richard Gispert}

\address{Institut d'Astrophysique Spatiale, CNRS-Universit\'{e} Paris-Sud,
Orsay, France}

\begin{abstract}

As Cosmic Microwave Background (CMB) measurements are becoming more ambitious,
the issue of foreground contamination is becoming more pressing. This is
especially true at the level of sensitivity, angular resolution and for the
sky coverage of the planned space experiments \maps and \planck. We present in
this paper an indicator of the accuracy of the separation of the CMB
anisotropies from those induced by foregrounds.

Of course, the outcome will depend on the spectral and spatial characteristics
of the sources of anisotropies. We thus start by summarising present knowledge
on the spectral and spatial properties of Galactic foregrounds, point sources,
and clusters of galaxies. This information comes in support of a modelling of
the microwave sky including the relevant components. The accuracy indicator we
introduce is based on a generalisation of the Wiener filtering method to
multi-frequency, multi-resolution data. While the development and use of this
indicator was prompted by the preparation of the scientific case for the
\plancks satellite, it has broader application since it allows assessing the
effective capabilities of an instrumental set-up once foregrounds are fully
accounted for, with a view to enabling comparisons between different
experimental arrangements.

The real sky might well be different from the one assumed here, and the analysis
method might not be in the end Wiener filtering, but this work still allow
meaningful {\em comparative} studies. As a matter of examples, we compare the
CMB reconstruction errors for the \maps and \plancks space missions, as well
as the robustness of the \plancks outcome to possible failures of specific
spectral channels or global variations of the detectors noise level across
spectral channels.

\end{abstract}

\end{frontmatter}

%******************************************************************************
\section{Introduction}
%******************************************************************************

Future Cosmic Microwave Background (hereafter CMB) anisotropy experiments
should tightly constrain the parameters of any theory for the origin of those
anisotropies. In the case of the \plancks space mission of ESA (formerly known
as {\sc COBRAS/SAMBA}), for inflation generated fluctuations, the value of the
cosmological parameters should be constrained at the percent level
\citep{1997MNRAS.291L..33B, 1997ApJ...488....1Z, EfstathiouBond98}. These
estimates generally rely on assuming that the measurements at some frequencies
are dominated by the primary cosmological signal, the other being dominated by
foregrounds emissions and used to ``clean'' the former measurements from any
substantial remaining contamination. Which frequency measurements to consider
appears as a matter of art.

In this paper, we propose a scheme for assessing quantitatively the effect of
foregrounds from an assumed model of the microwave sky, a description of the
experiment, and by supposing Wiener filtering is used to analyse the
data. While linearly optimal, this separation method has also difficulties
which we shall discuss later. Other non-linear methods are likely to be better
suited for recovering the non-Gaussian spatial distributions of the
foregrounds (and possibly of the CMB itself). But the linear nature of Wiener
filtering allows straightforward calculations leading in particular to an
effective spatial window (or beam profile) and an effective noise level for
the experiment, once foregrounds have been accounted for. This allows putting
previous cosmological parameters accuracy forecasts on a firmer ground, and it
is a useful tool for optimising experimental set-ups.

Undoubtly new aspects of the foreground emission will be discovered in future
experiments, and in particular by \maps and \plancks themselves, but we hope
that the present observations and understanding of foregrounds is already
sufficient to allow building a plausible model of the microwave sky. The more
so that we shall only need a rough statistical description to test the ability
of planned experiments to disentangle foregrounds from true CMB anisotropies.

This paper has three main parts. The first one is concerned with deriving a
plausible microwave sky model adapted to our present purposes. The next one
describes the Wiener filtering method in the context of CMB experiments and
leads to the definition of a ``quality factor'' which assesses the effect of
foreground contamination on the outcome of an experiment. The last part offers
practical examples in the case of \maps and \plancks by making use of our sky
model and analysis method. More precisely, section \ref{sec:gal_fore}
discusses the Galactic foregrounds, while section \ref{sec:sources} discusses
the extragalactic foregrounds due to infrared galaxies, radio--sources, and
the Sunyaev-Zeldovich effect on clusters of galaxies. These various
contributions are put into perspective with respect to CMB anisotropies in
\S~\ref{sec:comparing}. Section \ref{sec:separation} discusses component
separation methods, Wiener filtering, and derives the properties of the
proposed quality factor. Practical applications are considered in
\S~\ref{sec:experiments}.

Before we go on, let us first establish notations. The amplitude of any scalar
field (in particular the temperature anisotropy pattern) on the sphere,
$A(\theta, \phi )$, can be decomposed into spherical harmonics as
\begin{equation}
    A(\theta, \phi ) = \sum_{\ell, m} a_{\ell m}\, Y_{\ell m}(\theta, \phi) .
    \label{sph_harm}
\end{equation}
The multipole moments, $a_{\ell m}$, are independent for a statistically
isotropic field, \ie $\VEV{a_{\ell m}^2} = C_\ell\, \delta_{\ell\ell\prim}\,
\delta_{m m\prim}$; $C_\ell$ is called the angular power spectrum and it is
related to the variance of the distribution by
\begin{equation}
    \sigma^2 = \VEV{A^2} = \sum_{\ell} \frac{2\ell +1}{4\pi} C_\ell .
\end{equation}
The angular power spectrum characterises completely a Gaussian field.

%******************************************************************************
\section{Galactic Foregrounds \label{sec:gal_fore} }
%******************************************************************************

The Galactic emissions are associated with dust, free-free emission from
ionised gas, and synchrotron emission from relativistic electrons. We are
still far from an unified model of the galactic emissions. On the other hand,
the primary targets for CMB determinations are regions of weakest galactic
emission, while stronger emitting regions are rather used for studies of the
interstellar medium. The approach taken here is thus to find a model
appropriate for the best half of the sky from the point of view of a CMB
experiment. This is based on work done during the preparation of the
scientific case for \plancks (see \citet{Morionda96,phaseArep,aao}).

\subsection{Dust emission}
%*************************

This restriction to the best half of the sky results in considerable
simplification for modelling the dust emission, since there is now converging
evidence that the dust emission spectrum from {\em high} latitude regions with
{\em low} \HI\ column densities can be well approximated with a single dust
temperature and $ \nu^{2} $ emissivity, with no clear evidence for a very cold dust
component in these regions.

The spectrum of the dust emission at millimetre and sub-millimetre wavelengths
has been measured by the Far--Infrared Absolute Spectrophotometer (\firas)
aboard the {\sc Cosmic Background Explorer} (\cobe) with a 7 degree beam. In
addition, several balloon experiments have detected the dust emission at high
galactic latitude in a few photometric bands with angular resolution between
30 arcminutes and 1 degree (see for example \cite{\FischerEtal95} and
\cite{\DebernardisEtala92}). The spectrum of the Galaxy as a whole
\cite[]{\WrightEtal91,\ReachEtal95} cannot be fitted with a single dust temperature
and emissivity law. These authors have proposed a two temperatures fit
including a very cold component with $\rm T \sim 7\,$K.  The interpretation of
this result for the Galactic plane is not straightforward because dust along
the line of sight is expected to spread over a rather wide range of
temperatures just from the fact that the stellar radiation field varies widely
from massive star forming regions to shielded regions in opaque molecular
clouds. On the other hand, a consistent picture for the high galactic latitude
regions has recently emerged.

\citet{\BoulangerEtal96} analysed the far--IR \& sub-mm dust spectrum by
selecting the fraction of the \firass map, north of $- 30^\circ $ declination,
where the \HI\ emission is weaker than $250$\,K.km/s (36\% of the sky).  The
declination limit was set by the extent of the \HI\ survey made with the
Dwingeloo telescope \cite[]{\BurtonHartmann94}. For an optically thin emission
this threshold corresponds to $\rm N(\HI)=5 \times 10^{20}$\,cm$^{-2}$. Below
this threshold, the correlation between the \firass and Dwingeloo values are
tight. For larger \HI\ column density the slope and scatter increase, which is
probably related to the contribution of dust associated with molecular
hydrogen since this column density threshold coincides with that inferred from
UV absorption data for the presence of H$_2$ along the line of sight
\cite[]{\SavageEtal77}. \citet{\BoulangerEtal96} found that the spectrum of the
\HI-correlated dust emission seen by \firass is well fitted by one single dust
component with a temperature $\rm T = 17.5$\,K and an optical depth $\rm \tau
/ N_H = 1.0 \times 10^{-25}(\lambda/250 \mu \mathrm{m})^{-2}\,
\mathrm{cm}^{2}$.

The residuals to this one temperature fit already allow to put an upper limit
on the sub-mm emission from a very cold component almost one order magnitude
lower than the value claimed by \citet{\ReachEtal95}. Assuming a temperature
of 7 K for this purported component one can set an upper limit on the optical
depth ratio $\rm \tau(7K)/\tau(17.5K) \sim 1$. \citet{\DwekEtal96} have also
derived the emission spectrum for dust at Galactic latitudes larger than
$40^\circ $, by using the spatial correlation of the \firass data with the
$100 \mu\mathrm{m}$ all-sky map from the Diffuse Infrared Background
Experiment (\dirbe). They compared this spectrum with a dust model not
including any cold component. The residuals of this comparison shows a small
excess which could correspond to emission from a very cold component at a
level just below the upper limit set by the \citet{\BoulangerEtal96}
analysis. Both studies thus raise the question of the nature of the very cold
component measured by \citet{\ReachEtal95}.

\citet{\PugetEtal96} found that the sub-mm residual to the one temperature fit
of \citet{\BoulangerEtal96} is isotropic over the sky. To be Galactic, the
residual emission would have to originate in a halo large enough ($> 50$\,kpc)
not to contradict the observed lack of Galactic latitude and longitude
dependence. Since such halos are not observed in external galaxies,
\citet{\PugetEtal96} suggested that the excess is the Extragalactic Background
Light due to the integrated light of distant IR sources. Since then this
cosmological interpretation has received additional support. 
First, the main uncertainty in these analysis was the difficulty to establish
the contribution from dust in the ionised gas. \cite{LagacheThesis98} has extracted
a galactic component likely to be associated with the ionised gas. Second, the
analysis was redone (\cite{GuiderEtal97,LagacheThesis98,LagacheEtal98}) in
 a smaller fraction of the sky where the \HI\ column density is even 
smaller ($N_\HI < 1 \times 10^{20}$\,H\,cm$^{-2}$). 
In these regions, the \HI-correlated dust emission is
essentially negligible and the ``residual'' is actually the dominant
component. This leads to a much cleaner determination of the background
spectrum which is slightly stronger than the original determination. As we
shall see below in \S~\ref{sec:irrad_sources}, the cosmological interpretation
fits well with the results of recent IR searches for the objects that cause
this background. Finally, the recent analyses by \citet{\SchlegelEtal97} and
\citet{\HauserEtal98} also detected a residual significantly higher than the
one obtained by \citet{GuiderEtal97}, \citet{LagacheThesis98}, \citet{1998ApJ...508..123F} and  \citet{LagacheEtal98}. The determination of the extragalactic component in
\citet{\SchlegelEtal97} is not very accurate from their own estimate. The difference
with \citet{\HauserEtal98} is mostly due to the fact that these authors do not find
any evidence for the emission associated with ionised gas.

%The
%spectrum presented by \citet{\PugetEtal96} for the residuals of the \firas--
%\HI\ correlation in the regions of lowest \HI\ column densities contains a
%fraction of the emission in the far--IR from dust emission associated with the
%diffuse ionised gas seen at high latitude. 

In summary, there is now converging evidence that the dust emission spectrum
from most of the {\em high} latitude regions with low \HI\ column densities
can be well approximated with a single dust temperature and $ \nu^{2} $
emissivity with no evidence for a very cold dust component. The sub-mm excess
seen in the \firass data is rather of extragalactic origin. Of course, this
only holds strictly for the emissions averaged over the rather large lobe of
the \firass instrument, and for regions of low column density. In denser areas
like the Taurus molecular complex, there is a dust component seen by \irass at
$100\mu$m and not at $60\mu$m. This cold \irass component is well correlated
with dense molecular gas as traced by $^{13}$CO emission
\cite[]{\LaureijsEtal93,\AbergelEtal94, 1998A&A...333..709L}. 
In the following, we assume that the
CMB data analysis will be performed only in the simpler but large connected
region of low \HI\ column density.

Concerning the scale dependence of the amplitude of the fluctuations,
\cite{\GautierEtal92} found that the power spectrum of the 100\mic\ \iras\
data is decreasing approximately like the 3$^{rd}$ power of the spatial
frequency, $\ell$, down to the \irass resolution of 4 arc minutes. More
recently, \cite{\Wright97} analysed the \dirbes data by two methods, and
concluded that both give consistent results, with a high latitude dust
emission spectrum, $C(\ell)$, also $\propto \ell^{-3}$ in the range $2< \ell <
300$, \ie\ down to $\theta \sim 60/\ell \sim 0.2\deg$ (with $C(10)^{1/2} \sim
.22$\,MJy/sr at $|b| > 30$).

\subsection{Free-free emission}
%******************************

Observations of H$_\alpha $ emission at high Galactic latitude and dispersion
measurements in the direction of pulsars indicate that low density ionised gas
(the Warm Ionised Medium, hereafter the WIM) accounts for about 25\% of the
gas in the Solar Neighbourhood \cite[]{\Reynolds89}.  The column density from
the mid-plane is estimated to be in the range 0.8 to $1.4 \times 10^{20}\,
\mathrm{cm}^{-2}$. Until recently, little was known about the spatial
distribution of this gas but numerous H$_\alpha $ observing programs are
currently in progress.  An important project is the northern sky survey
started by Reynolds (WHAM Survey).  This survey consist of H$_\alpha$ spectra
obtained with a Fabry-Perot through a $1^\circ $ aperture.  The spectra will
cover a radial velocity interval of 200 km/s centred near the LSR with a
spectral resolution of 12 km/s and a sensitivity of 1 cm$^{-6}$ pc ($5 \sigma
$).  Several other groups are conducting H$_\alpha$ observations with wide
field camera ($10^\circ$) equipped with a CCD and a filter
\citep{1998IAUS..190E..58G}.  These surveys with an angular resolution of a
few arc minutes should be quite complementary to the WHAM data and should
allow covering about 90 \% of the sky.  From these H$_\alpha$ observations one
can directly estimate the free-free emission from the WIM (since both scale in
the same way with the emission measure $\propto \int n_e^2 dl$, see
\cite{1998PASA...15..111V} for a careful discussion).

\citet{\KogutEtal96} have found a correlation at high latitudes between the
\dmrs emission after subtraction of the CMB dipole and quadrupole and the
\dirbes $240 \mu$m map. The amplitude of the correlated signal is plotted in
Figure~\ref{fig:kogut}.  The observed change of slope of the correlation
coefficients at 90 GHz is just what is expected from the contribution of the
free-free emission as predicted by \cite{\BennettEtal94}.  The spectrum of
this emission may be described by $I_\nu \propto \nu^{-0.16}$ (the index is
the best fit value of \citet{\KogutEtala95}). More recently
\citet{\OliveiraEtal97} cross-correlated the Saskatoon data
\citep{1997ApJ...474L..77T} with the \dirbes data and also found a correlated
component, with a normalisation in agreement with that of
\citet{\KogutEtal96}. This result thus indicates that the correlation found by
\citet{\KogutEtal96} persists at smaller angular scales.

\citet{VeeraraghavanDavis97} used $H\alpha$ maps of the North Celestial Pole
(hereafter NCP) to determine more directly the spatial distribution of
free-free emission on sub-degree scale. Their best fit estimate is $C^{ff} =
1.3_{-0.7}^{+1.4}\ \ell^{-2.27\pm0.07}$ $\mu\mathrm{K}^2$ at 53 GHz, if they
assume a gas temperature $\sim 10^4$ K. While this spectrum is significantly
flatter than the $\ell^{-3}$ dust spectrum, the normalisation is also
considerably lower than that deduced from \citet{\KogutEtal96}. Indeed, their
predicted power at $\ell =300$ is then a factor of 60 below that of the \cobe\
extrapolation if one assumes an $\ell^{-3}$ spectrum for the free-free
emission. Additionally \citet{\LeitchEtal97} found a strong correlation
between their observations at 14.5 and 32 GHz towards the NCP and \irass
$100 \mu$m emission in the same field. However, starting from the corresponding
H$_\alpha$ map, they discovered that this correlated emission was {\em much
too strong} to be accounted for by the free-free emission of $\sim 10^4$ K
gas. These and other results suggest that only part of the microwave emission
which is correlated with the dust emission traced by \dirbe\ at $240 \mu$m may
be attributed to free-free emission as traced by $H\alpha$. The missing part
could be attributed to free-free emission uncorrelated with the dominating
component at $240 \mu$m.

\citet{\DraineLazarian97} recently proposed a different interpretation for the
observed correlation. They propose that the correlated component is produced
by electric dipole rotational emission from very small dust grains under
normal interstellar conditions (the thermal emission at higher frequency
coming from large grains). Given the probable discrepancy noted above between
the level of the correlated component detected by \citet{\KogutEtal96} and the
lower level of the free-free emission traced by $H\alpha$, it appears quite
reasonable that a substantial fraction of this correlated component indeed
comes from spinning dust grains. The spectral difference between free-free and
spinning dust grains shows up at frequencies $\simlt 30$ GHz, \ie\ outside of
the range probed by \maps and \planck. They are thus spectrally equivalent for
the 2 experiments. However the spatial properties may be different. Indeed the
spinning dust grains should be well mixed with the other grains resulting in
tight correlations persisting at all spatial scales. The free-free emission
might instead decorrelate from the dust emission at very small scales (since
$H\,II$ regions might be mostly distributed in ``skins'' surrounding dense
\HI\ clouds as proposed by \citet{1977ApJ...218..148M}, a model supported by
comparisons of \irass and H$_\alpha$ maps).

In the following we shall assume for simplicity that in the best half of the
sky (therefore avoiding dense \HI\ clouds) the correlated microwave emission
with $I_\nu \propto \nu^{-0.16}$ in the \maps and \plancks range is well
traced at all relevant scales by the dust emission which dominates at high
frequency and is well traced by \HI. Still there may well be an additional
free-free emission which is not well traced by the \HI-correlated dust
emission. Indeed the correlated emission detected by \citet{\KogutEtal96}
might be only part of the total signal with such a spectral signature that
\dmrs detected. While all of this signal may be accounted for by the
correlated component, the error bars are large enough that about half of the
total signal might come from an uncorrelated component\footnote{At 53 GHz, on
the 10$^\circ$ scale, the total free-free like signal is $\Delta T_{ff} = 5.2
\pm 4.2 \mu$K, while the correlated signal is $\Delta T_{cor} = 6.8 \pm 1.6
\mu$K. If we assume $\Delta T_{cor} = 5.2 \mu$K and $\Delta T_{uncor} = \Delta
T_{cor}$, we have $\Delta T_{ff} = 7.3 \mu$K which is well within the allowed
range.}. This is what we shall assume below.

\subsection{Synchrotron emission}
%********************************

Away from the Galactic plane region, synchrotron emission is the dominant
signal at frequencies below $\sim$5 GHz, and it has become standard practice
to use the low frequency surveys of \cite{\HaslamEtal82} at 408 MHz and
\cite{\ReichReich88} at 1420 MHz to estimate by extrapolation the level of
synchrotron emission at the higher CMB frequencies.  This technique is
complicated by a number of factors. The synchrotron spectral index varies
spatially due to the varying magnetic field strength in the Galaxy
\cite[]{\LawsonEtal87}. It also steepens with frequency due to the increasing
energy losses of the electrons. Although the former can be accounted for by
deducing a spatially variable index from a comparison of the temperature at
each point in the two low frequency surveys, there is no satisfactory
information on the steepening of the spectrum at higher frequencies. As
detailed by \cite{\DaviesEtala96}, techniques that involve using the 408 and
1420 MHz maps are subject to many uncertainties, including errors in the zero
levels of the surveys, scanning errors in the maps, residual point sources and
the basic difficulty of requiring a very large spectral extrapolation (over a
decade in frequency) to characterise useful CMB observing frequencies.
Moreover, the spatial information is limited by the finite resolution of the
surveys: $0.85\dg$ FWHM in the case of the 408 MHz map and $0.6\dg$ FWHM in
the case of the 1420 MHz one. But additional information is available from
existing CMB observing programs.

%Additional information is available from existing CMB observing programs in
%the frequency range 5--15 GHz on $\sim 7\dg$ and $\sim 1\dg$ angular
%scales. Using data at frequencies higher than 408 or 1420 MHz improves the
%spectral leverage. For instance, the Jodrell Bank 5 GHz interferometer has
%been used to make a high sensitivity survey of the northern sky in the
%declination range $35\dg-45\dg$ on a scale of $\sim 2\dg$.  
%A full analysis of
%the data has been undertaken by \cite{\JonesEtal96}, and we consider here the
%results for the quietest region of the sky as identified from the 1420 MHz map
%and from the Green Bank discrete source catalogue \cite[]{\CondonEtal89}.  A
%comparison of the (Jodrell Bank) observed \rms signal levels and those at 1420
%MHz yields a best fit spectral index between 1.4 and 5 GHz: over the 800
%square degrees of sky included in the analysis, the best fit spectral index is
%$\alpha=-0.9 \pm 0.3$.

For instance the Tenerife CMB project \cite[]{\HancockEtal94,\DaviesEtalb95}
has observed a large fraction of the northern sky with high sensitivity at
10.4 and 14.9 GHz (and also at 33 GHz in a smaller region). A joint likelihood
analysis of all of the sky area implies a residual \rms signal of $24\mu$K at
10 GHz and $20 \mu$K at 15 GHz. Assuming that these best fit values are
correct, one can derive spectral indices of $\alpha=-1.4$ between 1.4 and 10.4
GHz and $\alpha=-1.0$ between 1.4 and 14.9 GHz.  These values, which apply on
scales of order $5\dg$, are in agreement with those obtained from other
observations at 5 GHz on $\sim 2\dg$ scales \citep{\JonesEtal96}.
 
At higher frequency, the lack of detectable cross-correlation between the
Haslam data and the \dmrs data leads \cite{\KogutEtala95} to impose an upper
limit of $\alpha =-0.9$ for any extrapolation of the Haslam data in the
millimetre wavelength range at scales larger than $\sim 7\dg$. In view of the
other constraints at lower frequencies, it seems reasonable to assume that
this spectral behaviour also holds at smaller scales.

The spatial power spectrum of the synchrotron emission is not well known and
despite the problems associated with the 408 and 1420~MHz maps it is best
estimated from these. We have computed the power spectrum of the 1420~MHz map
for the sky region discussed above.  The results show that at $\ell \simgt
100$ the power spectrum falls off approximately as $\ell^{-3}$ (i.e. with the
same behaviour as the dust emission).

\subsection{A simple galactic model}
%***********************************

The galactic emission is strongly concentrated toward the galactic plane. But
what is the geography of the galactic fluctuations? Do we expect large
connected patches with low levels of fluctuations?  What is the amplitude of
fluctuations typical of the best half of the sky?

To answer these questions, we created a galactic model valid at scale $\simgt
1\dg$ from spectral extrapolations of spatial templates taken from existing
observations. The 408 MHz full-sky map of Haslam is our template for the {\em
synchrotron} emission, extrapolated to other frequencies with a spectral index
of $\alpha = -0.9$.

We use the \dirbes $240 \mu$m map as a template for the \HI-correlated dust and
the associated free-free emission detected by \citet{\KogutEtal96}. As
mentioned earlier, the global free-free like emission show that the \HI\
correlated emission accounts for most (at least 50\%), and maybe all, of the
global signal. We conservatively assume that there may be a second,
\HI-uncorrelated, component accounting for 5\% of the total dust and 50\% of
the free-free emission. There is of course no known template for such a
component, but we assume that it should have the same ``texture'' than the
correlated component: we simulate it by using again the \dirbes 240\mic\ map,
but North/South inverted (in galactic coordinates). While arbitrary, this
choice preserves the expected latitude dependence of the emission, and the
expected angular scale dependence of the fluctuations. The dust spectral
behaviour is modelled as a single temperature component with $T_d$ = 18K and
$\nu^{2} $ emissivity. We assume that the global free-free like emission will
behave according to $I_\nu \propto \nu^{-0.16}$, and normalise it to give a
total {\em rms} fluctuation level of 7.3 $\mu$K at 53 GHz.
% ------------------------   6.2=5.3*2/2.45*sqrt 2 ---------------------------
%\com{why : 6.2=5.3*2/2.45*sqrt 2??? I understand the *sqrt(2) but /2.45?
%scaling l-3 for  7deg -> 1deg = facteur 2.65}
Since the weakest emissions in \dirbes are at the 1 MJy/sr level (and at the
10K level for Haslam), we immediately see that the corresponding fluctuations
should be at the level of a few $\mu$K around 100 GHz.

Using this model, we can now create full sky maps of the total galactic
emission converted to an equivalent thermodynamical temperature at any chosen
frequency. Note that at this resolution, extragalactic point sources and the
SZ effect should contribute negligible signal.  In order to obtain {\em local}
estimates of the level of fluctuations, we have computed their {\em rms}
amplitude ($\sigma_T$), over square patches of 3 degrees on each side
(i.e. containing about 11 beams of 1 degree FWHM). Thus, this estimator only
retains perturbations at angular scales roughly between 1 and 3 degrees,
equivalent to restricting the contribution to the variance to a range of
angular modes, with $\ell$ between $\sim$ 20 and 60. Figure~\ref{fig:varm}
shows an example of such a map at $\lambda =2.4$ mm (125 GHz). The two darkest
shades of grey ($\sigma_T \simgt 10$\microK) delimit the area where galactic
fluctuations would be comparable to those of a \cobe-normalised CDM
model\footnote{The parameters of this model are $\Omega_b =0.05$ for the
baryonic abundance, $\Omega_c = 0.95$ for that of CDM, $\Lambda = 0$, $H_0 =
50$\, Km/s/Mpc, no reionisation, and only scalar modes}; they only represent
20\% of the total sky area at this scale. This map provides direct graphical
evidence that a large fraction of the sky is quite ``clean'' around the degree
scale \cite[]{\BouchetEtalc95,\BouchetEtalb95}.

%++++++++++++++++++++++++++++++++++++++++++++++++++++++++++++++++++++++++++++++
% Maps of fluctuation levels
%++++++++++++++++++++++++++++++++++++++++++++++++++++++++++++++++++++++++++++++
\begin{figure}[htbp] \medskip \centering \centerline{ 
%\vbox{
%\psfig{file=aitoff_2.41mm.ps,angle=90,width=\textwidth} \\ \vspace{-3truecm}
%\psfig{file=aitoff_fluct_cdm.ps,angle=90,width=\textwidth}
\vspace{10cm}
maps available in the full version at
$ftp://ftp.iap.fr/pub/from\_users/bouchet/wiener7.ps.gz$
\vspace{2cm}
%} 
}
\caption[]{Top: map of the temperature fluctuations due to the galactic
foregrounds at $\lambda =2.4$ mm. The fluctuation levels, denoted by 5 shades
of grey, are obtained by computing the local variance over square patches with
3 degrees on a side (see text).  From the lightest (small $\sigma_T$) to the
darkest regions close to the galactic plane, the levels correspond to
$\sigma_T < 1.0, 3.16, 10.0, 31.1 \mu$K, and $\sigma_T > 31.1 \mu$K.  As one
can see on a map produced with the same method on a CMB fluctuations maps (in
a standard CDM model), one would have $\sigma_T \simgt 10$\microK, i.e. the
galactic fluctuations should be at least 10 times smaller than the CMB ones
(in that theory) in $\sim$ 58\% of the sky (at 2.4 mm).}
\label{fig:varm}
\end{figure}
%++++++++++++++++++++++++++++++++++++++++++++++++++++++++++++++++++++++++++++++

%++++++++++++++++++++++++++++++++++++++++++++++++++++++++++++++++++++++++++++++
% Cumulative Distribution Functions
%++++++++++++++++++++++++++++++++++++++++++++++++++++++++++++++++++++++++++++++
\begin{figure*}[htbp]
\hbox{
\hskip -20pt \psfig{figure=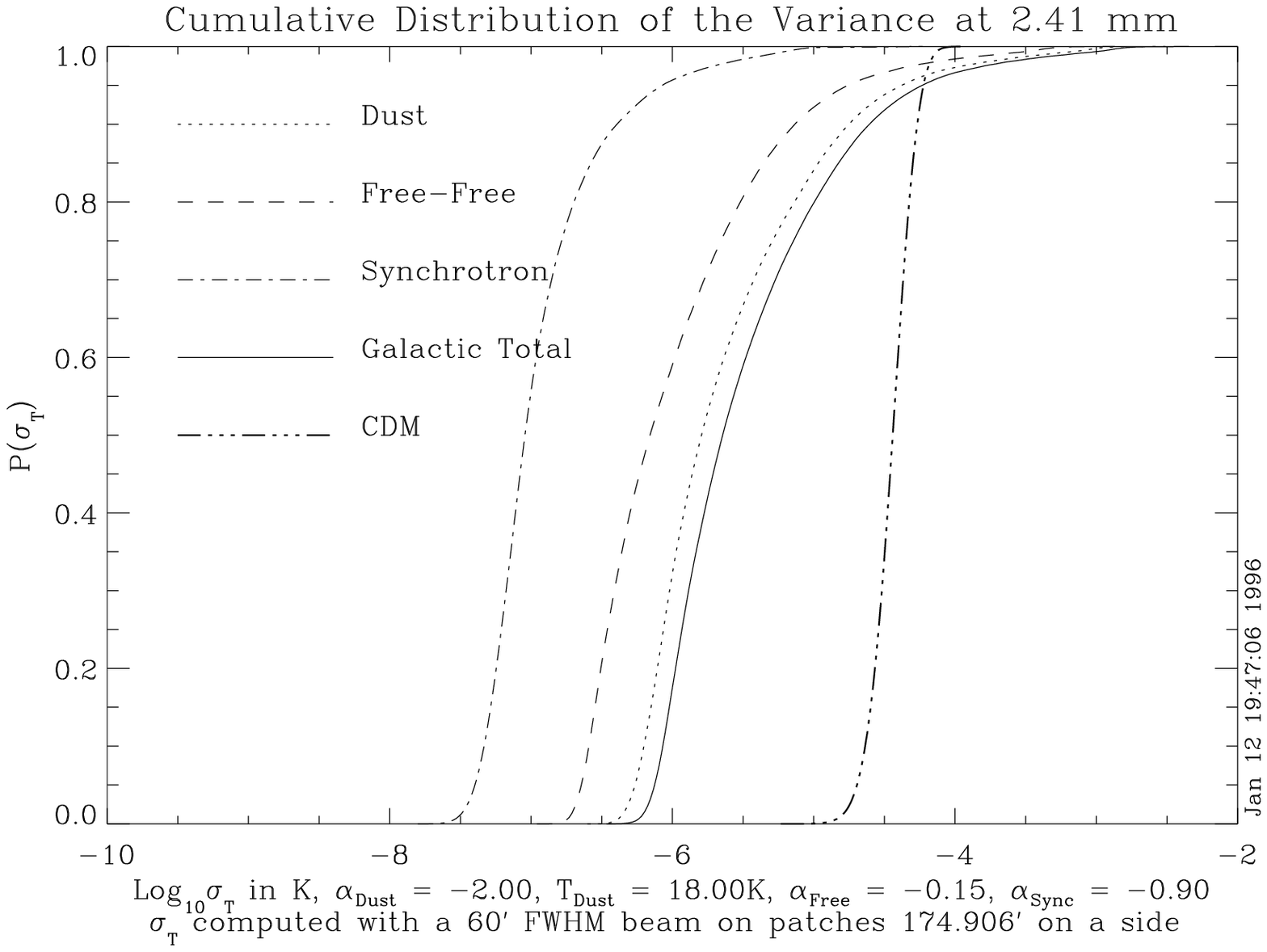,width=0.525\textwidth}
\hskip -10pt \psfig{figure=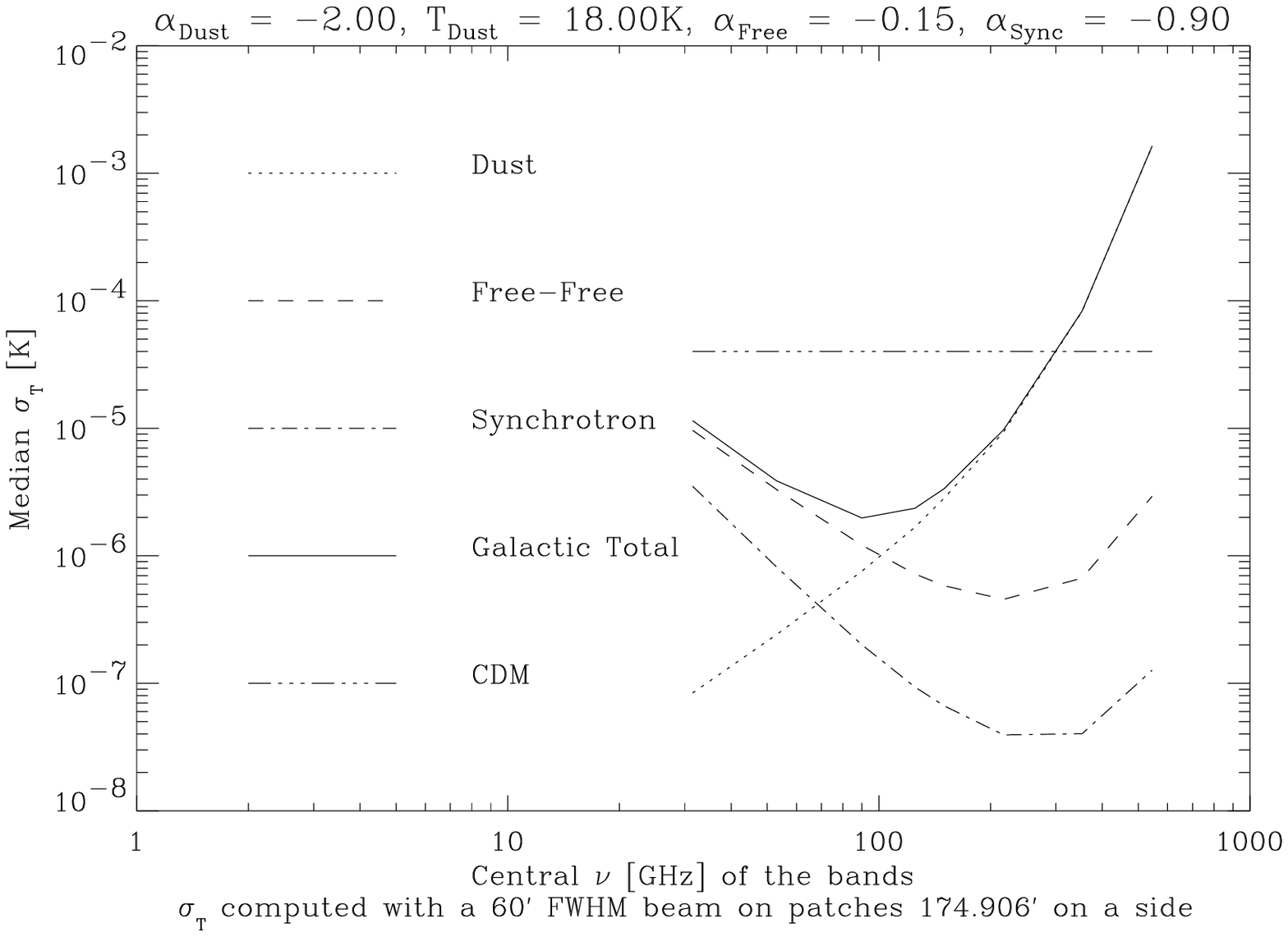,width=0.525\textwidth} }
%\vbox{\vskip 10cm}
\caption[]{a) Cumulative distribution functions of the galactic foreground
fluctuations at $\lambda=2.4$~mm (125~GHz) in 3 degree patches for a 1 degree
FWHM Gaussian beam. It gives directly the the fraction of the sky in which the
rms fluctuation level is lower than a certain value. The CDM curves gives for
comparison the expected range of fluctuations for a \cobes normalised CDM
model. b) Median fluctuation values (i.e. corresponding to 50\%\ of the sky)
for the galactic components around the degree scale as a function of
frequency. The galactic fluctuations have a clear minimum around 100~GHz.}
\label{fig:cpdf-med}
\end{figure*}
%++++++++++++++++++++++++++++++++++++++++++++++++++++++++++++++++++++++++++++++

Analysing figure~\ref{fig:varm} more quantitatively,
figure~\ref{fig:cpdf-med}.a shows the fraction of the sky for which the {\it
rms} fluctuations of each of the foregrounds and of their total contribution
are less than a given value at $\lambda= 2.4$ mm. These curves are steep for
low fluctuation levels; the best half of the sky (median value) is only a
factor of $\sim 4$ noisier than the best regions, but a factor $\sim 10$ below
the CDM level. The high level ``plateaus'' correspond to low galactic latitude
regions. Indeed the maps tell us that that the ``clean'' sky essentially
corresponds to the areas at high galactic latitudes, excepting only a few hot
spots like the Magellanic clouds.  Figure~\ref{fig:cpdf-med}.b, which plots
the median fluctuation values as a function of frequency, shows a clear
minimum around 100 GHz, at a level lower than 10 \% of the expected level of
the CMB anisotropies. Thus measurements restricted to that frequency would
already be sufficient to achieve better than 10\% accuracy around the degree
scale (provided a few high latitude ``hot or cold spots'' are flagged out with
higher and lower frequency channels).

%++++++++++++++++++++++++++++++++++++++++++++++++++++++++++++++++++++++++++++++
% Plot of galactic model at 7 deg versus Kogut
%++++++++++++++++++++++++++++++++++++++++++++++++++++++++++++++++++++++++++++++
\begin{figure}[htbp]
\centering \centerline{
\hskip -15pt \psfig{file=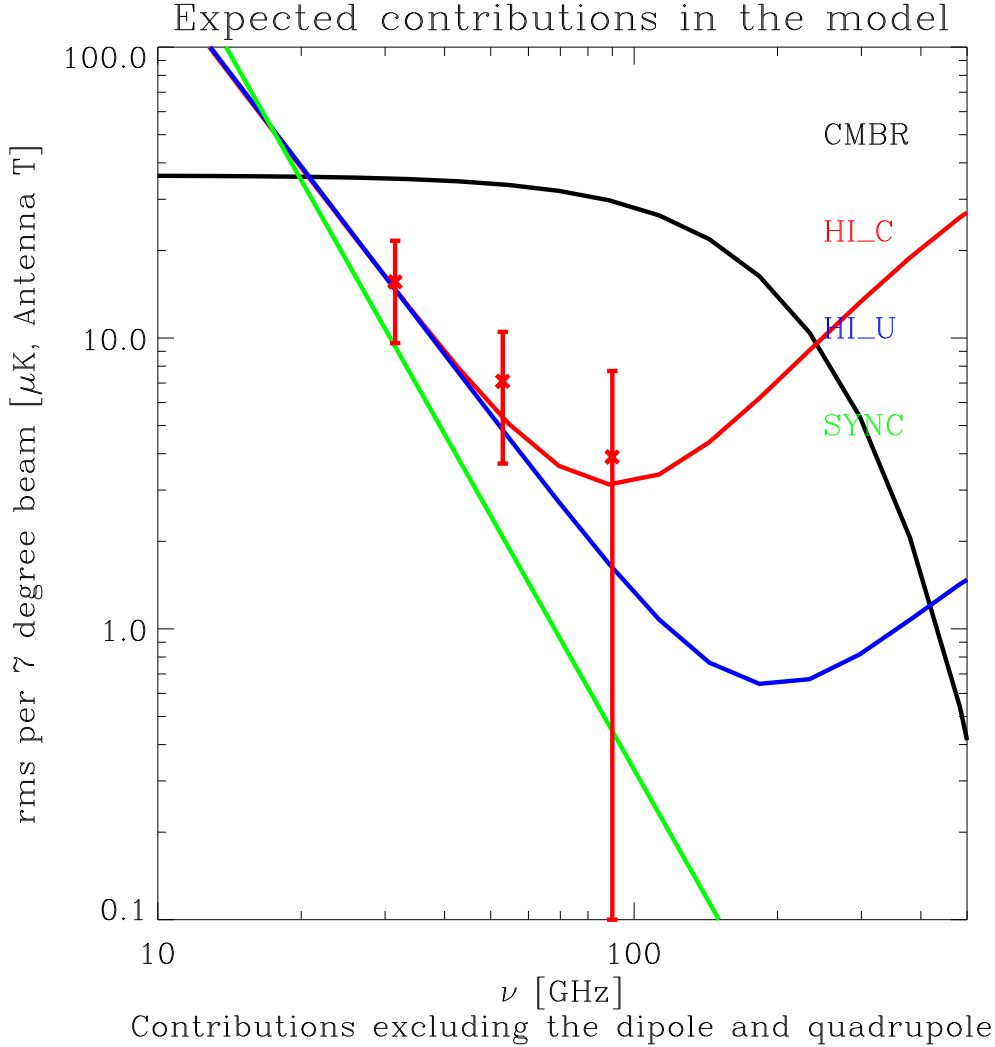,width=\textwidth} }
\caption[ ]{Standard deviation of the fluctuations in the synchrotron, \HI
correlated and uncorrelated emissions in our galactic model, as compared to
the CMB fluctuations whose level was set by the DMR measurement, and the data
points for the DMR-DIRBE correlated component found by Kogut et al. 95 (see
text). A Gaussian beam of 7 degree FWHM was used.}
\label{fig:kogut}
\end{figure}
%++++++++++++++++++++++++++++++++++++++++++++++++++++++++++++++++++++++++++++++

As was shown above, the angular power spectra $C(l)$ of the galactic
components all decrease strongly with $\ell$, approximately as $C(\ell)
\propto \ell^{-3}$. Smaller angular scales thus bring increasingly small
contributions per logarithmic interval of $\ell$ to the variance,
$\ell(\ell+1) C(\ell)$: the galactic sky get smoother on smaller angular
scales. We set the normalisation constants of the spectra to obtain a level of
fluctuations representative of the best half of the sky. In practice, this was
achieved by generating an artificial map with $C(\ell) = \ell^{-3}$. We then
measured its $\sigma_T$ using our previous procedure of measuring the variance
in 3 degree patches after convolving with a one degree beam. This gives the
required normalisation factor of the power spectra by comparisons with the
median values plotted in fig.\ref{fig:cpdf-med}.b. We find that
\begin{equation}
    \ell\, C_\ell^{1/2} = c_X\, \ell^{-1/2}\ \mu{\mathrm K} ,
\label{pow_gals}
\end{equation}
with $c_X$ given by $c_{sync}=2.1$, $c_{free}=13.7$, and $c_{dust}=13.5$ at
100 GHz. In addition, one has $c_{HI-U}=8.5$ and $c_{HI-C}=20.6$ for the
Dust+Free-free components, uncorrelated and correlated (respectively) with
\HI.

Of course, by their very construction, these normalisations are only
appropriate for intermediate and small scales, about $\ell \simgt 10$. The
corresponding fluctuation levels at 3.0 mm and 1.4 mm are shown in
figure~\ref{fig:contribs}. By using these normalisations and our spectral
model, we can compute the {\em rms} per 7 degree beam at every frequency and
check that the \HI-correlated component indeed provides a good fit with the
results of the \cite{\KogutEtal96} analysis, as is shown in
figure~\ref{fig:kogut}.

%******************************************************************************
\section{Extragalactic foregrounds\label{sec:sources}}
%******************************************************************************

Many detailed studies have been devoted to effects that may blur the
primordial signature of the CMB anisotropies. Gravitational lensing by mass
concentrations along the light ray paths may for instance alter the detailed
map patterns and add a stochastic component. Or photons passing through a fast
evolving potential well might be redshifted (Rees-Sciama effect). Secondary
fluctuations might be generated during a reionisation phase of the
Universe. For all this processes, the answer \citep{ReesSciama68,
\OstrikerVishniac86, \Seljaka96, JaffeKamion98} is that their impact is quite
small at scales corresponding to $\ell \simeq 1/\theta \simlt 1000$, and can
easily be accounted for at the analysis stage of CMB maps.  Further
fluctuations may also be imprinted, in case of a strongly inhomogeneous
reionisation; see \eg\ \cite{\AghanimEtalq96} for further discussion. We shall
ignore these effects in the following.

\subsection{Infrared and radio sources background\label{sec:irrad_sources} }
%****************************************************************

Here we attempt to evaluate what might be the contribution from the Far
Infrared Background produced by the line of sight accumulation of
extragalactic sources, when seen at the much higher resolution foreseen for
future CMB experiments.

\begin{figure}[htbp] \centering \centerline{ 
\psfig{file=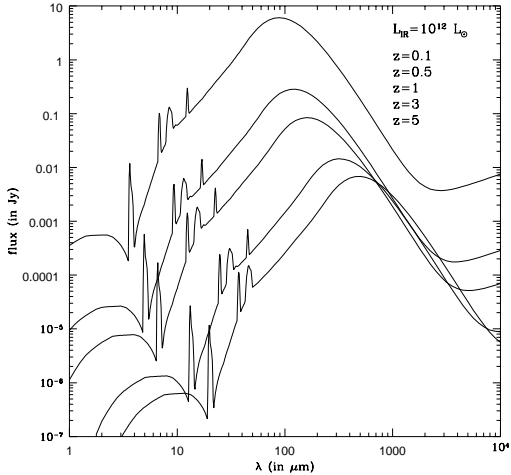, width=0.5\textwidth} }
\caption[]{Effect of redshifting a template starburst galaxy spectrum
(left). Longward of $\sim 300$\mic, the flux of such galaxies are similar at
all $z\simgt 0.5$. }
\label{fig:galspec}
\end{figure}

Given the steepness of rest-frame galaxy spectra longward of $\sim 100$\mic,
predictions in this wavelength range are rather sensitive to the assumed
high-$z$ history. Indeed, as can be seen from figure~\ref{fig:galspec}.a, a
starburst galaxy at $z=5$ might be more luminous than its $z=0.5$ counterpart
because the redshifting of the spectrum can bring more power at a given
observing frequency than the cosmological dimming, at least in the $\lambda
\simgt $800\mic\ range \cite[\eg\ ][]{\BlainLongair93} (precisely the range
explored by the \hfis instrument aboard \planck). This ``negative
K-correction'' means that i) predictions might be relatively sensitive to the
redshift history of galaxy formation ii) variations of the observing frequency
should imply a partial decorrelation of the galaxy contribution, \ie one
cannot {\em stricto sensu} describe this contribution by spatial properties
(\ie a spatial template) and a frequency spectrum (but see below). This is
compounded by the fact that this part of the spectrum is not well known
observationally, even at $z \simeq 0$, and very little is known about the
redshift distribution of faint infrared sources.

%++++++++++++++++++++++++++++++++++++++++++++++++++++++++++++++++++++++++++++++
% Redshifting an IR galaxy spectrum
%++++++++++++++++++++++++++++++++++++++++++++++++++++++++++++++++++++++++++++++
%++++++++++++++++++++++++++++++++++++++++++++++++++++++++++++++++++++++++++++++

Estimates of the contribution of radio-sources and infrared galaxies to the
anisotropies of the microwave sky have till recently relied on extrapolations
from redshift $z =0$ all the way to a (large) assumed galaxy formation onset
redshift. Given the uncertainties described above, this makes the reliability
of the predictions of this type of modelling hard to assess in the $\nu >
100$\,GHz range.  On the other hand, predictions of galaxy formation models in
the UV and optical bands have received a lot of attention in the last few
years; they rely on substantially more involved semi-analytical models of
galaxy formation which provide a physical basis to the redshift history of
galaxies \citep{\WhiteFrenk91}. In short, one starts from a matter power
spectrum (typically a standard, \cobe-normalised CDM), and estimates the
number of dark matter halos as a function of their mass at any redshift (\eg\
using the Press-Schechter approach). Standard cooling rates are used to
estimate the amount of baryonic material that forms stars. The stellar energy
release is obtained from a library of stellar evolutionary tracks and spectra;
computations may be then performed to compare in detail with
observations. This approach, despite some difficulties, has been rather
successful, and many new observations fit naturally in this framework.
%\cite[\eg]{kauffman}.

This type of physical modelling has recently been extended to the far-infrared
range by \citet{\GuiderdoniEtal98}.  Given the stellar energy release,
\citet{\GuiderdoniEtal98} estimated the fraction reradiated in the microwave
range using a simple geometrical model, which yields an infrared luminosity
function at all redshifts. This, together with an assignment of synthetic
spectra to a given infrared luminosity, allows predictions of the numbers and
fluxes of faint galaxies at any frequency. To constrain the fraction of
heavily reddened objects as a function of redshift, they used the $z\sim 0$
\irass data at 60\mic\ \citep{\LonsdaleEtal90} to normalise their model, and
they further assumed that the isotropic background discovered by
\citet{\PugetEtal96} is indeed the long-sought Cosmological Infrared
Background (CIBR) due to the accumulated light of infrared sources.

\begin{figure}[htbp] \vbox{
\psfig{file=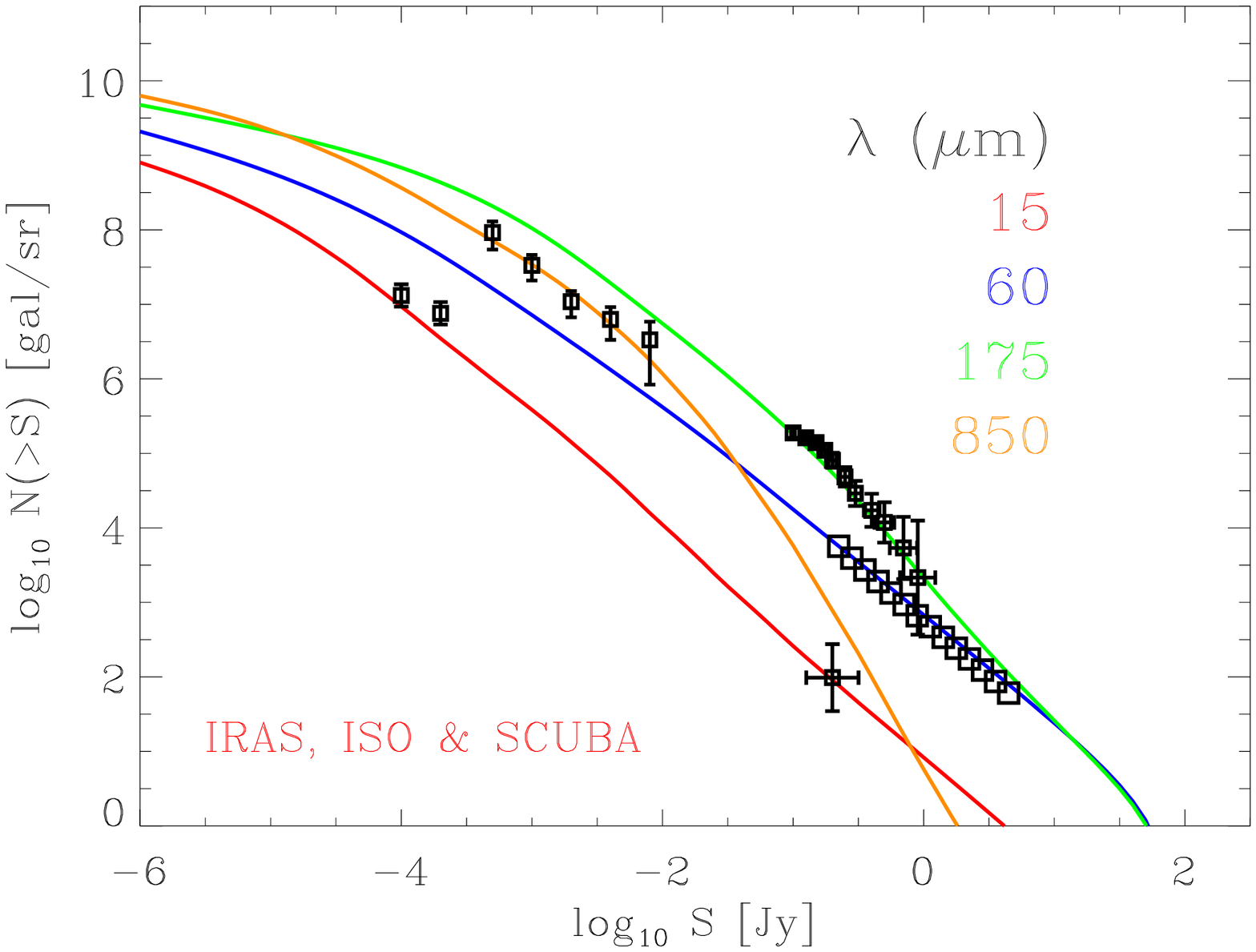, angle=90, width=0.5\textwidth,
height=0.4\textwidth}
\psfig{figure=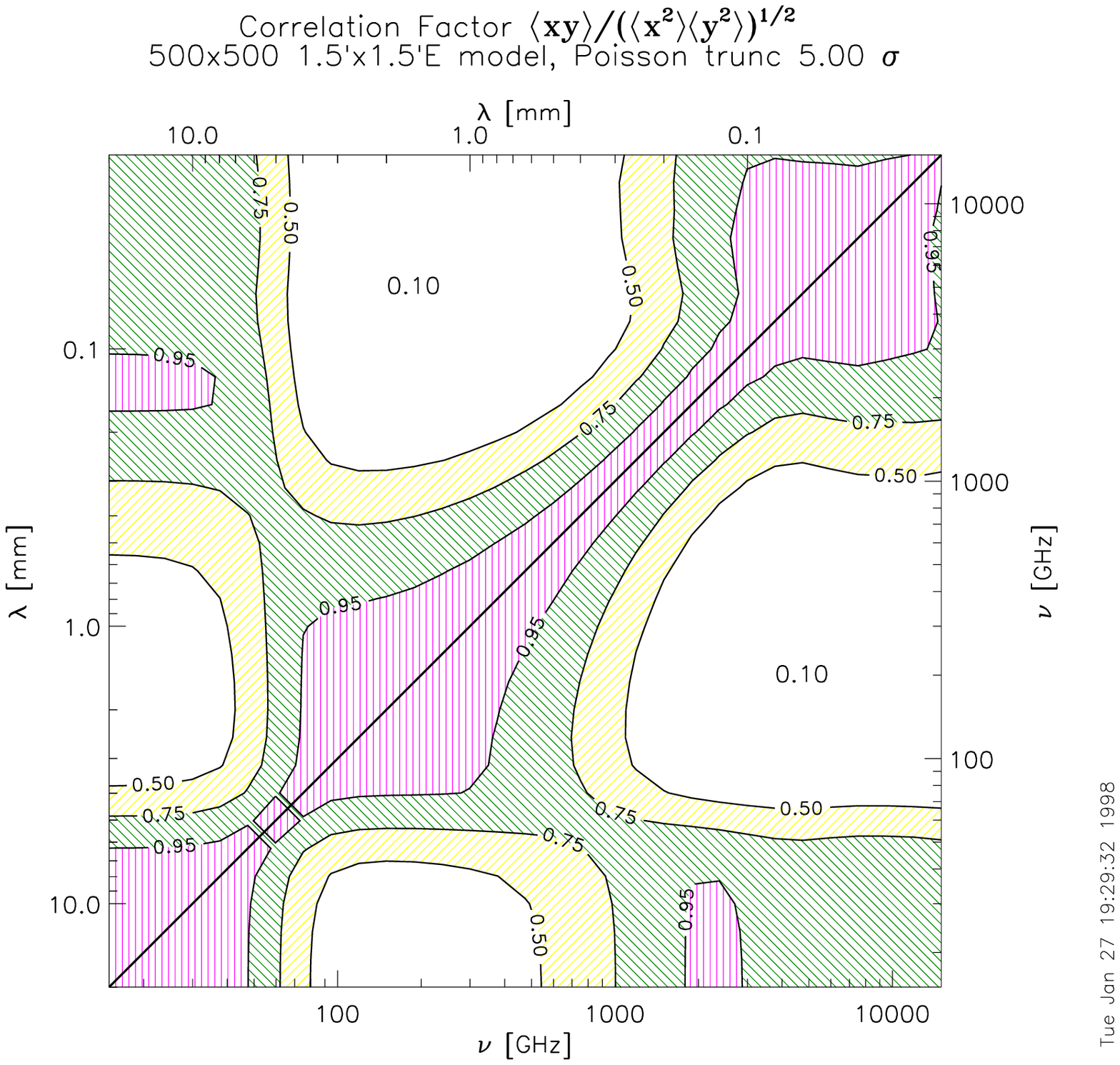,width=0.425\textwidth}
}
\caption{\small a) Number counts prediction from the E model of
\cite{GuiderEtal97} \& the available data (see text). b) Cross-correlation
coefficients between maps based on the E model.}
\label{fig:galflux_pix}
\end{figure}

\citet{GuiderEtal97} selected a small set of possible redshift histories
satisfying this (integral) constraint, and predicted that ISO observations at
175\mic\ could help breaking the remaining degeneracy in the model. Such
observations were very recently done
\citep{KawaraEtal97, \ClemensEtal98, 9902122} and the
number of sources found is best described by one model of this family
(their model E) 
corresponding to a fairly large fraction of very obscured objects. As shown by
figure~\ref{fig:galflux_pix}.a, the ISO Hubble deep field data at 15\mic\
\citep{\OliverEtal97} also agree with the predictions of this model. Even the
first {\sc SCUBA} determination \citep{1997ApJ...490L...5S, 9812412}
at 850\mic\ seems 
to be reasonably fitted by model E. In short, this model is successful in
predicting the latest source counts over a broad range of frequencies. This
gives us confidence that the contribution of infrared galaxies to CMB
measurements performed in the same frequency range (that of the \hfi, $\nu >
100$\,GHz) can be assessed with decent accuracy. The situation is different
for the redshift distribution of the sources which is more sensitive than the
number counts to the details of the evolutionary scenarios. The current
observational constraints are so far limited to the spectroscopic follow-up of
IRAS sources at 60 $\mu$m which essentially highlights the IR properties of
galaxies in the local universe. It is only when redshift follow-up of these
new far-infrared catalogs will have been completed that new generations of
models can be developed and refined assessment of the infrared source
contamination can be made.

In what follows, we shall assume that resolved sources have been removed, and
we focus on the spatial and spectral properties of the remaining {\em
unresolved} background. That can be done only by assuming a specific
instrumental configuration giving the level of detector noise and a
geometrical beam for each channel. Indeed the variance of the shot noise
contributed by $N$ sources of flux $S$ randomly distributed on the sky is
$\sigma^2 = N S$ and the corresponding power spectrum is $C(\ell) = N/(4\pi)
S^2$. For a population of sources of flux distribution per solid angle
$dN/dS/d\Omega$ the power spectrum is, after removing all sources brighter
than $S_c$
\begin{eqnarray}
	C(\ell) = \frac{\sigma_{\rm conf}^2}{\Omega_{\rm beam}} = \int_0^{S_c}
        \frac{dN}{d\Omega dS} S^2 dS .
\end{eqnarray}
If the cut $S_c$ is defined as $S_c = q \left(\sigma^2_{\rm conf} +
\sigma^2_{\rm dust} + \sigma^2_{\rm instrum} + \sigma^2_{\rm CMB}
\ldots\right)^{1/2}$ with $q$ fixed to 5, this formula allows to derive
iteratively the confusion limit $\sigma_{\rm conf}$. This is the standard
deviations of the fluctuations of the unresolved galaxy background, once all
sources with $S > 5 \sigma_{\rm tot}$ (i.e. including the other sources of
fluctuations, from cirrus, $\sigma_{\rm dust}$, detector noise $\sigma_{\rm
instrum}$, the CMB, $\sigma^2_{\rm CMB}$) have been removed.

%++++++++++++++++++++++++++++++++++++++++++++++++++++++++++++++++++++++++++++++
% predicted source counts on all the sky
%++++++++++++++++++++++++++++++++++++++++++++++++++++++++++++++++++++++++++++++
\begin{table}[htb] \begin{center} \scalebox{1.2}[0.9]{ 
\begin{tabular}{|c||r|r|r|r|r|r|r|}
\hline \multicolumn{8}{|c|}{NUMBER COUNTS WITH \planck\ \hfi\ } \\ \hline
\hline $\nu$ & $\theta_{FWHM}$ & $\sigma_{ins}$ & $\sigma_{cir}$ &
$\sigma_{CMB}$ & $\sigma_{conf}$ & $\sigma_{tot}$ & $N(>5\,\sigma_{tot})$ \\
GHz & arcmin & mJy & mJy & mJy & mJy & mJy & sr$^{-1}$ \\ (1) & (2) & (3) &
(4) & (5) & (6) & (7) & (8) \\ \hline \hline 857 & 5 & 43.3 & 64 & 0.1 & 146 &
165 & 954 \\ 545 & 5 & 43.8 & 22 & 3.4 & 93 & 105 & 515 \\ 353 & 5 & 19.4 &
5.7 & 17 & 45 & 53 & 398 \\ 217 & 5.5 & 11.5 & 1.7 & 34 & 17 & 40 & 31 \\ 143
& 8.0 & 8.3 & 1.4 & 57 & 9.2 & 58 & 0.36 \\ 100 & 10.7 & 8.3 & 0.8 & 63 & 3.8
& 64 & 0.17 \\ \hline \end{tabular} } \end{center}
\caption{{\small Theoretical estimates from model E of
\cite{GuiderEtal97}. (1) \hfi\ wavebands central frequency in GHz.  (2) Beam
full width half maximum in arcmin.  (3) $1\,\sigma$ instrumental noise for 14
month nominal mission.  (4) $1\,\sigma$ fluctuations due to cirrus at $N_{HI}
= 1.3~10^{20}$ cm$^{-2}$ (level of the cleanest 10 \% of the sky). The
fluctuations have been estimated following Gautier {\it et al.} (1992) with
$P(k) \propto k^{-2.9}$ and $P_{0, 100 \mu m} = 1.4~10^{-12} B_{0, 100 \mu
m}^3$.  (5) $1\,\sigma$ CMB fluctuations for $\Delta T /T =10^{-5}$.  (6)
$1\,\sigma$ confusion limit due to FIR sources in beam $\Omega \equiv
\theta_{FWHM}^2$, defined by $\sigma_{conf} = (\int_0^{S_{lim}} S^2 (dN/dS)
dS\ \Omega)^{1/2}$.  The values $\sigma_{conf}$ and $S_{lim}=q\,\sigma_{tot}$
have been estimated iteratively with $q=5$.  (7) $\sigma_{tot}=(\sigma_{ins}^2
+ \sigma_{conf}^2 + \sigma_{cir}^2 + \sigma_{CMB}^2)^{1/2}$. Here
$\sigma_{cir}$ is for $N_{HI} = 1.3~10^{20}$ cm$^{-2}$. (8) Surface density of
FIR sources for $S_{lim}=5\,\sigma_{tot}$. }}
\label{tab:counts}
%\vspace{-12pt}
\end{table}
%++++++++++++++++++++++++++++++++++++++++++++++++++++++++++++++++++++++++++++++
%  used by Hivon
%  freq  sigma_conf(10%)  sigma_conf(50%) 

%HFI
%  857.0  145.9            153.9
%  545.0   93.3             95.0
%  353.0   45.4             45.6
%  217.0   17.1             17.1
%  143.0    9.2              9.2
%  100.0    3.8              3.8
%LFI
%  100.0    3.5              3.5
%   70.0    2.2              2.2
%   44.0    2.2              2.2
%   30.0    3.6              3.6

Table~\ref{tab:counts} summarises the E-model results for the \hfis instrument
\cite{Guiderpriv}. The main feature is the very high level of the confusion
limit, which translates in modest number of detected sources. One should
realize though that this estimate is very pessimistic since it assumes a very
na\"{\i}ve source removal scheme by a simple thresholding. In practice one
would at least use a compensated filter (\eg by doing aperture photometry)
which removes the contributions to the variance of all long
wavelength\footnote{A compensated filter (whose integral is naught)
essentially nulls the contribution of fast decreasing power spectra
contributors, like dust, whose contribution to the variance is concentrated at
large scales. And it also decreases the amount of white noise from the
detector and the unresolved sources themselves, since then sources have to
stand out above a reduced threshold.}.  A much larger number of sources would
then be removed since the counts are very steep (as a matter of example,
\cite{1998MNRAS.295..877G} predict with the same model more than 120 000
sources at 350 $\mu$m (860GHz) at flux $>$ 100 mJy, a detection threshold
suggested by the recent work of \cite{HobsonEtal98b}). One should thus regard
the confusion level derived above as an upper limit which very likely exceeds
the real residual unresolved background from sources.

The spectral dependence of the confusion limit can be approximately modelled
as that of a modified black body with an emissivity $\propto \nu^{0.70}$ and a
temperature of 13.8 K. A more precise fit (at the $~$ 20\% level) yields at
$\nu >$ 100 GHz for the equivalent temperature fluctuation power spectrum
(using relative bandwidths of 0.25) is given by
\begin{equation}
	\ell C(\ell)^{1/2} \simeq \frac{7.1\ 10^{-9}}{e^{x/2.53}-1} (1 -
	\frac{0.16}{x^4}) \frac{\sinh^2 x}{x^{0.3}}\ \ell \ {\rm [K]} ,
	\label{eq:hivon1}
\end{equation}
with $x=h\nu/2kT_0 = \nu/ (113.6 {\rm GHz})$. This gives a value of
$C(\ell)^{1/2} = 0.005$ \microK\ at 100GHz. At lower frequencies ($\nu <$ 100
GHz), this model yields instead
\begin{equation}
        \ell C(\ell)^{1/2} \simeq 6.3\ 10^{-9} \left( 0.8 - 2.5x +
        3.38x^2 \right) \frac{\sinh^2 x}{x^{4}}\ \ell \ {\rm [K]}
        \label{eq:hivon2}
\end{equation}

%++++++++++++++++++++++++++++++++++++++++++++++++++++++++++++++++++++++++++++++
% Plots of galactic pixels spectra
%++++++++++++++++++++++++++++++++++++++++++++++++++++++++++++++++++++++++++++++
%++++++++++++++++++++++++++++++++++++++++++++++++++++++++++++++++++++++++++++++

Figure~\ref{fig:galspec} shows that the relative fluxes in different
observing bands can vary depending on the redshift of the object. Conversely,
different bands do not weight equally different redshift intervals. The
previous analysis thus does not tell us whether the fluctuation pattern at a
given frequency is well correlated with the pattern at another frequency. Of
course, the answer will depend on the resolution of the maps and the noise
level. To answer this question for the \cite{aao}, maps of the galaxy
contribution were generated at 30 different frequencies, as in
\cite{countsDwek} and \cite{HivonEtal96}, but for model E. As shown in
figure~\ref{fig:galflux_pix}.b, the
cross-correlation coefficients of the maps (once the 5$\,\sigma_{tot}$ sources
have been removed) is better than 0.95 in the 100-350 GHz range, better than
0.75 in the 350-850GHz, and it is still $\simgt 0.60$ in the full 100-1000 GHz
range. This indicates that one can treat the IR background from unresolved
sources reasonably well as just another template to be extracted from the
data, with a fairly well defined spectral behaviour\footnote{Note though that
the model may not do full justice to the diversity of spectral shapes of the
contributing IR galaxies; if this turns out to be the case, we could for
instance treat this contribution as an additional noise contribution to the
detectors, as was conservatively assumed in the \plancks Phase A study.}, at
least in the range probed by the \hfi.

While the model above might well be the best available for bounding the
infrared sources properties, it does not take into account the contribution
from low-frequency point sources like blazars, radio-sources, etc\ldots which
are important at the lower frequencies probed by \maps and the \lfis. Using
only that model would under-estimate the source contribution at low
frequencies. On the other hand, \citet{\ToffolattiEtal97} made detailed
predictions in the \lfis range which give
\begin{equation}
        \ell C(\ell)^{1/2} \simeq \frac{ 5.7 \sinh^2(\nu/113.6)}{
        (\nu/1.5)^{(4.75- 0.185 \log(\nu/1.5) )} }\ \ell\ {\rm [K]}.
        \label{eq:toffo}
\end{equation} 
This yields $C(\ell)^{1/2}$=0.02 \microK\ at 100GHz. We shall also use that
prediction when analysing the expected performances of \map\footnote{Although
in that case the remaining unresolved background should be somewhat higher due
to the lower sensitivity and angular resolution of \map.}. Here again, we
assume that this unresolved background has well-defined spectral properties.

In order to obtain a pessimistic estimate of the total contribution from
sources, we shall consider in the following {\em two} uncorrelated backgrounds
from sources, the one contributed by IR sources, as described by the
equations~\eqref{eq:hivon1} \& \eqref{eq:hivon2}, and the one contributed by
radio-sources, as described by equation\eqref{eq:toffo}. The corresponding
levels are compared with the other sources of fluctuations in
figure~\ref{fig:contribs} and they are quite low (at least at $\ell \simlt
1000$) as compared to the expected fluctuations from the CMB or from the
Galaxy.

%\com{quelque chose sur les correlations...}

\subsection{Clusters of galaxies\label{sec:inventory_sz} }
%********************************************************

Hot ionised gas along the line of sight generate distortions of the CMB
fluctuations when the incoming photons scatter off electrons and get a shift
in frequency. This translates in an apparent temperature decrement in the low
frequency (Rayleigh-Jeans) side of the spectrum, and a temperature excess in
the Wien side. This is called the Sunyaev-Zeldovich (hereafter SZ) effect
\citep{ZeldovichSunyaev69}. The intensity variation of the CMB, $\Delta
I_{\nu}/I_{\nu} = y \times f(x)$ is controlled by the Compton parameter
\begin{equation}
        y = \frac{k\sigma_T}{m_ec^2}\int T_e(l)n_e(l) dl , \label{eq:y}
\end{equation}
where $T_e$ and $n_e$ stand for the electron temperature and density. The
spectral form factor in the non relativistic limit depends only on the
adimensionnal frequency $x=h\nu/kT_{CMB}$ according to
\begin{equation}
        f(x)=\frac{xe^x}{(e^x-1)} \left[ x\left( \frac{e^x+1}{e^x-1} \right) -
        4 \right] .  \label{eq:ynu}
\end{equation}
During the \plancks phase A study, \cite{\AghanimEtalq96} devised a model for
generating maps of the Sunyaev-Zeldovich effect (both thermal and kinetic) and
analysed the capabilities of \plancks in detecting clusters of galaxies
\cite[][see their Figure 1.6]{\AghanimEtala97}. Since then, the model was
improved by using better estimates of the counts (and of the peculiar
velocities), and generalised to encompass various cosmological models. The
counts were derived from the Press-Schechter mass function
\citep{PressSchechter74}, normalised using the X-ray temperature distribution
function derived from \citet{\HenryArnaud91} data as in
\citet{\VianaLiddle96}.  The following power spectrum of the fluctuations due
to the Sunyaev-Zeldovich thermal (hereafter SZ) effect from clusters of
galaxies was evaluated by analysing maps of the Compton parameter $y$
generated with these counts.

%++++++++++++++++++++++++++++++++++++++++++++++++++++++++++++++++++++++++++++++
\begin{figure}[htbp] \centering \centerline{ 
\psfig{file=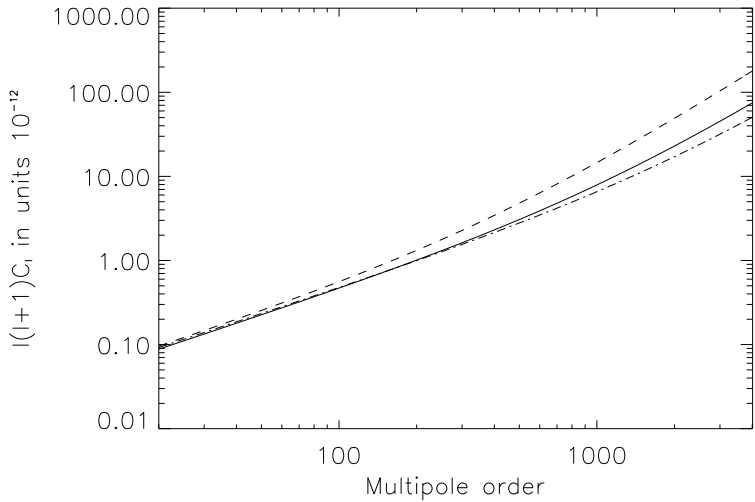, width=0.75\textwidth} }
\caption[]{Power spectra of the Sunyaev-Zeldovich thermal effect for different
cosmological models: standard CDM (solid line), open CDM (dashed line) and
lambda CDM (dotted line) are shown.}
\label{fig:pow_SZ}
\end{figure}
%++++++++++++++++++++++++++++++++++++++++++++++++++++++++++++++++++++++++++++++

For the standard CDM model, it was found that the $y$ fluctuations spectrum is
well-fitted in the range $20 < \ell < 4000$ by
\begin{equation}
        \ell(\ell+1)\, C_\ell= a_{ysz}\, \ell \left[ 1 + b_{ysz}\,\ell\right]
        , \label{eq:pow_sz}
\end{equation}
with $a_{ysz} = 4.3 \times 10^{-15}$ and $b_{ysz} = 8.4 \times 10^{-4}$. Of
course, the values of the fitting parameters in equation\eqref{eq:pow_sz}
depend on the assumed cosmological scenario. For an open model
($\Omega_0=0.3$), one finds instead $a_{ysz} = 4.6 \times 10^{-15}$ and
$b_{ysz} = 2.2 \times 10^{-3}$. These are quite small differences, as
illustrated by figure~\ref{fig:pow_SZ}. This is natural since most of the
contributions come from relatively low redshifts while the counts have
precisely been normalised via $z\sim 0$
observations. Equation\eqref{eq:pow_sz} for standard CDM yields
\begin{equation}
        \ell\, C_\ell^{1/2}= 0.27\,\left[\ell\, (1 + 8.4 \times
        10^{-4}\,\ell)\right]^{1/2}\,\mu{\mathrm K}
\end{equation}
for the temperature fluctuation spectrum at 100 GHz.

On small angular scales (large $l$), the power spectrum of the SZ thermal
effect exhibits the characteristic dependence of the white noise. This arises
because at these scales the dominant signal comes from the point--like
unresolved clusters. On large scales (small $l$), the contribution to the
power comes from the superposition of a background of unresolved structures
and extended structures. The transition between the two regimes occurs for
$l\simeq1/b_{sz}$ that is when the angular scale is close to the pixel size of
the simulated maps. One should note though that this approach neglects the
effect of the incoherent superposition of lower density ionised regions like
filaments. While the contributions in the linear regime are easy to compute
\citep{1995ApJ...442....1P}, the non-linear contributions must be evaluated
using numerical techniques \citep{1995ApJ...442....1P,
1996AAS...189.1303P}. While the modelisation above is quite sufficient for the
purposes of this paper, detailed simulations of the component separation
should benefit from simulated maps of the SZ effect with more realistic low
contrast patterns.

The kinetic SZ effect due to the Doppler effect from clusters in motion along
the line of sight (hereafter {\em l.o.s}) cannot, of course, be spectrally
distinguished from the primary fluctuations. This additional contribution is
about an order of magnitude smaller than the thermal SZ effect
\citep{1997AA...325....9A, AghanimEtalb97}. We shall thus neglect this
contribution to the temperature fluctuations in the following analysis.

%******************************************************************************
\section{Comparing contributions to the microwave sky\label{sec:comparing}}
%******************************************************************************

\subsection{Description of the detectors noise \label{sec:sky_noise}}
%--------------------------------------------------------------------

In the course of the measuring process, detector noise is added to the total
signal {\em after} the sky fluctuations have been observed, which can be
described by a convolution with the beam pattern. In order to directly compare
the astrophysical fluctuations with those coming from the detectors, it is
convenient to derive a fictitious noise field ``on the sky'' which, once
convolved with the beam pattern and pixelised, will be equivalent to the real
one \cite[]{\Knox95}. Modelling the angular response of channel $i$,
$w_i(\theta)$, as a Gaussian of FWHM $\theta_i$, and assuming a white noise
detector noise power spectrum, the sky or ``unsmoothed'' noise spectrum is
then simply the ratio of the constant white noise level by the square of the
spherical harmonic transform of the beam profile
\begin{equation}
        C_i(\ell) = c_{noise}^2 \exp\left(- {(\ell+\frac{1}{2})^2 \over 2
        (\ell_i+\frac{1}{2})^2} \right) \simeq c_{noise}^2 \exp(-{\ell^2
        \theta_i^2 \over 2\sqrt{2 \ln 2}}),
\label{pow_nois}
\end{equation}
with
\[
        (\ell_i+\frac{1}{2})^{-1} = 2 \sin\left({\theta_i \over 2\sqrt{8 \ln
        2}}\right) \label{eq:C_noise}
\]
and $c_{noise}^2 = \sigma_i^2 \Omega_i = \sigma_i^2 \times 2 \pi [ 1 -
\cos(\theta_i/2) ]$, if $\sigma_i$ stands for the 1--$\sigma$ $\Delta T/T$
sensitivity per field of view. The values of $c_{noise}$ for different
experiments of interest can be found in Table~\ref{tab:perfs}.

While convenient, the noise description above is quite oversimplified. Indeed
detector noise is neither white, nor isotropic. The level of the noise depends
in particular on the total integration time per sky pixel, which is unlikely
to be evenly distributed for realistic observing strategies. In addition,
below a (technology-dependent) ``knee'' frequency, the white noise part of the
detector spectrum becomes dominated by a steeper component (typically in
$1/f$, if the frequency $f$ is the Fourier conjugate of time). This extra
power at long wavelength will typically translate in ``stripping'' of the
noise maps, a common disease. Of course, redundancies can be used to lessen
this problem (\cite{Wright96}), or even reduce it to negligible levels if the
knee frequency is small enough \citep{Janssen96, \Delabrouille98}.

%++++++++++++++++++++++++++++++++++++++++++++++++++++++++++++++++++++++++++++++
% Tables of performances
%++++++++++++++++++++++++++++++++++++++++++++++++++++++++++++++++++++++++++++++
\newcommand{\hf}{\hfill}
\begin{table}[htb] \begin{center} \scalebox{0.795}[0.695]{
\begin{tabular}{|c||p{32pt}|p{32pt}|p{32pt}|p{32pt}|p{32pt}||p{32pt}|p{32pt}|p{32pt}|p{32pt}|p{32pt}|}
\hline \hline \multicolumn{6}{|c||}{\maps (as of January 1998) } &
\multicolumn{5}{|c|}{BoloBall} \\ \hline \hline $\nu$ & \hf 22 & \hf 30 & \hf
40 & \hf 60 & \hf 90 & \hf 143 & \hf 217. & \hf 353 & \hf - & \hf - \\
$\theta_{FWHM}$ & \hf 55.8 & \hf 40.8 & \hf 28.2 & \hf 21.0 & \hf 12.6 & \hf
8.0 & \hf 5.5 & \hf 5 & \hf - & \hf - \\ $\Delta T$ & \hf 8.4 & \hf 14.1 & \hf
17.2 & \hf 30.0 & \hf 50.0 & \hf 35 & \hf 78 & \hf 193 & \hf - & \hf - \\
$c_{noise}$& \hf 8.8 & \hf 10.8 & \hf 9.1 & \hf 11.8 & \hf 11.8 & \hf 11.3 &
\hf 17.4 & \hf 39.0 & \hf - & \hf - \\ \hline
\end{tabular} }
\vspace{8truept} \scalebox{0.795}[0.795]{
\begin{tabular}{|c||p{32pt}|p{32pt}|p{32pt}|p{32pt}||p{32pt}|p{32pt}|p{32pt}|p{32pt}|p{32pt}|p{32pt}|}
\hline \hline & \multicolumn{4}{|c||}{Proposed
\lfi}&\multicolumn{6}{|c|}{Proposed \hfi} \\ \hline\hline $\nu$ & \hf 30 & \hf
44 & \hf 70 & \hf 100 & \hf 100 & \hf 143 & \hf 217. & \hf 353 & \hf 545 & \hf
857 \\ $\theta_{FWHM}$ & \hf 33 & \hf 23 & \hf 14 & \hf 10 & \hf 10.7 & \hf
8.0 & \hf 5.5 & \hf 5.0 & \hf 5.0 & \hf 5.0 \\ $\Delta T$ & \hf 4.0& \hf 7.0 &
\hf 10.0 & \hf 12.0 & \hf 4.6 & \hf 5.5 & \hf 11.7& \hf 39.3 & \hf401 & \hf
18182 \\ $c_{noise}$& \hf 2.5 & \hf 3.0 & \hf 2.6 & \hf 2.3 & \hf 0.9 & \hf
0.8 & \hf 1.2 & \hf 3.7 & \hf 38 & \hf 1711 \\ \hline
\end{tabular} }
\end{center}
\caption[Summary of experimental characteristics used for comparing
experiments]{{ Summary of experimental characteristics used for comparing
experiments. Central band frequencies, $\nu$, are in Gigahertz, the FWHM
angular sizes, $\theta_{FWHM}$, are in arcminute, and $\Delta T$ sensitivities
are in $\mu\mathrm{K}$ per $\theta_{FWHM} \times \theta_{FWHM}$ square pixels;
the implied noise spectrum normalisation $c_{noise} = \Delta T
(\Omega_{FWHM})^{1/2}$, is expressed in $\mu\mathrm{K.deg}$.}}
\label{tab:perfs}
\end{table}
%++++++++++++++++++++++++++++++++++++++++++++++++++++++++++++++++++++++++++++++

\subsection{Angular Scale Dependence of the Fluctuations}
%*******************************************************

\begin{figure}[htbp] \vbox{ \hbox{
\psfig{figure=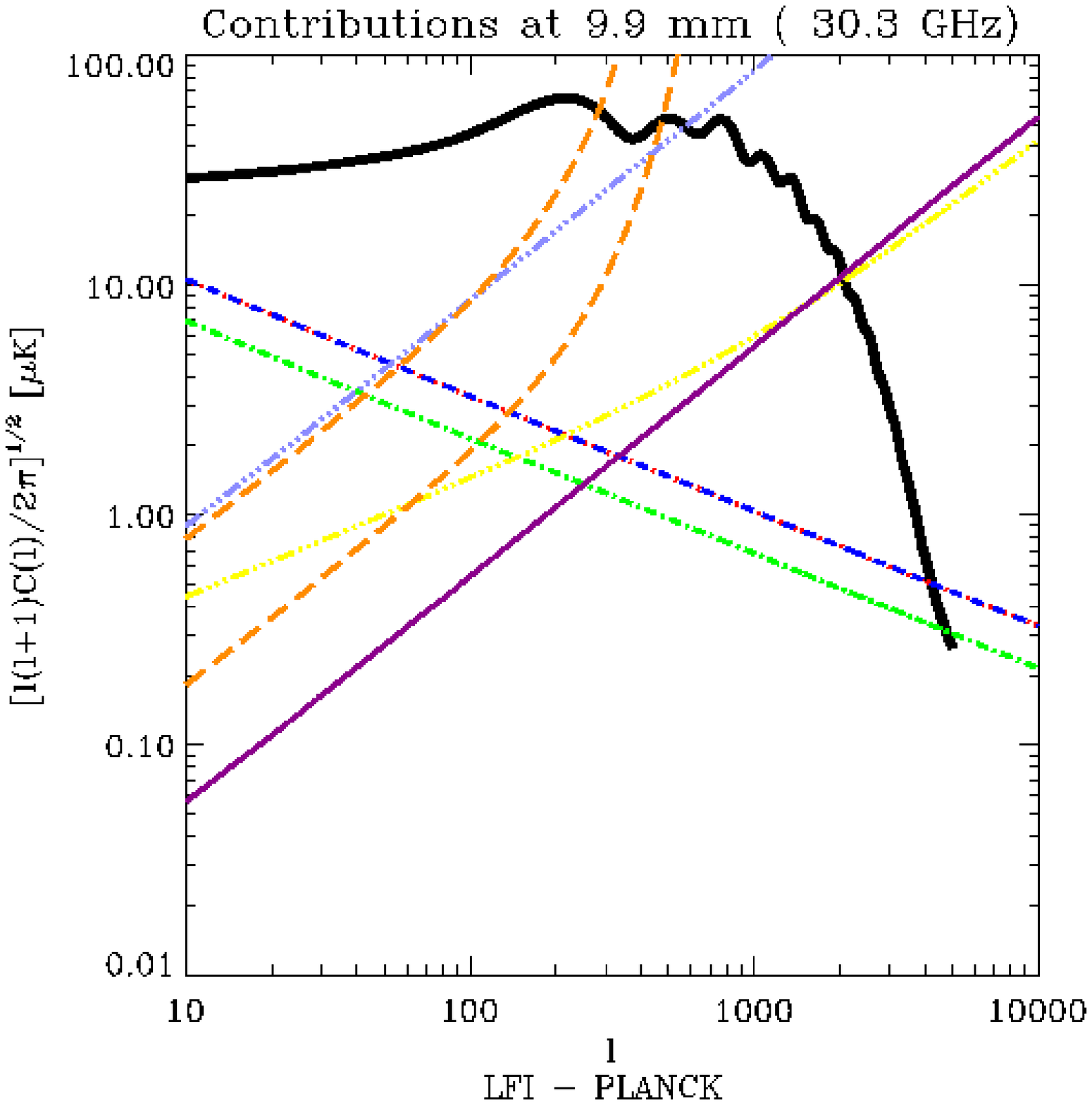,width=0.5\textwidth}
\psfig{figure=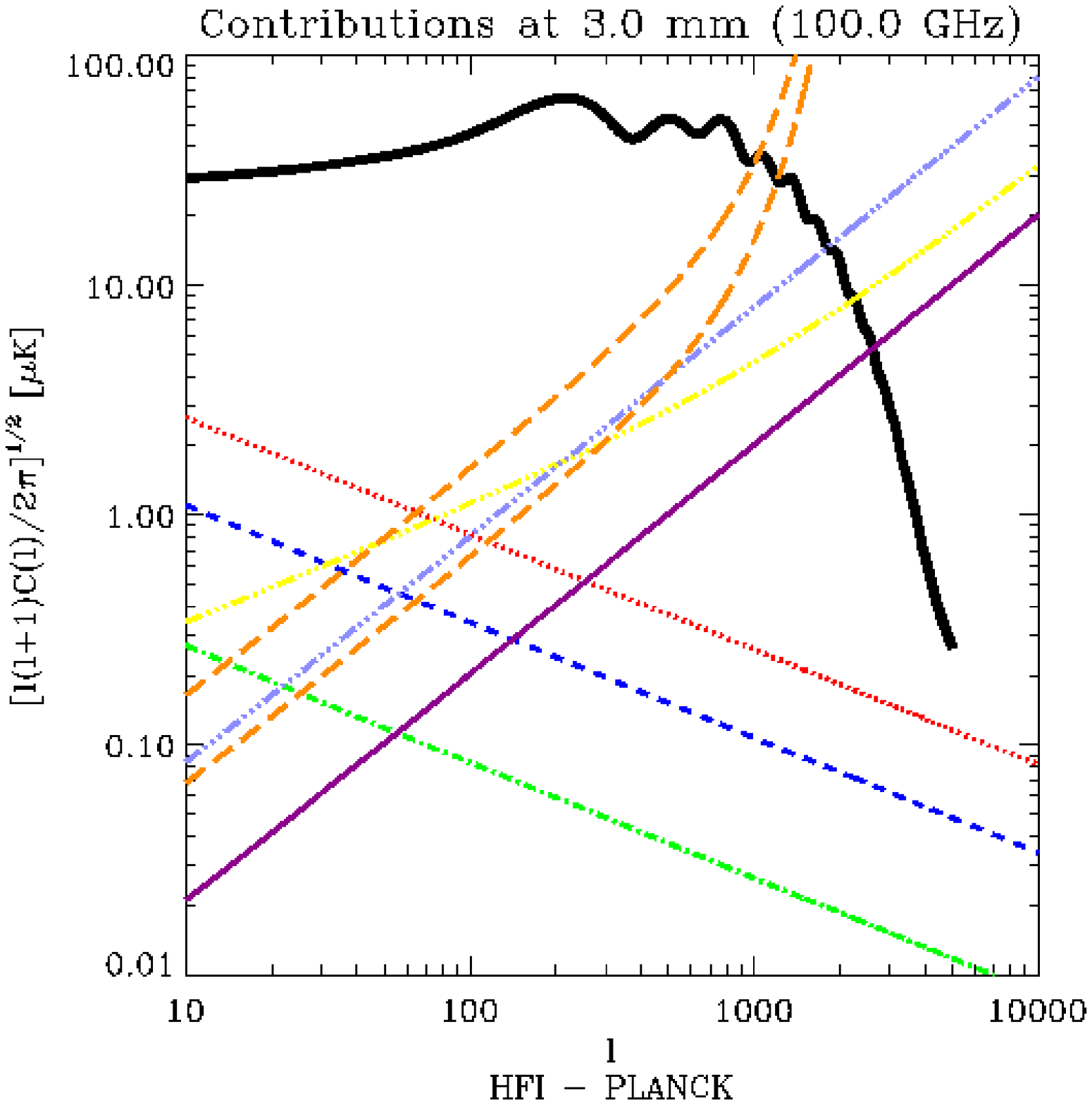,width=0.5\textwidth} } \hbox{
\psfig{figure=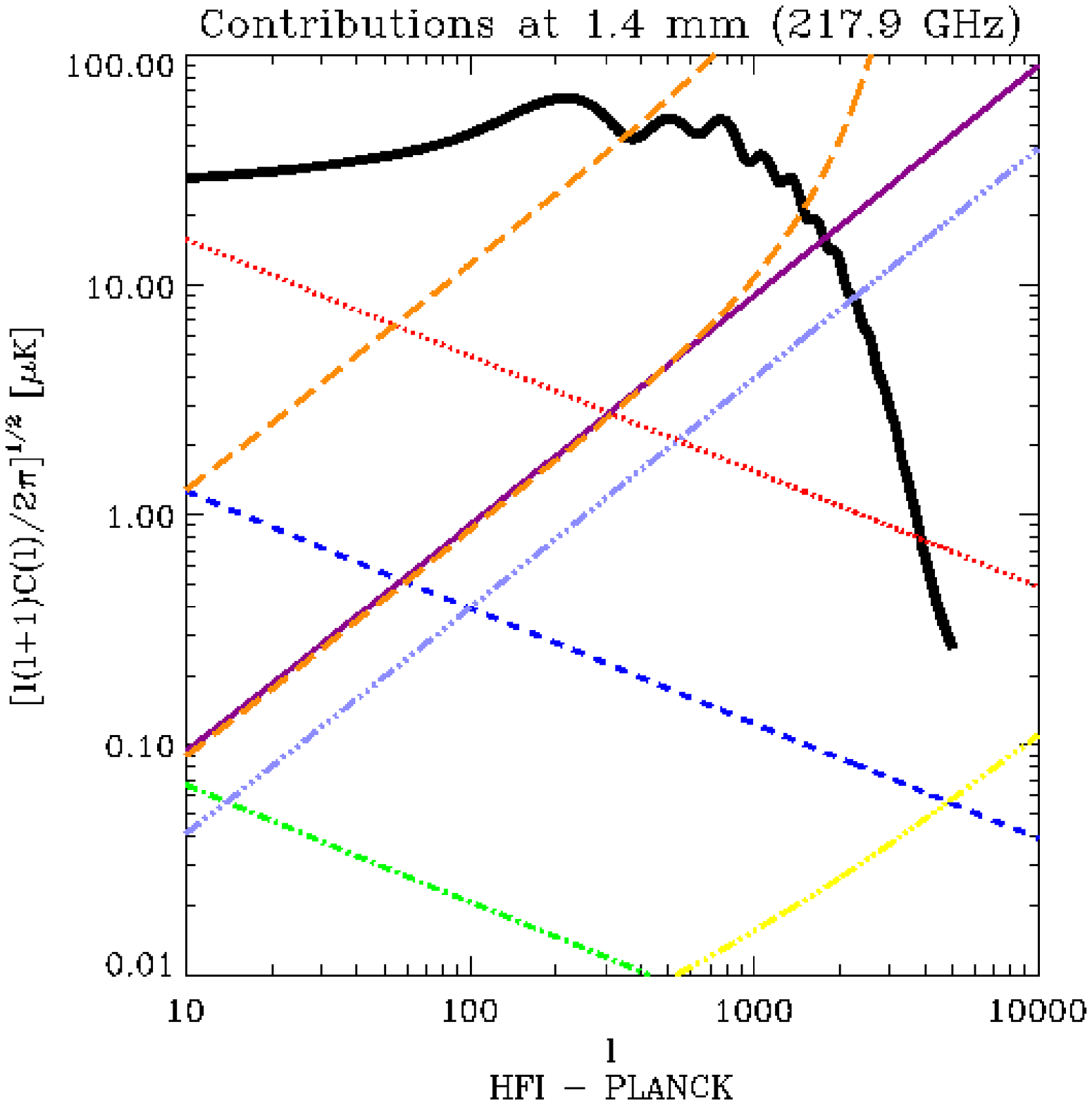,width=0.5\textwidth}
\psfig{figure=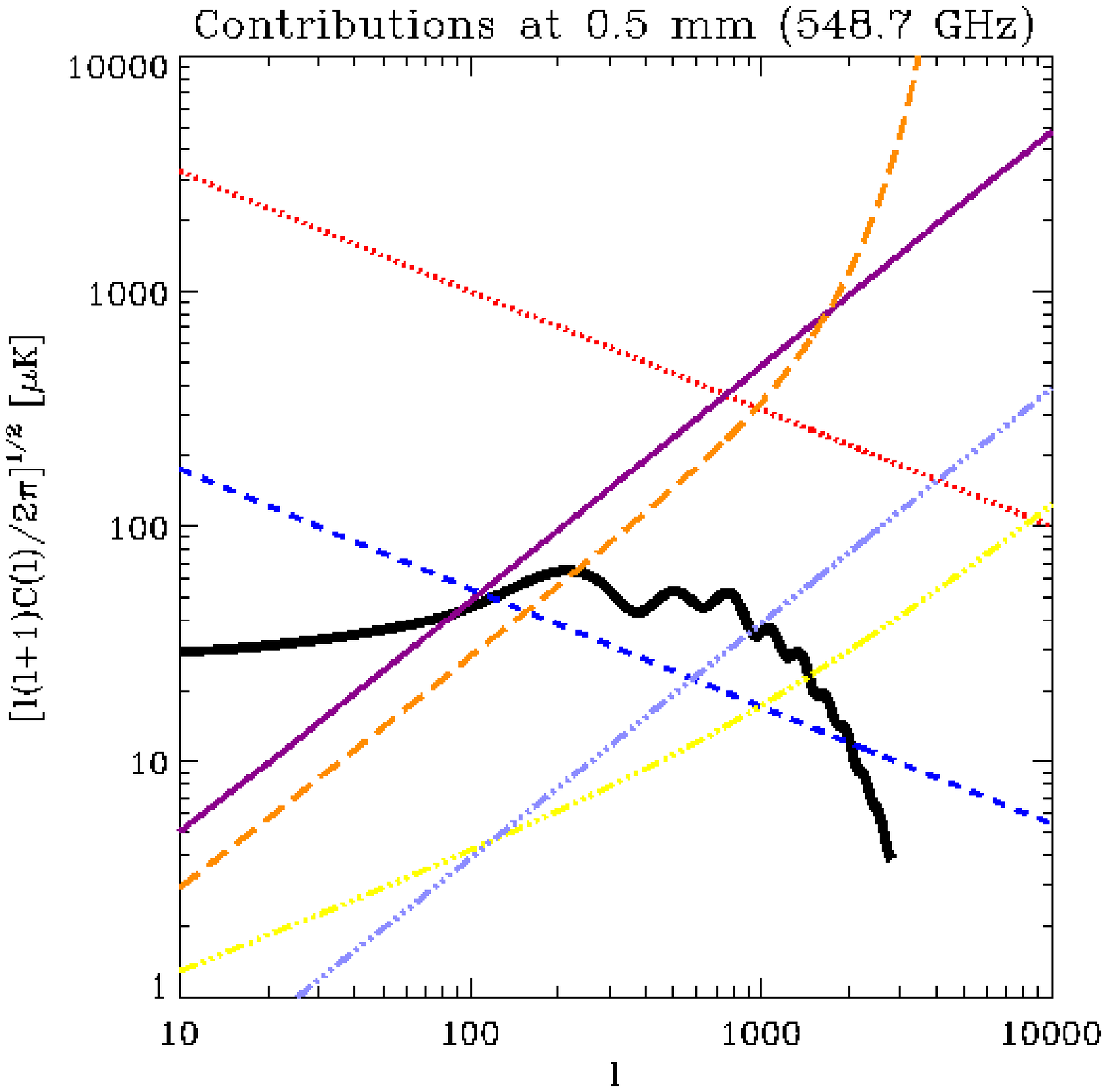,width=0.5\textwidth} } }
\caption[]{Contributions to the fluctuations of the various components, as a
function of angular scale, at different frequencies. The thick solid black
line corresponds to the CMB fluctuations of a \cobe-normalised CDM model. The
dots, dashes and dot-dashes (in red, blue, and green) refer respectively to
the HI correlated, HI uncorrelated, and synchrotron emissions of the
galaxy. The light blue triple dots-dashes displays the contributions from
unresolved radio sources, while the purple line corresponds to the unresolved
contribution from infrared sources.  Long dashes show ``on-sky'' noise level
(see text, eq.~\eqref{pow_nois}). On the 30 GHz plot, the highest noise level
corresponds to \map, and the lower one to the \lfi. On the 100 GHz plot, the
highest noise level corresponds to the \lfi, and the lower one to the \hfi. In
the two remaining plots, the noise level corresponds to the \hfis case.}
\label{fig:contribs}
\end{figure}

Figure~\ref{fig:contribs} compares at 30, 100, 217 and 857 GHz the power
spectrum of the expected primary anisotropies (in a standard CDM model) with
the power contributed by the galactic emission (eq.~[\ref{pow_gals}]), the
unresolved background of radio and infrared sources (eq.~[\ref{eq:hivon1},
\ref{eq:hivon2}, \ref{eq:toffo}]), the S-Z contribution from clusters
(eq.~[\ref{eq:pow_sz}]), and on-sky noise levels (eq.~[\ref{pow_nois}])
corresponding to the \planck\ mission. It is interesting to note that even at
100 GHz, the dust contribution might be stronger than the one coming from the
synchrotron emission, at least for levels typical of the best half of the
sky. Since point processes have flat (``white noise'') spectra, their
logarithmic contribution to the variance, $\ell(\ell+1) C(\ell) \propto
\ell^2$, increases and becomes dominant at very small scales.

%++++++++++++++++++++++++++++++++++++++++++++++++++++++++++++++++++++++++++++++
% Plots of contributions
%++++++++++++++++++++++++++++++++++++++++++++++++++++++++++++++++++++++++++++++
%++++++++++++++++++++++++++++++++++++++++++++++++++++++++++++++++++++++++++++++

\subsection{Frequency Scale Dependence of the Fluctuations}
%**********************************************************

\begin{figure}[htbp]
\centering \centerline{ \psfig{file=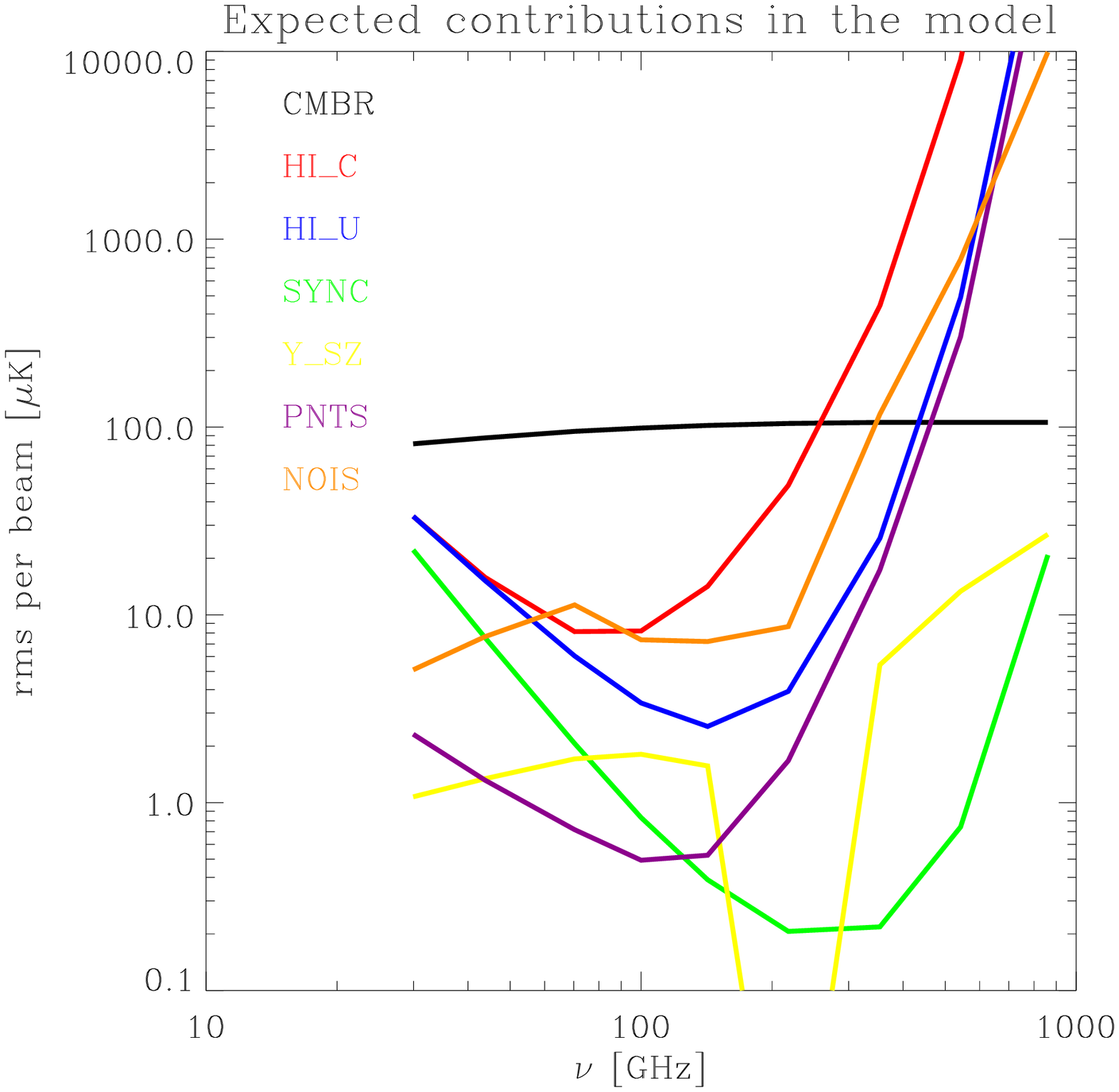, width=\textwidth} }
\caption[ ]{Standard deviation per beam versus frequency for all the relevant
model components in the \plancks case.}
\label{fig:rmsmodel}
\end{figure}

Given the power and frequency spectra of the microwave sky model, one can then
compare the {\em rms} contributions per beam at any frequency $\nu_i$ for any
experiment by
\begin{equation}
        \sigma(\nu_i) = \sum_\ell (2\ell+1)\ C_\ell(\nu_i)\ w_i(\ell)^2 ,
\end{equation}
where $w_i$ stands for the transform of the beam profile at the frequency
$\nu_i$. Figure~\ref{fig:rmsmodel} gives an example for the \plancks case.

%++++++++++++++++++++++++++++++++++++++++++++++++++++++++++++++++++++++++++++++
% Rms vs nu for all the model components
%++++++++++++++++++++++++++++++++++++++++++++++++++++++++++++++++++++++++++++++
%++++++++++++++++++++++++++++++++++++++++++++++++++++++++++++++++++++++++++++++

\subsection{Angular-Frequency Dependence of the Fluctuations \label{sec:nu_l} }
%******************************************************************************

%++++++++++++++++++++++++++++++++++++++++++++++++++++++++++++++++++++++++++++++
% Plots of Nu-L
%++++++++++++++++++++++++++++++++++++++++++++++++++++++++++++++++++++++++++++++
\begin{figure}[htbp]
\centering \centerline{ \hbox{
\psfig{figure=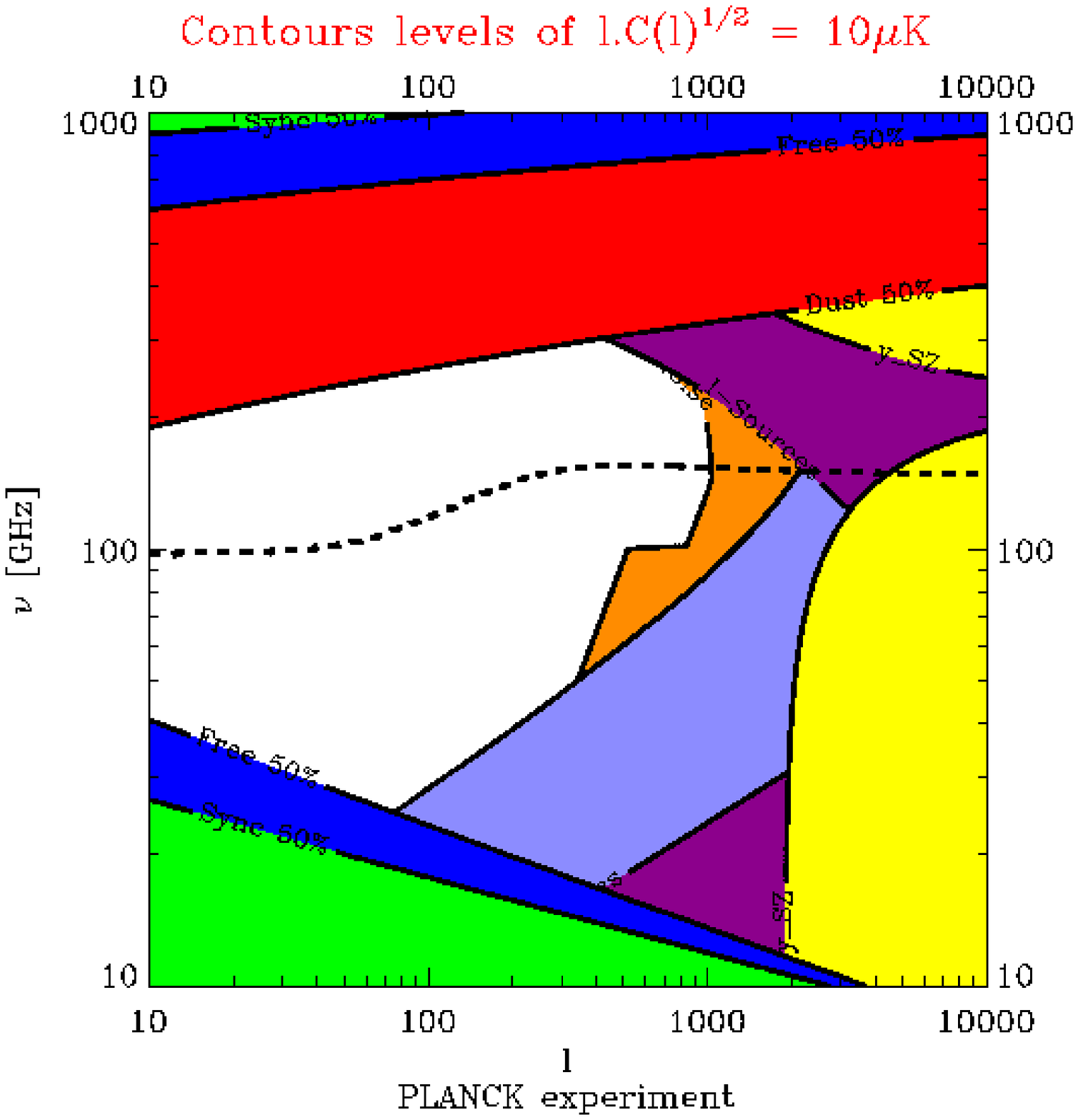,width=0.5\textwidth}
\psfig{figure=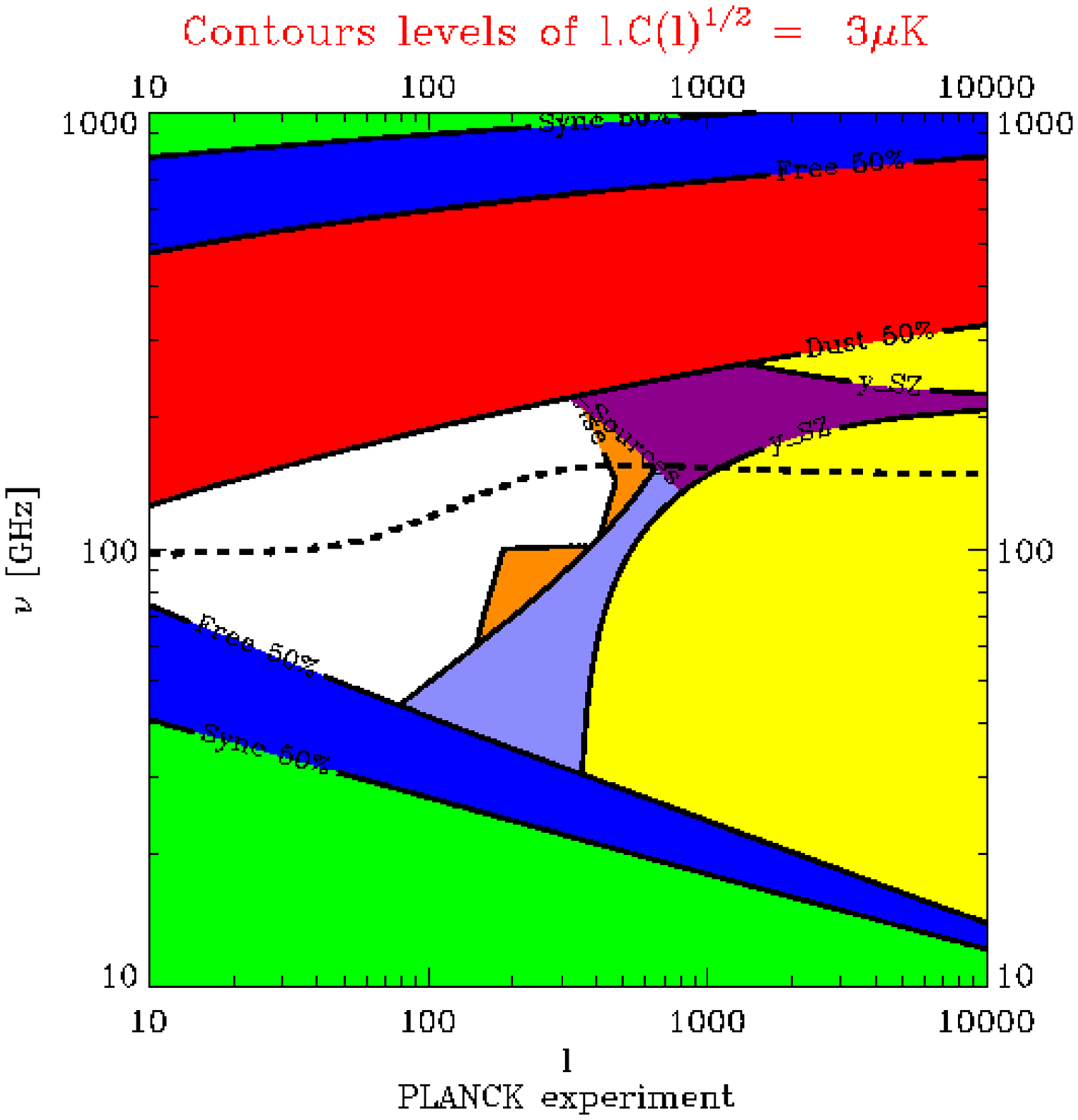 ,width=0.5\textwidth} } }
%\vbox{
%\psfig{figure=contours10_pla.ps,width=0.62\textwidth} \\ \vspace{6pt}
%\psfig{figure=contours3_pla.ps ,width=0.62\textwidth} } }
\caption[]{ Contour levels of the different components of our model in the
angular scale--frequency plane. The levels indicate when the power spectra of
the various components reach $(10\mu$K)$^2$ and $(3\mu$K)$^2$ - respectively
one tenth and one hundredth of the level of the \cobes detection; the filled
side of the lines shows where the corresponding process is higher than the
level line (with the same colour coding than in previous plots, but for the HI
correlated and uncorrelated components which have been replaced here by the
total dust and free-free emissions). Thus the central area is the region where
a CMB signal at 1/10 or 1/100 of the \cobes level will dominate over all other
components. The central dashed line materialises the locus of weakest overall
contamination by foregrounds. As before the galactic components corresponds to
the median sky level. The noise level shown is for \planck.}
\label{fig:nul_conts}
\end{figure}
%++++++++++++++++++++++++++++++++++++++++++++++++++++++++++++++++++++++++++++++

An extensive series of angular power spectra at many frequencies (such as
those in figure~\ref{fig:contribs}), or of amplitudes of fluctuations as a
function of frequency for a series of angular scales (such as
figure~\ref{fig:cpdf-med}), provides a detailed picture of the behaviour of
the various components. Still, it is also illuminating to retain the full
dependence of the fluctuation amplitudes on frequency and angular scale to
build a more synthetic view of the fluctuations ``landscape''.

Figure~\ref{fig:nul_conts} shows the contours in the $\nu-\ell$ plane where
the fluctuations, as estimated from $\ell(\ell+1) C(\ell)/2\pi$, reach one
tenth (left) or one hundredth (right) of the large scale \cobes level
[$\ell(\ell+1) C(\ell)/2\pi = (30\mu\mathrm{K})^2/10 \sim
(10\mu\mathrm{K})^2$]. These contours map the three dimensional topography of
the fluctuations of individual components in the $\nu-\ell$ plane. The
synchrotron component (when expressed in equivalent temperature fluctuations)
defines a valley which opens towards large $\ell$, since $\ell^2 C(\ell)$
decreases with $\ell$ as $1/\ell$. The \HI-uncorrelated component defines a
shallower and gentler valley (see also fig.\ref{fig:cpdf-med}.b), while the
\HI-correlated emission creates a high frequency cliff. The large $\ell$ end
of the valley is barred by the point sources ``dam''. In any case, even in the
case of \planck, the noise level\footnote{The noise levels plotted were
computed using a linear interpolation between the columns of the summary
table~\ref{tab:perfs}.} is higher than the contribution from the unresolved
sources background.

The heavy black line shows the path followed by a stream lying at the bottom
of the valley, i.e. it traces the lowest level of total fluctuations (but
noise). Its location confirms that $\nu \simeq 100\,$GHz is the best frequency
for low-$\ell$ measurements. At $\ell \simgt 200$, the optimal frequency moves
to higher values and is determined by the minimum of the fluctuations from
unresolved sources and clusters; indeed, the dotted line shows the minimum
location if clusters' fluctuations are ignored. Taking them into account shows
that the optimal frequency is $\simlt 200$ GHz for high resolution work around
$\ell \sim 1000$, rather close to the zero of the Sunyaev-Zeldovich effect at
217 GHz. Of course the exact optimal value depends on the relative weight of
the unresolved backgrounds from sources and that from clusters, and cannot be
very precise at this stage. Still, the optimal range is $\simeq 150-200$ GHz,
which is within the range accessible to the most sensitive detectors available
today (typically bolometers can span the 100 to 1000 GHz range). In the case
of \planck, one can expect that the most stringent constraints on the
high-$\ell$ part of the CMB spectrum will come from the 143 \& 217 GHz
channel, not the lower frequency ones of lower angular resolution.

%++++++++++++++++++++++++++++++++++++++++++++++++++++++++++++++++++++++++++++++
% Plots of Nu-L weights
%++++++++++++++++++++++++++++++++++++++++++++++++++++++++++++++++++++++++++++++
\begin{figure*}[htbp]
%\hbox{ 
%\psfig{figure=nul_weights_mp+_Z.ps,width=0.49\textwidth}
%\psfig{figure=nul_weights_pla_Z.ps,width=0.49\textwidth} }
\vspace{2cm}
Figures available with full version at \\
$ftp://ftp.iap.fr/pub/from\_users/bouchet/wiener7.ps.gz$
\medskip
\caption[]{Each hatched areas delimits the regions where a given component
accounts for 50\% of the total contribution to the variance including noise,
in a particular experiment. Each panel has been restricted to the frequency
range that the experiment span. Inner contours show when a component accounts
for 90\% of the signal, and the central contour (horizontal hatches) delimits
the region where the CMB anisotropies should contribute 99\% of the signal. a)
on the left is shown the \maps case. b) the right panel corresponds to
\planck.}
\label{fig:nul_fracs}
\end{figure*}
%++++++++++++++++++++++++++++++++++++++++++++++++++++++++++++++++++++++++++++++

In order to better identify the dominant contributions in a given experiment,
Figure~\ref{fig:nul_fracs} shows the regions of the $\nu - \ell$ plane where
the contribution from a given process $\ell\, C_p(\ell)$ accounts for half of
the total including the noise, $\ell\, \sum_p C_p(\ell)$ (i.e. of the
contribution per logarithmic interval centred around that scale $\ell$ to the
variance of the measurement at frequency $\nu$). The inner contours show when
that process accounts for 90\% of the total (and 99\% for the CMB). This
figure thus shows the areas in the $\nu$-$\ell$ plane where a given component
will be dominating the measurements.

Two cases were considered for illustration. On the left, the noise level and
the frequency range correspond to the \maps experiment which uses passively
cooled HEMT detectors, while the \plancks case which combines actively cooled
HEMTs and bolometers is shown on the right. In the latter case, reaching 10\%
accuracy on the CMB anisotropies over all modes between a few and a thousand,
in the cleanest 50\%\ of the sky, should be easily accessible. A 1\% accuracy
can only be obtained though in a substantially smaller angular range, if no
foreground subtraction is performed. We now turn to that issue.

%******************************************************************************
\section{Separation of components \label{sec:separation} }
%******************************************************************************

Our goal in this section is to introduce a simple tool, which we denote as the
``quality factor'' of an experiment, that will allow in
\S\ref{sec:experiments} a comparison of experimental set-ups, once the effect
of foregrounds is included. We start by establishing notations for describing
our model of the sky (\S\ref{sec:phys_mod}) and derive in \S\ref{sec:inv_prob}
a formulation of the separation problem. General properties of linear
inversion methods are described in \S\ref{sec:lin_inv}. The Wiener filtering
method is described in \S\ref{sec:polish}, and \S\ref{sec:char_wien} gives the
properties of this method, while \S\ref{sec:quality} shows that a simple
indicator allows capturing most aspects of the inversion. Simple examples and
more technical points are discussed in the final sub-section.

\subsection{Physical model \label{sec:phys_mod} } 
%************************************************

We note $\calF(\nu, \be)$ the flux at a frequency $\nu$ in the direction on
the sphere referenced by the unit vector $\be$. We make the hypothesis that
$\calF(\nu, \be)$ is a linear superposition of the contributions from $N_c$
components, each of which can be factorised as a spatial template
$\calT_p(\be)$ at a reference frequency (e.g. at 100 GHz), times a spectral
coefficient $g_p (\nu)$
\begin{equation}
        \calF(\nu, \be) = \sum_{p=1}^{N_c} g_p(\nu)\, \calT_p(\be) .
\end{equation}
This factorisation is not {\em stricto sensu} applicable to the background due
to unresolved point sources, since at different frequencies different redshift
ranges dominate, and thus different sources. But even at the \planck-\hfi\
resolution, this is, as we showed earlier, a weak effect which can be safely
ignored for our present purposes.

Let us also note that the factorisation assumption above does not restrict the
analysis to components with a spatially constant spectral behaviour. Indeed,
these variations are expected to be small, and can thus be linearised too. For
instance, in the case of a varying spectral index whose contribution can be
modelled as $\nu^{\alpha(\be)} \calT_p(\be)$, we would decompose it as
$\nu^{\bar{\alpha}}[1 + (\alpha(\be) - \bar{\alpha}) \ln\nu] \calT_p(\be)$. We
would thus have two spatial templates to recover, $\calT_p(\be)$ and
$(\alpha(\be)-\bar{\alpha})\calT_p(\be)$ with different spectral behaviours,
$\propto \nu^{\bar{\alpha}}$ and $\propto \nu^{\bar{\alpha}} \ln\nu $
respectively. But given the low expected level of the high latitude
synchrotron emission, this is unlikely to be necessary. This simple trick may
still be of some use though to describe complex dust properties in regions
with larger optical depth than assumed here (see also the alternative way
proposed by \citet{1998ApJ...502....1T}).

\subsection{The separation problem \label{sec:inv_prob} } 
%*******************************************************

During an experiment, the microwave sky is scanned by arrays of detectors
sensitive to various frequencies and with various optical responses and noise
properties. Their response is transformed in many ways by the ensuing
electronics chains, by interactions between components of the experiments, the
transmission to the ground, etc. The resulting sets of time-ordered data (TOD)
needs then to be heavily massaged to produce well-calibrated pixelised maps at
a number of frequencies, each with an effective resolution, and well specified
noise properties. We assumed here that all this work has already been done
(for further details, see \citet{\JanssenGulkis92, Wright96, \Tegmark97,
1998A&AS..127..555D, 1998MNRAS.298..445D}).

We thus focus on the next step, \ie the joint analysis of noisy pixelised sky
maps, or signal maps $\bs_i$, at $N_\nu$ different frequencies in order to
extract information on the different underlying physical components. It is
convenient to represent a pixelised map as a vector containing the $N_p$ map
pixels values, the $j$-th pixel corresponding to the direction on the sky
$\be_j$. Let us denote by $\bt_p$ the unknown pixelised maps of the physical
templates $\calT_p$. We thus have to find the best estimates of the templates
$t_p(\be_j)$, which are related to the observed signal by the $N_\nu \times
N_p$ equations
\begin{eqnarray}
        s_i(\be_j) & = & \sum_{p=1}^{N_c}\ \int_{\nu = 0}^\infty d\nu\ v_i
        (\nu-\nu_i)\, g_p(\nu) \sum_{k=1}^{N_p} w_i (\be_j -\be_k)\ t_p(\be_k)
        + n_i(\be_j) \\ & = & \sum_{p=1}^{N_c} \left[ v_i \star g_p(\nu)
        \right]\, \left[ w_i \star t_p(\be) \right] + n_i(\be_j) ,
\end{eqnarray}
where $\star$ stands for the convolution operation. This system is to be
solved assuming we know for each signal map its effective spectral and optical
transmission, $v_i (\nu-\nu_i)$ and $w_i (\be)$ and that the covariance matrix
$\bN_i = \VEV{\bn_i \bn_i^T}$ of the (unknown) noise component $\bn_i$ has
been properly estimated during the first round of analysis.

Since the searched for templates $\bt_p$ are convolved with the beam
responses, $w_i$, it is in fact more convenient to formulate the problem in
terms of spherical harmonics transforms (denoted by an over-brace), in which
case the convolutions reduce to products. For each mode $\bl \equiv \{\ell,
m\}$, we can arrange the data concerning the $N_\nu$ channels as a complex
vector of observations, $\by(\bl) \equiv \{ \overbrace{s_1}(\bl),
\overbrace{s_2}(\bl), \ldots , \overbrace{s_{N_\nu}}(\bl) \}$, and we define
similarly a complex noise vector, $\bb(\bl) \equiv \{ \overbrace{n_1}(\bl),
\overbrace{n_2}(\bl), \ldots , \overbrace{n_{N_\nu}}(\bl) \}$, while the
corresponding template data will be arranged in a complex vector $\bx(\bl)$ of
length $N_c$. We thus have to solve for $\bx$, for each mode $\bl$
independently, the matrix equation
\begin{equation}
        \by(\bl) = \bA(\bl) \bx(\bl) + \bb(\bl) , \label{eq:model_ell}
\end{equation}
with $A_{i p}(\bl) = \left[ v_i \star g_p(\nu) \right]\, w_i(\bl)$. Note
though the independence of the $\bl$ modes would be lost in case one uses
other functions that spherical harmonics (\eg to insure orthogonality of the
basis functions when all the sky is not available for the analysis
\citep{\Gorski94}). Then instead of solving $N_p$ systems of $N_\nu$
equations, one will have to tract a much larger system of $N_p \times N_\nu$
equations (possibly in real space rather than in the transformed space). For
simplicity we treat each mode separately in the following.

\subsection{Linear Inversions \label{sec:lin_inv} }
%**************************************************

Any {\it linear} recovery procedure may be written as an $N_c \times N_\nu$
matrix $\bW(\ell)$ which applied to the $\by(\bl)$ vector yields an estimate
$\hat \bx(\bl)$ of the $\bx(\bl)$ vector, i.e.
\begin{equation}
       \hat \bx(\bl) = \bW(\bl)\, \by(\bl) .
\end{equation} 
In the following, we show that the matrix $\bR = \bW \bA$ plays in important
role in determining the properties of the estimated templates by the selected
method, whatever it might be. Indeed, since $\hat\bx = \bW \left[ \bA \bx +
\bb \right]$, we have
\begin{equation} 
        \VEV{\hat\bx} = \bR \VEV{\bx} \quad\mathrm{and}\quad \VEV{\hat\bx
        \hat\bx^\dagger} = \bR \VEV{\bx \bx^\dagger} \bR^\dagger + \bW \bB
        \bW^\dagger , \label{eq:Rprop}
\end{equation}
where $\bx^\dagger$ is the Hermitian conjugate of $\bx$ (\ie $\bx^\dagger$ is
the complex conjugate of the transpose, $x^T$, of the vector $\bx$), and $B =
\VEV{\bb \bb^\dagger}$ stands for the (supposedly known) noise covariance
matrix. We denote by $\VEV{Y}$ the average over an ensemble of statistical
realisations of the quantity $Y$. The covariance matrix of the reconstruction
errors, $\bE$, is given by
\begin{equation} 
        \bE \equiv \VEV{(\bx - \bxh) (\bx - \bxh)^\dagger} = \bC - \bR \bC -
        \bC \bR^\dagger + \VEV{\hat\bx \hat\bx^\dagger} ,
\end{equation}
where $\bC= \VEV{\bx \bx^\dagger}$ stands for the covariance matrix of the
underlying templates.

An estimate of this covariance matrix $\bC$ should of course be estimated from
the data itself. Since the elements of the covariance matrix of the {\em
recovered} templates (assuming they have zero mean) may be written as
\begin{equation} 
         \VEV{\hat x_p \hat x_q^\star} = R_{p p} \VEV{x_p x_q^\star} R_{q
         q}^\star + \sum_{r\ne p} \sum_{s \ne q} R_{p r} \VEV{x_r x_s^\star}
         R_{s q}^\dagger + \left[\bW \bB \bW^\dagger \right]_{p q} ,
\end{equation} 
we define the intermediate variable $\tilde x_p = \hat x_p / R_{p p}$ in order
to write
\begin{equation} 
        \VEV{\tilde x_p \tilde x_q^\star} = \VEV{x_p x_q^\star} + \calB ,
\end{equation} 
where $\calB$ is then an additive bias. This shows that an unbiased estimate
of the covariance matrix of the unknown templates, $\hat\bC$, may be obtained
by
\begin{equation}
        \hat\bC \equiv \hat{\VEV{\bx \bx^\dagger}} = {1 \over 2\ell +1}
        \sum_{m=-\ell}^{m=+\ell} \tilde\bx \tilde \bx^\dagger - \calB .
\end{equation}
This expression is only formal though, since $\calB$ contains some of the
unknowns. Still, this form allows to generalise the calculation of
\cite[]{\Knox95} of the expected error (covariance) of the power spectra
estimates derived in such a way from the maps. Indeed,
\begin{eqnarray}
        \Cov{\hat C_{p q}} & = & \VEV{\hat C_{p q}^2} - \VEV{\hat C_{p q}}^2
        \\ & = & \VEV{ \left[{1 \over 2\ell +1} \sum_m \tilde x_p \tilde
        x_q^\star - \calB_{p q} \right]^2} - C_{p q}^2 .
\end{eqnarray}
Once this expression is developed, there are many terms with fourth moments of
the form $\VEV{x_r(\ell, m)\, x_s(\ell, m)^\star\, x_t(\ell, m\prim)\,
x_u(\ell, m\prim)}$.

{\em If we assume that all the $x_p$ are Gaussian variables}, then we can
express these fourth moments as products of second moments, and recontract the
sums to finally obtain
\begin{eqnarray}
        \Cov{\hat C_{p q}} & = & (\hat C_{p q} + \calB_{p q})^2 + {2\over
        2\ell+1}(\hat C_{p q} + \calB_{p q})^2 \\ & - & 2\calB_{p q} ((\hat C
        _{p q} + \calB_{p q}) + \calB_{p q}^2 - \bC_{p q}^2 \\ & = & {2\over
        2\ell+1}( C_{p q} + \calB_{p q})^2 = {2\over 2\ell+1} \VEV{ {\hat x_p
        \over R_{p p}} {\hat x_q^\star \over R_{q q}^\star}}^2 .
        \label{eq:covC}
\end{eqnarray}
This expression shows the extra contribution of noise and foregrounds to the
cosmic variance. It can be re-expressed and simplified once a specific
inversion method has been specified.

\subsection{Wiener filtering \label{sec:polish} }
%************************************************

One particular choice of $\bW$ may be obtained by minimising the figure of
merit
\begin{equation}
        \chi^2(\bl) = (\by - \bA \hat \bx)^\dagger\ \bB^{-1}\ (\by - \bA
        \hat\bx) , \label{eq:chi2}
\end{equation}
for each mode $\bl$. The advantage of this simple inversion by $\chi^2$
minimisation is that it requires {\em no prior information on the unknown}
$\bx$, only some knowledge on the statistical properties of the noise maps
through their covariance matrix $\bB$. An assessment of this simple method
will be given in a sequel paper (hereafter paper II) concerned with actual
numerical simulations of the sky and its analysis.

Another possibility for linearly separating the components is to apply
``Wiener'' filtering techniques since \citet{\BouchetEtalc95} and
\citet{\TegmarkEfstathiou95} showed how to generalise this well-known
procedure (see \eg \cite{\BunnEtal94,Bond95} for an application to COBE-DMR
data, and \cite{\ZaroubiEtal95} for galaxy surveys data) to the case of
multi-frequency, multi-resolution, data. In the following, we rederive in a
self-contained way this Wiener filtering approach, with a number of extensions
which will prove useful for our purposes.

The inversion by $\chi^2$ minimisation is designed to best reproduce the
observed data set by minimising the residuals between the observations and the
model (eq.~[\ref{eq:chi2}]). Another possibility is to require a {\em
statistically minimal} error in the recovery, when the method is applied to an
ensemble of data sets with identical statistical properties. One then obtains
the ``Wiener'' matrix $\bW$, by demanding that the vector of differences
between the underlying and recovered templates, $\beps = \bx - \hat\bx$, be on
average of minimal norm
\begin{equation}
        \varepsilon^2(\bl) = \VEV{ |\bx - \hat\bx |^2} \equiv \VEV{(\bx -
        \bxh)^\dagger(\bx - \bxh)}
\label{eq:def_wien}
\end{equation}
In practice, we require that the derivatives of
\begin{eqnarray}
        \varepsilon^2(\bl) & = & \VEV{ |\bx - \bW (\bA \bx + \bb) |^2} \\ & =
        & ( W_{p\nu} A_{\nu p\prim} - \delta_{p p\prim} )( W_{p \nu\prim}
        A_{\nu\prim p\dprim} - \delta_{p p\dprim}) \VEV{ x_{p\prim}
        x_{p\dprim}^\star} + W_{p \nu} W_{p \nu\prim} \VEV{ b_{\nu}
        b_{\nu\prim}^\star }
\label{eq:get_wien}
\end{eqnarray}
versus all $W_{p\nu}$ (and their complex conjugate) be zero for each $\ell$
(with repeating indices implying summation in\eqref{eq:get_wien} and
\eqref{eq:sol_wien_gen}), which yields
\begin{equation}
        W_{p\nu} \left[ A_{\nu p\prim} \VEV{ x_{p\prim} x_{p\dprim}^\star }
        A_{\nu\prim p\dprim} + \VEV{ b_{\nu} b_{\nu\prim}^\star } \right] =
        A_{\nu\prim p\prim} \VEV{ x_{p\prim} x_{p}^\star } .
\label{eq:sol_wien_gen}
\end{equation}
In matrix form, the wiener matrix is then given by
\begin{equation}
        \bW = \bC \bA^T \left[ \bA \bC \bA^T + \bB \right]^{-1} ,
        \label{eq:sol_wien_mat}
\end{equation}
if we denote by $\bC$ the covariance matrix of the templates, $C_{p p\prim} =
\VEV{ x_{p} x_{p\prim}^\star }$. Alternatively, $\bW^\dagger = \left[ \bA \bC
\bA^T + \bB \right]^{-1} \bA \bC\ $ (since $\bA$ is real, $\bA^\dagger =
\bA^T$).

Since the Wiener matrix depends on $\bC$, \ie on the covariance matrix of the
templates, it is clear that this makes use of prior knowledge to construct
optimal weights.  One can thus think of Wiener filtering as a ``polishing''
stage of a first inversion (\eg as in \S~\ref{sec:lin_inv}) yielding an
estimate $\hat\bC$ of the covariance matrix needed to have optimal weights. A
practical implementation is described in paper II.

\subsection{Characteristics of the Wiener Separation \label{sec:char_wien} }
%***************************************************************************

The definition~(\ref{eq:sol_wien_gen}) of Wiener filtering implies that
\begin{equation}
        \VEV{\hat\bx\bx^\dagger} = \VEV{\bx \hat\bx^\dagger} =
        \VEV{\hat\bx\hat\bx^\dagger} .\label{eq:propW}
\end{equation}
As a result, the expression for the covariance matrix of the reconstruction
errors $\bE \equiv \VEV{\beps \beps^\dagger}$ reduces to the simple form
\begin{equation}
        \bE = \left[ \bI - \bR \right] \bC , \label{eq:covE}
\end{equation}
where $\bI$ stands for the identity matrix, and $\bR = \bW \bA$ as
before. Alternatively, $\bE = \bC \left[ \bI - \bR^\dagger \right]$.

The general expression of the covariance matrix of the templates given by
eq.~\eqref{eq:Rprop} may be simplified by using the specific Wiener matrix
properties of eqs.~(\ref{eq:propW}) and (\ref{eq:covE}), which yields
\begin{equation}
        \VEV{\hat\bx \hat\bx^\dagger} = \bR \VEV{\bx \bx^\dagger}
        . \label{eq:Rpows}
\end{equation}
Both equations\eqref{eq:covE} and\eqref{eq:Rpows} clearly show that the closer
$\bR$ is to the identity, the smaller are the reconstruction errors.

The diagonal elements of $\bE$ are of particular interest since they give the
residual errors on each recovered template,
\begin{equation}
        \varepsilon_p^2 = E_{p p} = (1-R_{p p}) C_{p p} .  \label{eq:res_tot}
\end{equation}
To break down this overall error into its various contributions, we start from
eq.~\eqref{eq:get_wien}
\begin{equation}
        \varepsilon_p^2 = (1-R_{p p})^2 C_{p p} + \sum_{p\prim \ne p \ne
        p\dprim} R_{p p\prim} R_{p p\dprim} C_{p\prim p\dprim} + W_{p \nu}
        W_{p \nu\prim} B_{\nu \nu\prim} .
\end{equation}
By using eq.~\eqref{eq:res_tot}, we get
\begin{equation}
        \varepsilon_p^2 = {1 \over R_{p p}} \sum_{p\prim \ne p \ne p\dprim}
        R_{p p\prim} R_{p p\dprim} C_{p\prim p\dprim} + W_{p \nu} W_{p
        \nu\prim} B_{\nu \nu\prim} , \label{eq:pow_leak}
\end{equation}
which describes the power leakage from the other processes and the co-added
noises.

In addition, the error on the deduced power spectra given by
eq.~\eqref{eq:covC} takes for Wiener filtering the simple form
\begin{equation}
        \Cov{\hat C_{p q}} \equiv \left[ \Delta \hat C_{p q} \right]^2 =
        {2\over 2\ell+1} \left[ { R_{p r} C_{r q} \over R_{p p} R_{q q}^\star
        } \right]^2 , \label{eq:cov_wien}
\end{equation}
which will prove particularly convenient to compare experiments.

\subsection{Case of uncorrelated templates \label{sec:quality} } 
%***************************************************************

In this theoretical analysis, we can always assume for simplicity that we have
decomposed the sky flux into a superposition of emissions from uncorrelated
templates, so that
\begin{equation}
        \VEV{x_{p} x_{p\prim}^\star} = \delta_{p p\prim}\, C_{p p}(\ell) =
        \delta_{p p\prim}\, \calC_p(\ell) .
\end{equation}
In our galactic model, it means in particular that we consider the templates
for the \HI-correlated and \HI-uncorrelated components rather than those for
the dust and free-free emissions. In actual implementations though it might
prove more convenient to relax this assumption and consider correlated
templates but with simpler spectral signatures (see Paper II). In the
uncorrelated case, the previous expressions (\ref{eq:Rpows}, \ref{eq:res_tot}
\& \ref{eq:pow_leak}, \ref{eq:cov_wien}) may be written as
\begin{eqnarray}
        \VEV{ | \hat x_p |^2} & = & Q_p \VEV{ | x_p |^2} = Q_p \calC_p(\ell)
\label{eq:QC} \\ \varepsilon_p^2 & = & (1-Q_p) \calC_p = 1 /Q_p
\sum_{p\prim \ne p} R_{p p\prim}^2 \calC_{p\prim} + W_{p \nu} W_{p \nu\prim}
B_{\nu \nu\prim}
\label{eq:Qerr} \\ \Delta \hat \calC_p & = & \sqrt{ {2 \over 2\ell +1} } {\calC_p \over Q_p}
\label{eq:QCov}
\end{eqnarray}
where $Q_p = \bR_{p p}$ stands for the trace elements of $\bW\bA$. Thus $Q_p$
tells us\begin{enumerate}

\item how the typical amplitude of the Wiener-estimated modes $\hat\bx$ are
damped as compared to the real ones (eq.~[\ref{eq:QC}]).

\item the spectrum of the residual reconstruction error in every map
(eq.~[\ref{eq:Qerr}]); note that this error may be further broken down in
residuals from each component in every map by using the full $\bR$ matrix.

\item the uncertainty added by the noise and the foreground removal ($\propto
1/Q_p -1$, eq.~[\ref{eq:QCov}]) to cosmic (or sampling) variance which is
given by $C_{p p}/\sqrt{2\ell +1}$ (this result only holds though under the
simplifying assumption of Gaussianity of all the sky components).

\end{enumerate}
Given these interesting properties, we propose to use $Q_p$ as a ``quality
factor'' to assess the ability of experimental set-ups to recover a given
process $p$ in the presence of other components; it assesses in particular how
well the CMB itself can be disentangled from the foregrounds.

Section \ref{sec:experiments} below is devoted to ``practical'' uses of the
quality factor, in conjunction with our assumed sky model.

\subsection{Discussion \label{sec:qual_disc} } 
%***********************************************

\subsubsection{Simple examples}
%------------------------------

This ``quality'' indicator generalises the real space ``Foreground Degradation
Factor'' introduced by \cite{\DodelsonStebbins94}. It may be viewed as an
extension of the usual window functions used to describe an experimental
setup. This can be seen most easily by considering a noiseless thought
experiment mapping directly the CMB anisotropies with a symmetrical beam
profile $w(\theta)$. Then the power spectrum of the map will be the real power
spectrum times the square of the spherical harmonic transform of the beam,
$<\hat x_p(\ell) > = w(\ell)^{2} <x_p(\ell)>$. The spherical transform of
$Q_p^{1/2}(\ell)$ is then the beam profile of a thought experiment directly
measuring CMB anisotropies. The shape of $Q_p(\ell)$ thus gives us a direct
insight on the real angular resolution of an experiment when foregrounds are
taken into account.

In addition, one often considers (e.g. for theoretical studies of the accuracy
of parameter estimation from power spectra) a somewhat less idealised
experiment which still maps directly the CMB, with a beam profile $w(\theta)$,
but including also detector noise, characterised by its power spectrum
$C_N$. Let us suppose this is analysed by Wiener filtering. We have to solve
$\by = \bA x + \bn$, with
\begin{equation}
        A^T = \{w, w, w, w,\ldots\}, \quad \VEV{x x^\dagger} = C_x, \quad \bB
        = C_N \times \boldsymbol{I} .  \label{eq:idealexp}
\end{equation}
The previous formulae then lead\footnote{By using $[\boldsymbol{I} + \beta
\boldsymbol{1}]^{-1}\ =\ \boldsymbol{I} - \beta / (1 + \beta N_c)\
\boldsymbol{1}$, if $\boldsymbol{1}$ stands for the $N_c \times N_c$ matrix
with all elements equal to 1, and $\boldsymbol{I}$ for the identity matrix.}
to
\begin{eqnarray}
        Q_x(\ell) & = & {C_x(\ell) \over C_x(\ell) + w^{-2} C_N(\ell)/N_c }
        \label{eq:SigNois} \\ \Delta \hat C_x & = & {2\over 2\ell+1}
        {C_x(\ell) \over Q_x} = {2\over 2\ell+1} \left(C_x(\ell) +
        w(\ell)^{-2}\, C_N(\ell) / N_c \right) ,
\end{eqnarray}

where $N_c$ stands as before for the number of channels (measurements).
Equation\eqref{eq:SigNois} tells us that the Wiener-estimated modes (as seen
by the ratio of the estimated to real spectra) are damped by the ratio of the
expected signal to signal + noise (the noise power spectrum being ``on the
sky'', \cf~\S\ref{sec:sky_noise}, with the noise power spectrum being the
noise power spectrum per channel divided by the number of channels). In brief,
what Wiener filtering does when the noise strength is getting larger than the
signal is to progressively set the estimate of the signal to zero in an
attempt to return only true features. And of course we recover the traditional
error on the estimated power spectrum added by noise found by
\citet{\Knox95}. Thus determining the quality factor allows to find the
equivalent ideal experiment often considered by theorists; it provides a
direct estimate of the final errors on the power spectrum determination from a
foreground model and the summary table of performance of an experiment.

\subsubsection{Reduced variables}
%--------------------------------

Note that the expressions found previously could be further simplified by
considering reduced templates, $\bx\prim$ which are uncorrelated and of unit
variance. Indeed, let us perform a Cholesky decomposition of $\bC$, \ie $\bC =
\bL \bL^T$ where $\bL$ is a lower triangular (real) matrix. The reduced
templates are then defined by $\bx\prim = \bL^{-1} \bx = \bL^T \bC^{-1} \bx$
(which implies, as desired that the reduced templates are ``orthonormalised'',
$\VEV{\bx\prim \bx^{\prime\dagger} } = \bL^T \bC^{-1} \VEV{\bx \bx^\dagger}
\bC^{-1\,T} \bL = \bI$).

Then the problem to solve may be written as
\begin{equation}
        \by = \bS \bx\prim + \bb ,
\end{equation}
where we have defined $\bS = \bA\bL$. Since $\VEV{\by \by^T} = \bS \bS^T +
\bB$, $\bS \bS^T$ is the useful signal covariance matrix (and $\bS$ may be
seen as it's Cholesky decomposition). In that case,
\begin{eqnarray}
        \bW\prim & = & \bS^\dagger \left[ \bS \bS^\dagger + \bB \right]^{-1}
        \\ \bE\prim & = & \VEV{(\bx\prim -\hat\bx\prim)(\bx\prim
        -\hat\bx\prim)^\dagger} = \bI - \bR\prim \\ \bR\prim & = & \bL^{-1}
        \bR \bL = \bS^\dagger \left[ \bS \bS^\dagger + \bB \right]^{-1} \bS
        \label{eq:snw}.
\end{eqnarray}
This form shows most clearly that the Wiener filter derived is indeed a
generalisation of the usual signal to noise weighting (eq. \eqref{eq:snw}). On
the other hand, this convenient form for theoretical analysis might not be so
practical since $\bC$ might be rather ill-determined from a first step of
analysis. Thus paper II will rather use the formulae obtained in the previous
sections.

%******************************************************************************
\section{Comparing experimental set-ups \label{sec:experiments} }
%******************************************************************************

\subsection{Examples of Wiener matrixes}
%***************************************

\begin{figure}[htbp] \centering \centerline{\vbox{ \hbox{ 
\psfig{figure=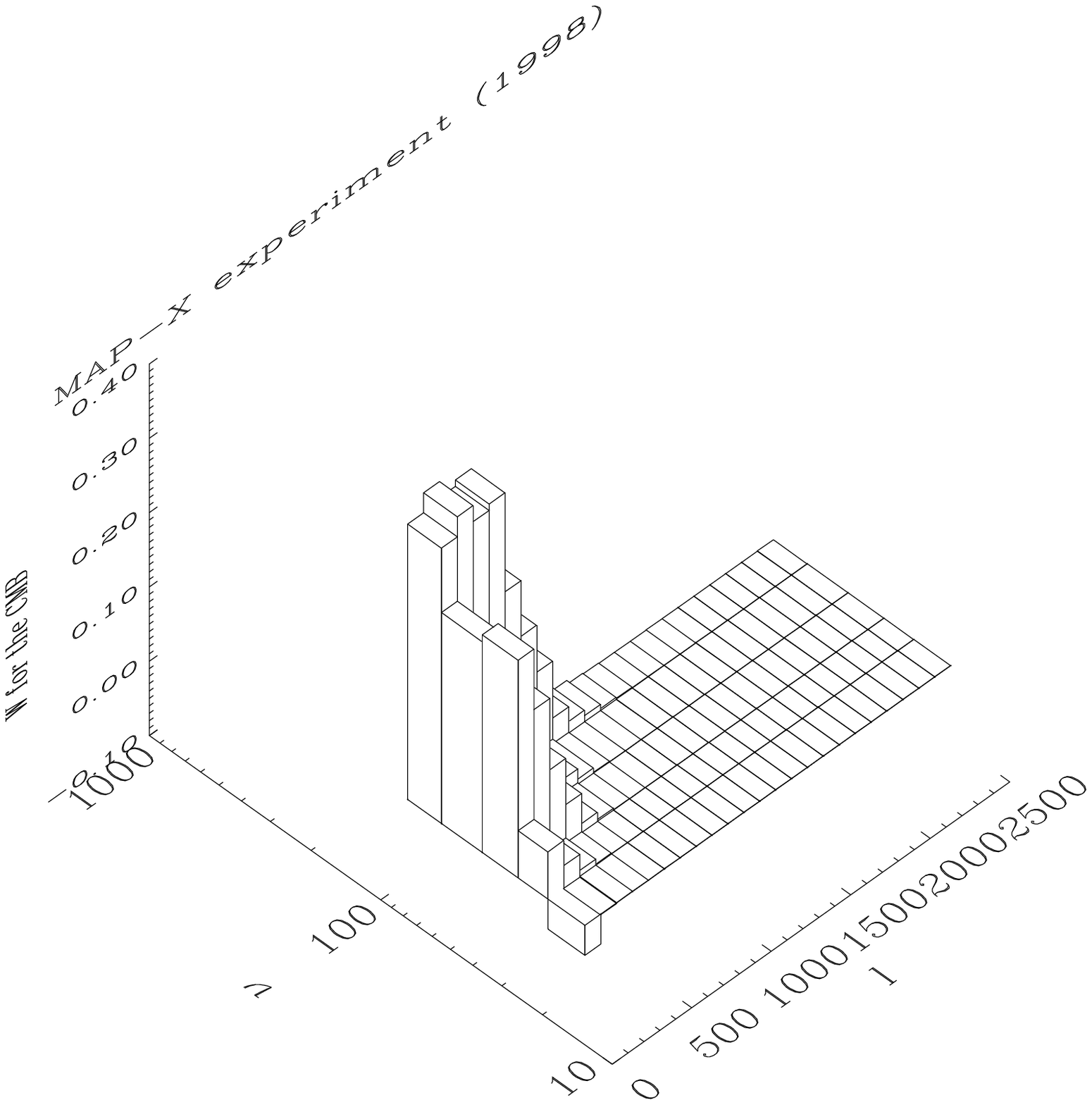,width=0.5\textwidth}
\psfig{figure=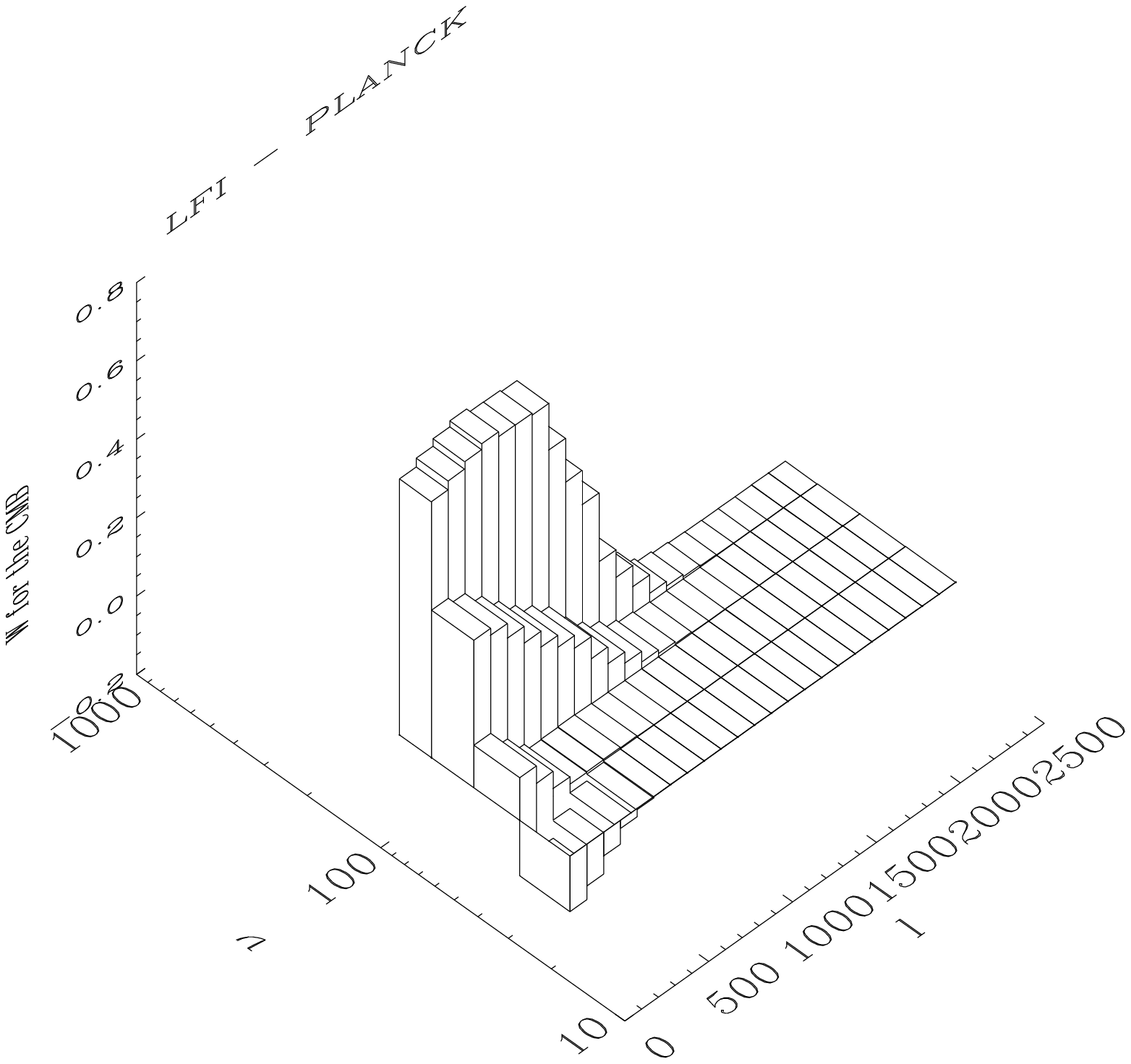,width=0.5\textwidth} } \hbox{
\psfig{figure=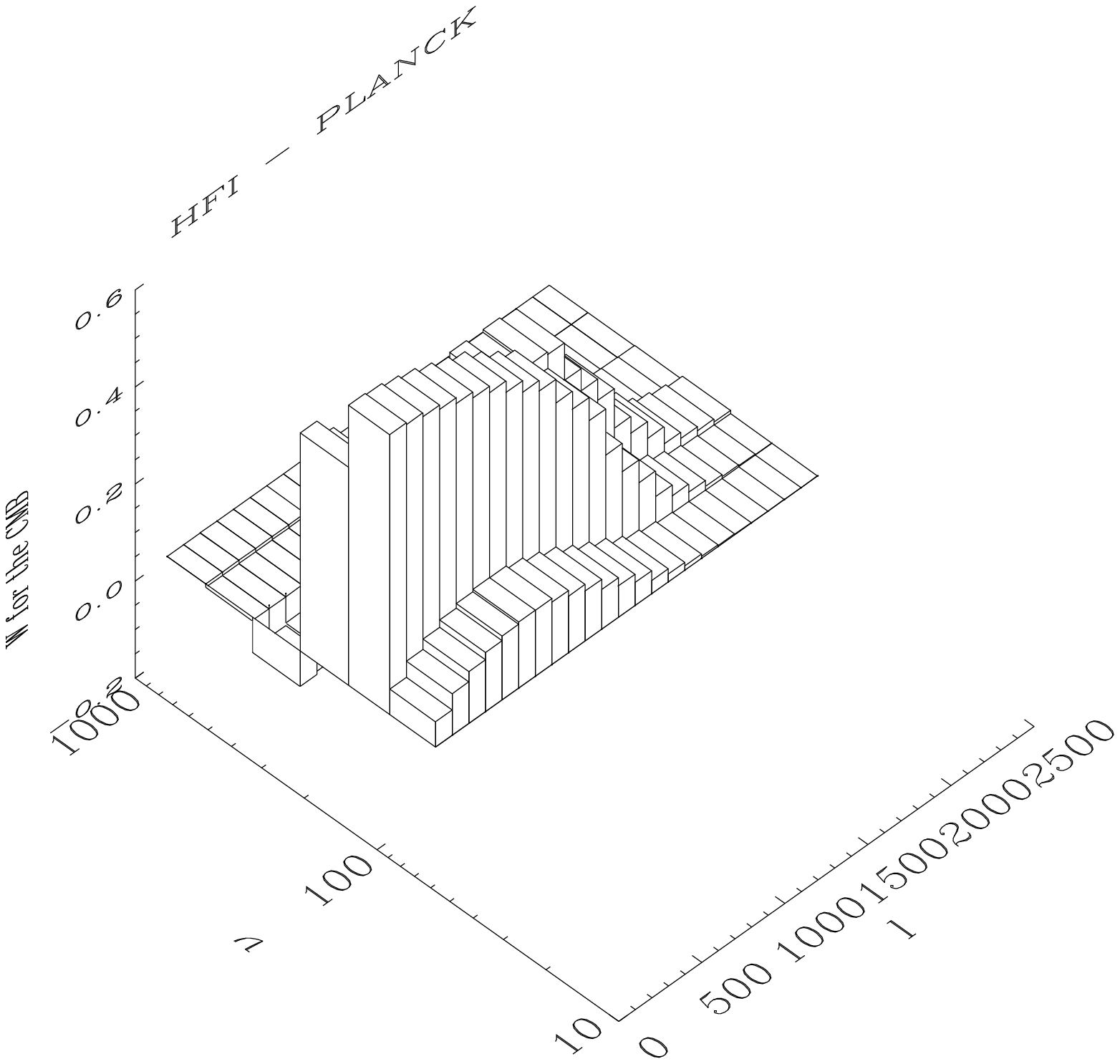,width=0.5\textwidth}
\psfig{figure=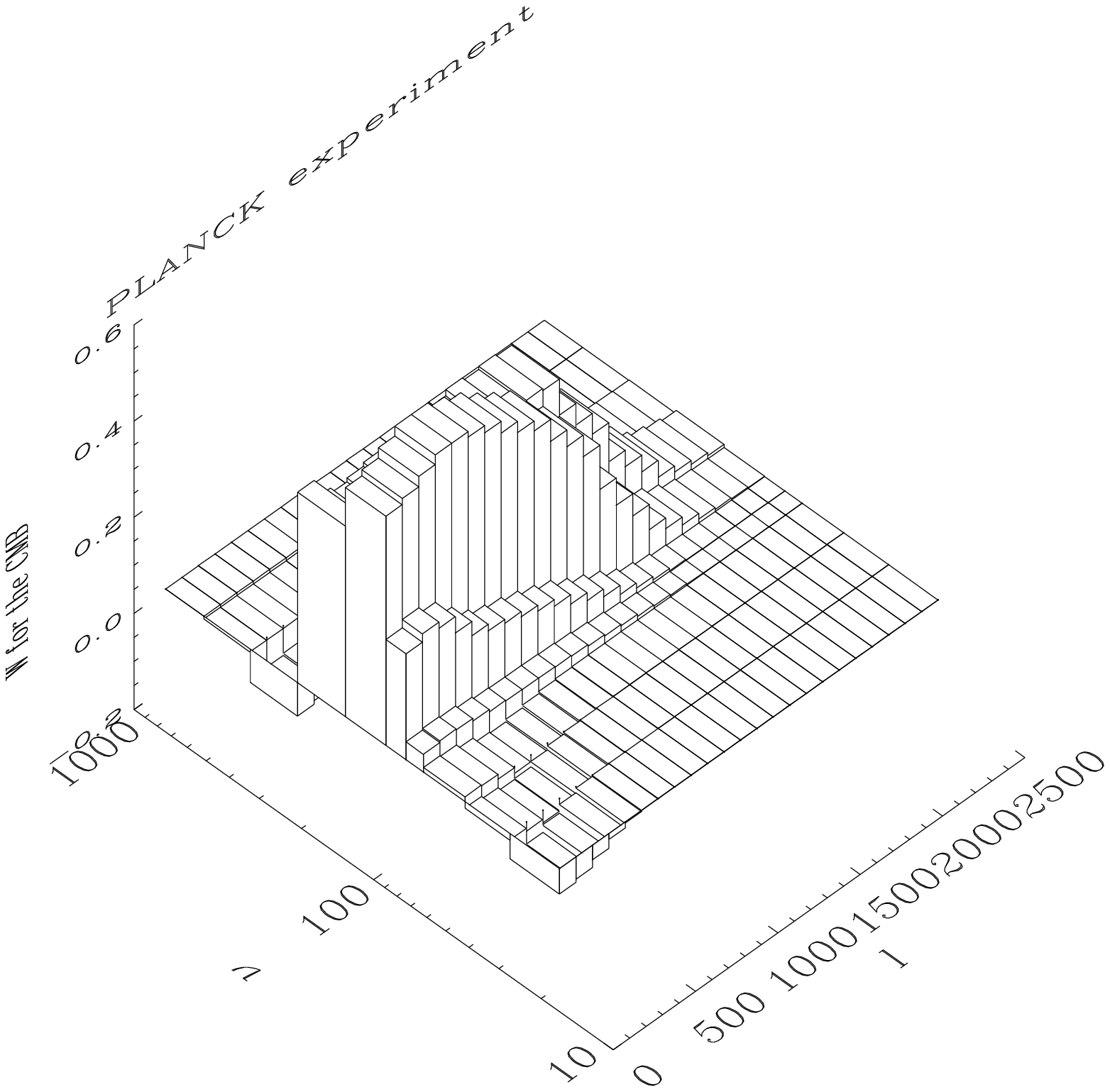,width=0.5\textwidth} } } }
\caption[]{CMB Wiener matrix elements at $\ell > 200$ for \map, the \lfi, the
\hfis and the full \plancks mission. Each bin in $\ell$ is the average over
one hundred contiguous values. The frequency bins are centred at the average
frequency of each channel, but their width in not representative of their
spectral width.}
\label{fig:wien_mat_cmb}
\end{figure}

Once the instrument, through the $\bA$ matrix, and the covariance matrix $\bC$
of the templates are known, the Wiener filter is entirely determined through
equation~\eqref{eq:sol_wien_mat}. Figure~\ref{fig:wien_mat_cmb} offers a
graphical presentation of the resulting values of the Wiener matrix
coefficients of the CMB component when we use our sky model (assuming
negligible errors in designing the filter). It shows how the different
frequency channels are weighted at different angular scales and thus how the
$\nu-\ell$ information gathered by the experiment is used.

In the \maps case, most weight is given to the 90 GHz channel. It is the only
one to gather CMB information at $\ell \simgt 600$, and it's own weight
becomes negligible at $\ell \simgt 1000$. In the \lfis case, the main CMB
channel is the 100 GHz one, which gathers information till $\ell \sim 1500$,
while the 70 GHz channel is of some help till $\ell \sim 1100$. The two other
channels only contribute at $\ell \simlt 500$. The situation is different in
the \hfis case, since no single channel dominates at all scales. The 143 GHz
channel is dominant till $\ell \sim 1400$ and useful till $\ell \sim 2000$,
while the 217 GHz channel becomes dominant at $\ell \simgt 1400$ and gather
CMB information till $\ell \sim 2300$. The 100 GHz channel contribution to the
CMB determination is only modest, and peaks at $\ell \sim 800$.

Once the information from the \lfis and the \hfis are considered jointly, the
previous \hfis weights are barely affected, and the \lfis weights appear to be
sizeable only in the 100 GHz channel, albeit at a fairly reduced level as
compared to the 100 GHz channel of the \hfis (which is itself only a modest
contributor to the CMB determination). As a result, this already suggests that
the impact of the \lfis on the high $\ell$ \hfis measurements ($\ell > 200$)
will be in controlling systematics and possibly in better determining the
foregrounds. Of course, one should note that this analysis assumes the
foregrounds are known well enough to design the optimal filter. Thus the \lfis
impact might be greater in a more realistic analysis when the filter is
designed with the only help of the measurements themselves.

%++++++++++++++++++++++++++++++++++++++++++++++++++++++++++++++++++++++++++++++
% Plots of Wiener matrices for the CMB
%++++++++++++++++++++++++++++++++++++++++++++++++++++++++++++++++++++++++++++++
%++++++++++++++++++++++++++++++++++++++++++++++++++++++++++++++++++++++++++++++

\subsection{Effective windows and beams}
%***************************************

%++++++++++++++++++++++++++++++++++++++++++++++++++++++++++++++++++++++++++++++
% Plots of Quality factor
%++++++++++++++++++++++++++++++++++++++++++++++++++++++++++++++++++++++++++++++
\begin{figure}[htbp] \centering \centerline{\vbox{\hbox{
\psfig{figure=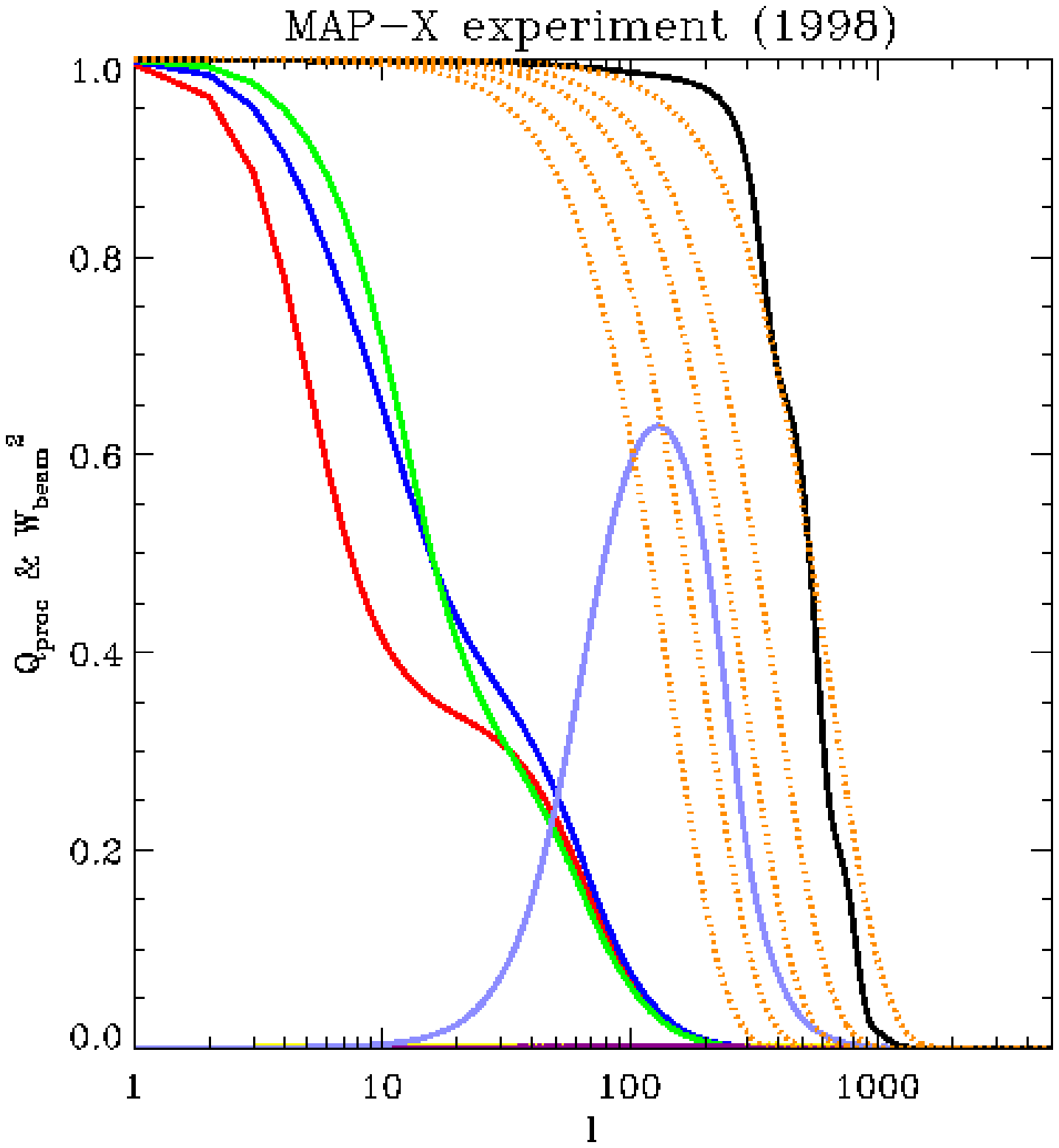,width=0.5\textwidth}
\psfig{figure=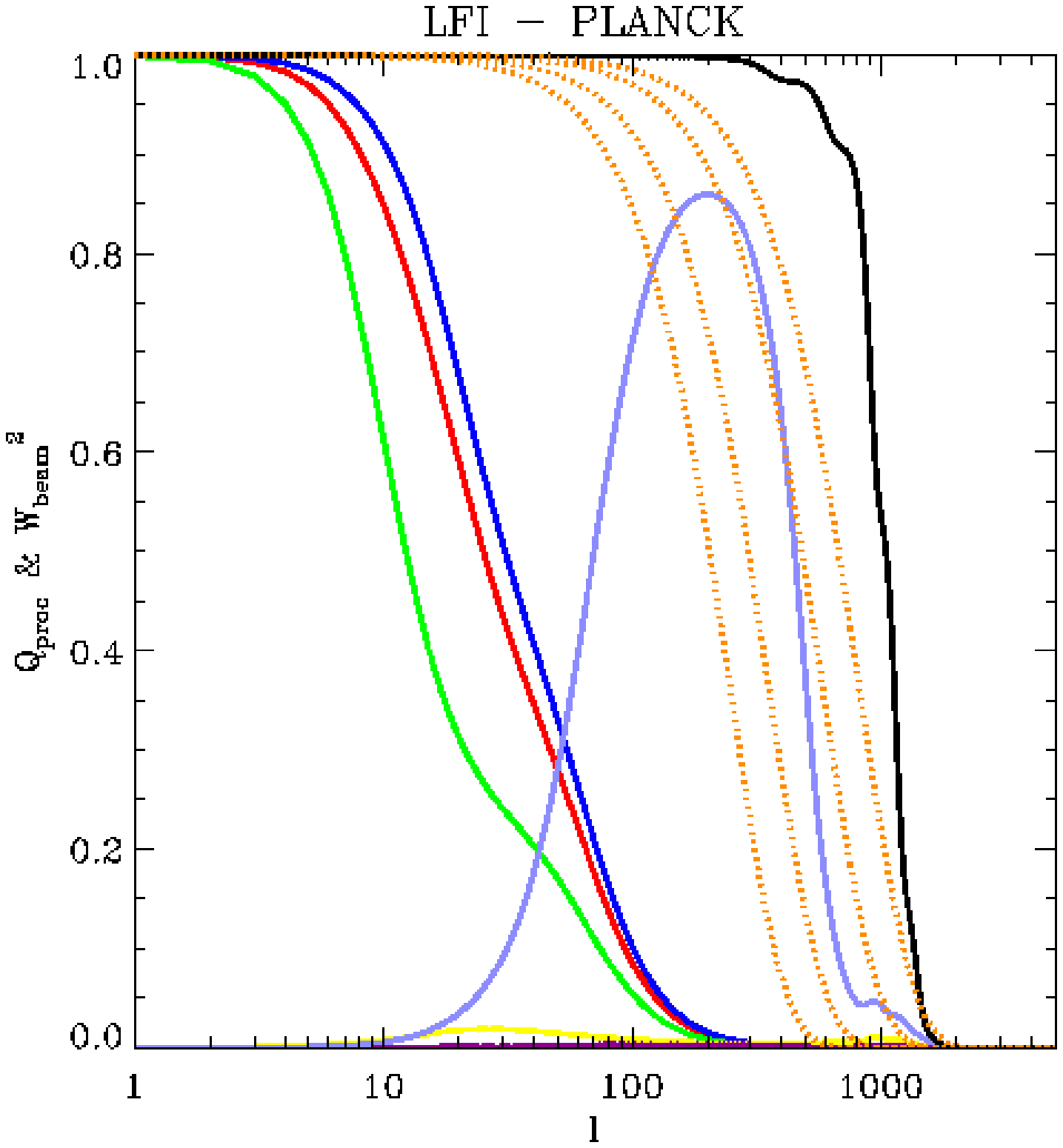,width=0.5\textwidth} }\hbox{
\psfig{figure=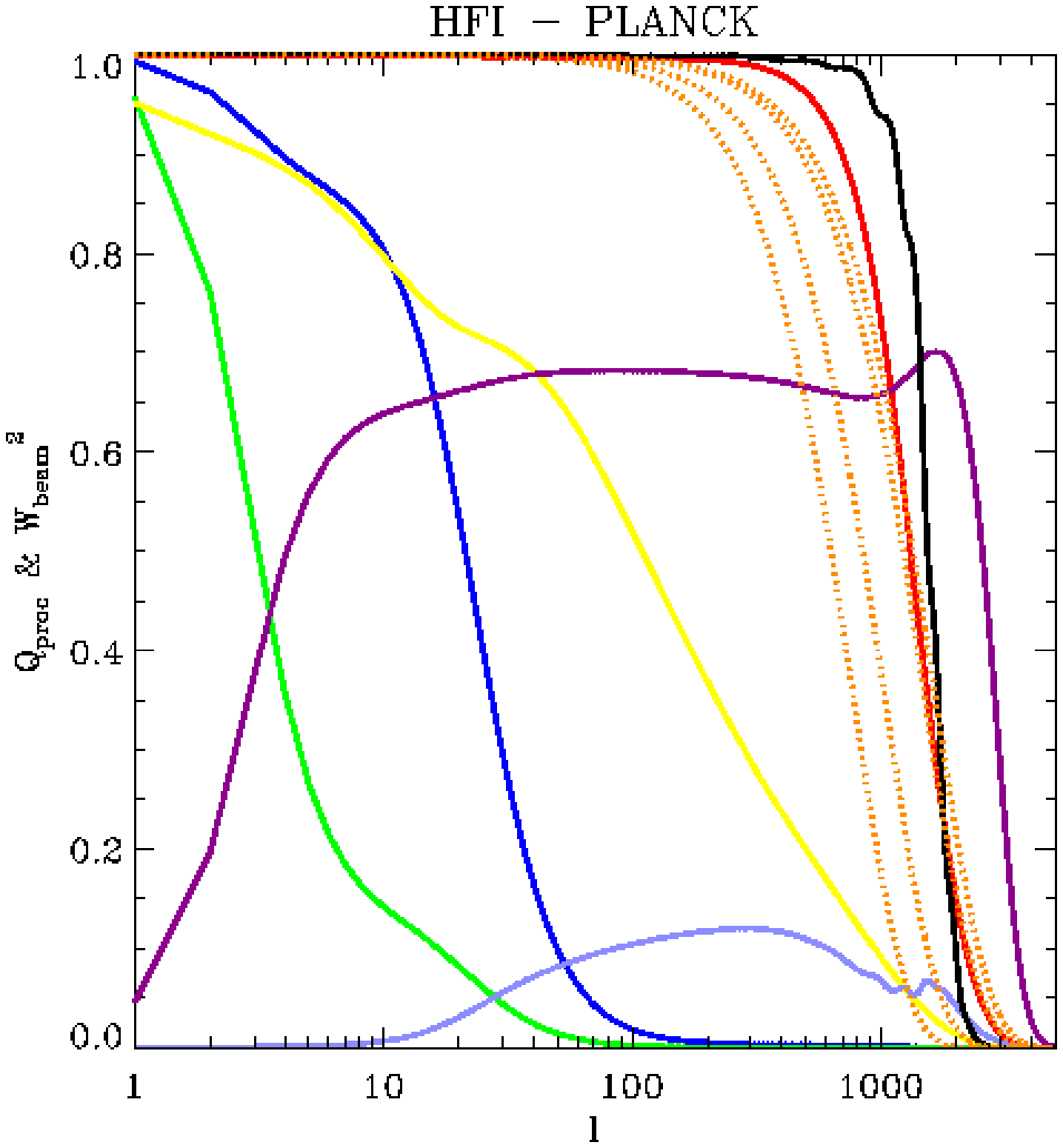,width=0.5\textwidth}
\psfig{figure=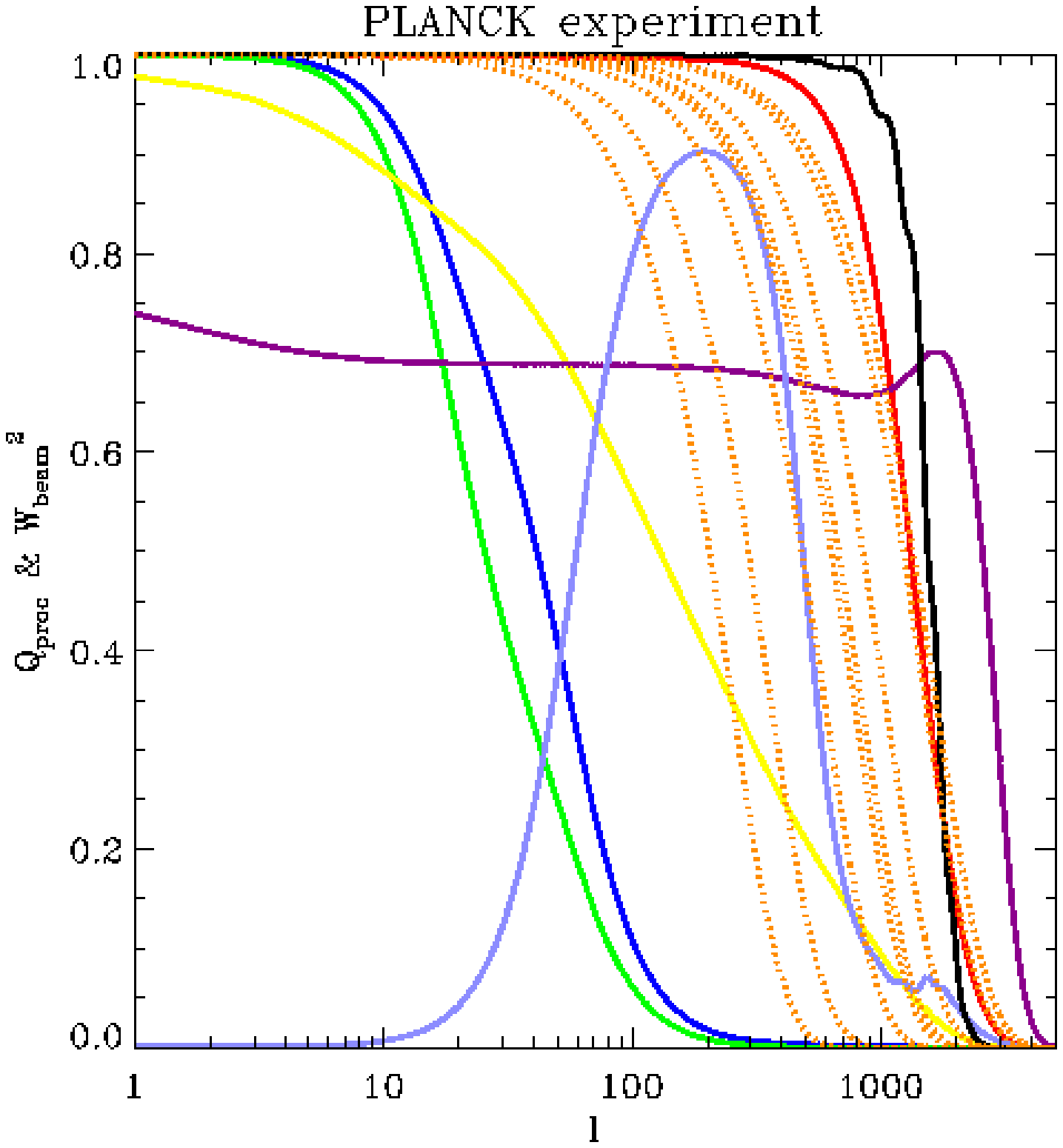,width=0.5\textwidth} } } }
\caption[]{Square of the $\ell$-space effective windows (\ie $Q_p$) for the
\lfi, \hfi, and \planck\. As usual, black is for the CMB, red, blue, and green
are for the Galactic components, yellow is for the SZ contribution. The
transform of the channel optical beams are shown for comparison as dotted
orange lines. }
\label{fig:qual_fac}
\end{figure}
%++++++++++++++++++++++++++++++++++++++++++++++++++++++++++++++++++++++++++++++

Figure~\ref{fig:qual_fac} gives the {\em effective} $\ell$-space window of the
experiment for each component, $Q_p(\ell)$, to be compared with the individual
windows of each channel, $w_i(\ell)$. For the CMB, it shows the gain obtained
by combining channels through Wiener filtering. For \map, the effective beam
is nearly equal to it's 90 GHz beam, while the improved noise level of the
\lfis allows it to do much better than it's ``best'' beam would suggest (in
particular at $\ell \sim 1000$). None of these two experiments can separate
the SZ or the infrared sources background contribution. As expected the
Galactic foregrounds are poorly recovered except on the largest scales (small
$\ell$) where their signal is strongest (we assumed spectra $\propto
\ell^{-3}$). The only exception is that of the \HI-correlated component which
is of course very well traced by the \hfi. When the \lfis and the \hfis are
combined, the main improvement appears to be on the SZ contribution thanks to
the greater spectral level arm offered to distinguish it from other
foregrounds.

%++++++++++++++++++++++++++++++++++++++++++++++++++++++++++++++++++++++++++++++
% Plots of Quality factor
%++++++++++++++++++++++++++++++++++++++++++++++++++++++++++++++++++++++++++++++
\begin{figure}[htbp] \centering \centerline{\hbox{
\psfig{figure=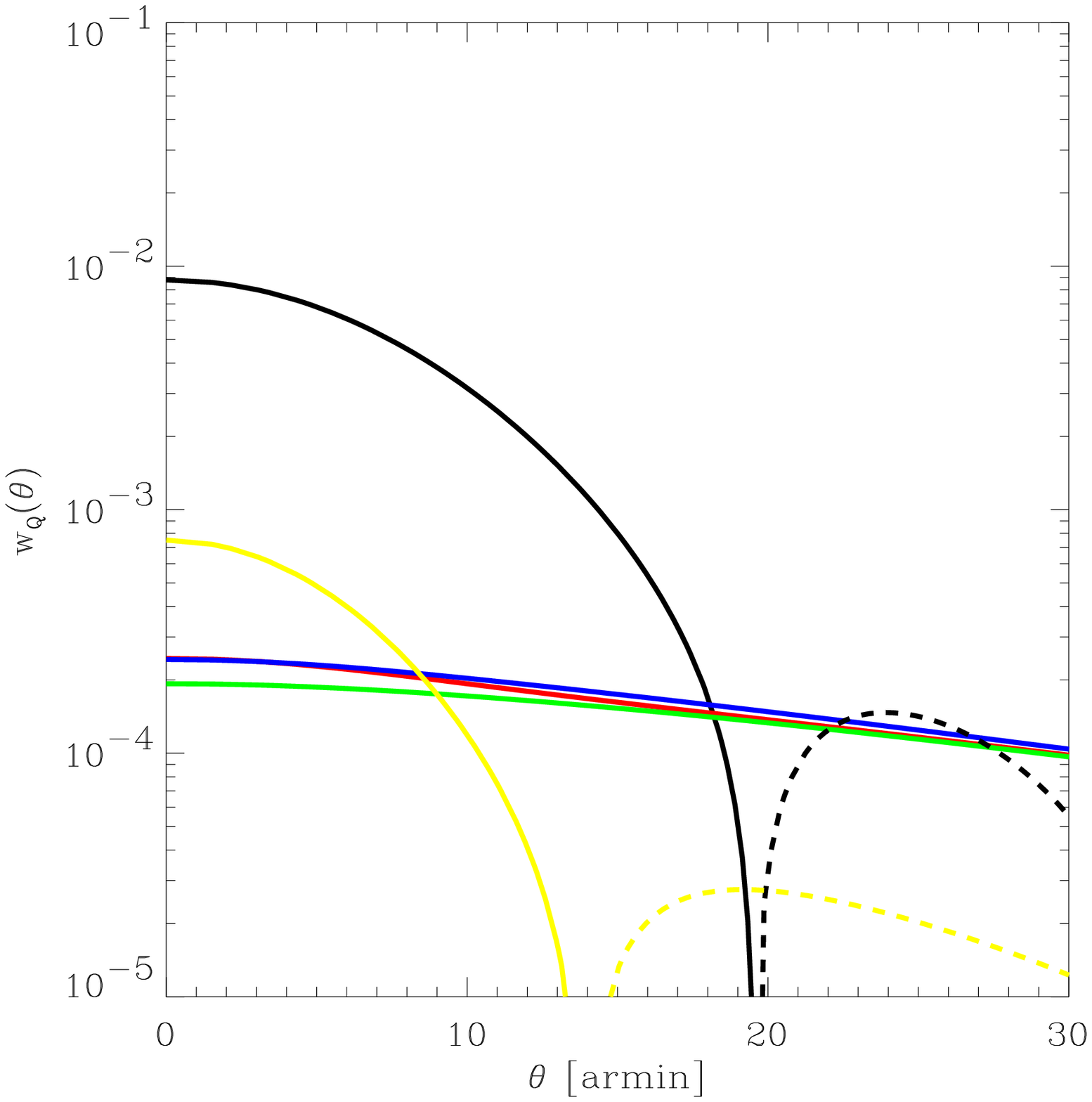,width=0.5\textwidth}
\psfig{figure=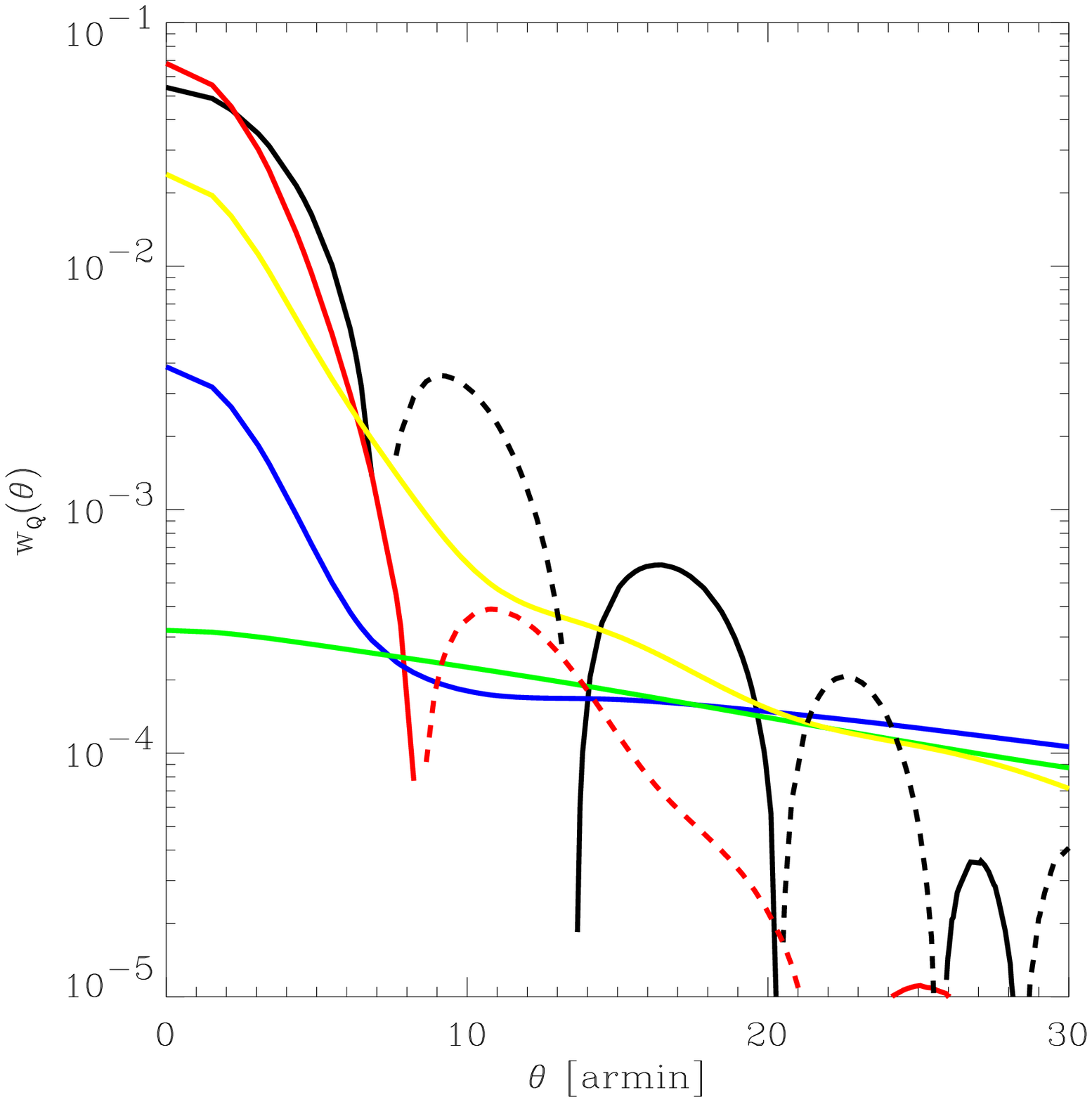,width=0.5\textwidth} } }
\caption[]{Corresponding real space effective beams (negative parts are
denoted by dashes). As usual, black is for the CMB, red, blue, and green are
for the Galactic components, yellow is for the SZ contribution.}
\label{fig:qual_beam}
\end{figure}
%++++++++++++++++++++++++++++++++++++++++++++++++++++++++++++++++++++++++++++++

Figure~\ref{fig:qual_beam} shows the inverse spherical harmonics transforms of
the $Q_p(\ell)$, \ie\ the {\em effective} angular beams. They show the
effective point spread function by which the underlying maps of the component
emissions would have been convolved once the analysis is complete. One can see
in particular that the FWHM of the CMB beam is $\simeq 16.5\prim$ for \maps
and $\simeq 7.5\prim$ for \plancks (and even better for the dust components).

\subsection{Maps reconstruction errors}
%**************************************

%++++++++++++++++++++++++++++++++++++++++++++++++++++++++++++++++++++++++++++++
% Plots of Errors
%++++++++++++++++++++++++++++++++++++++++++++++++++++++++++++++++++++++++++++++
\begin{figure}[htbp] \centering \centerline{ \vbox{ \hbox{
\psfig{figure=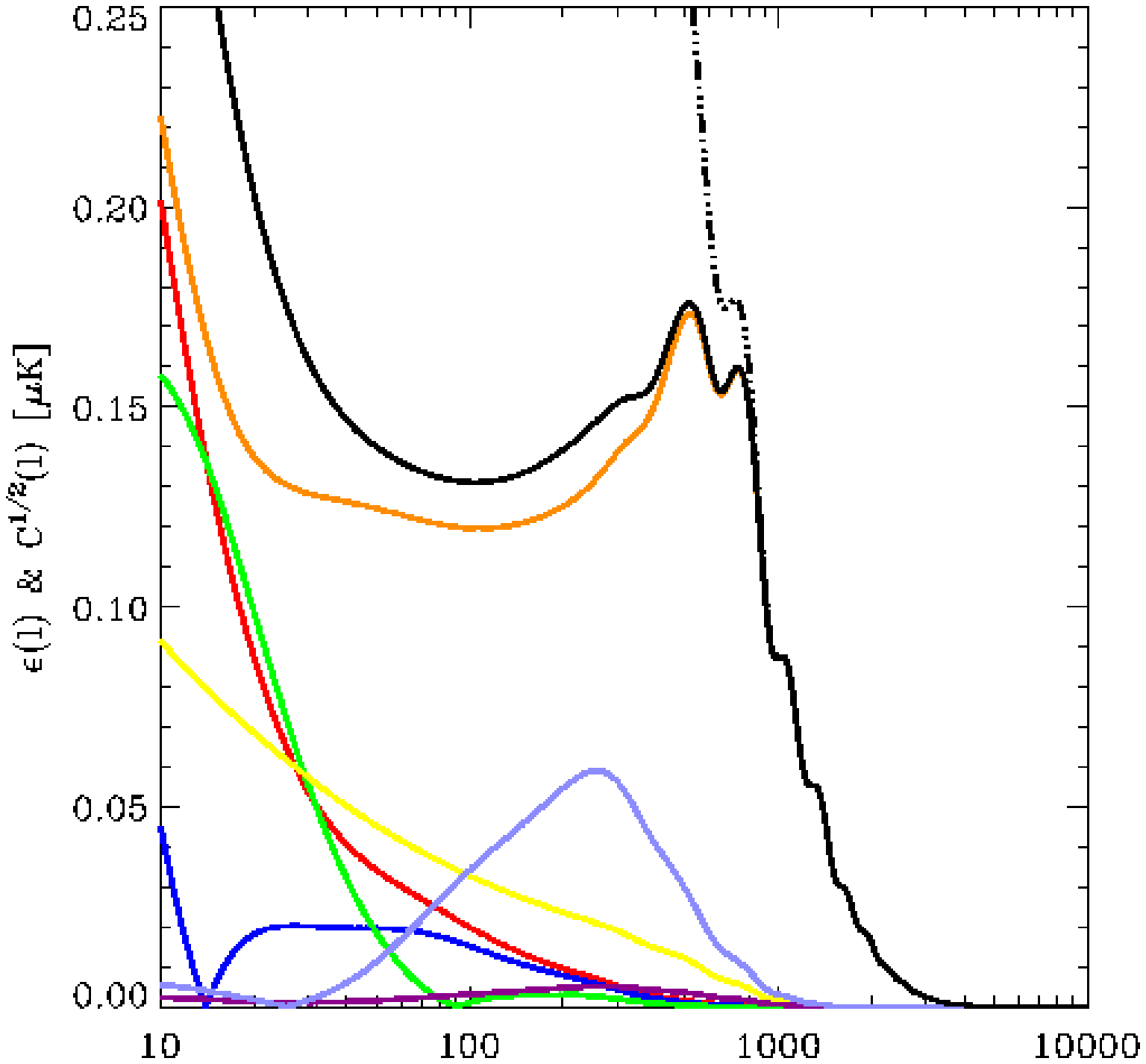,width=0.5\textwidth}
\psfig{figure=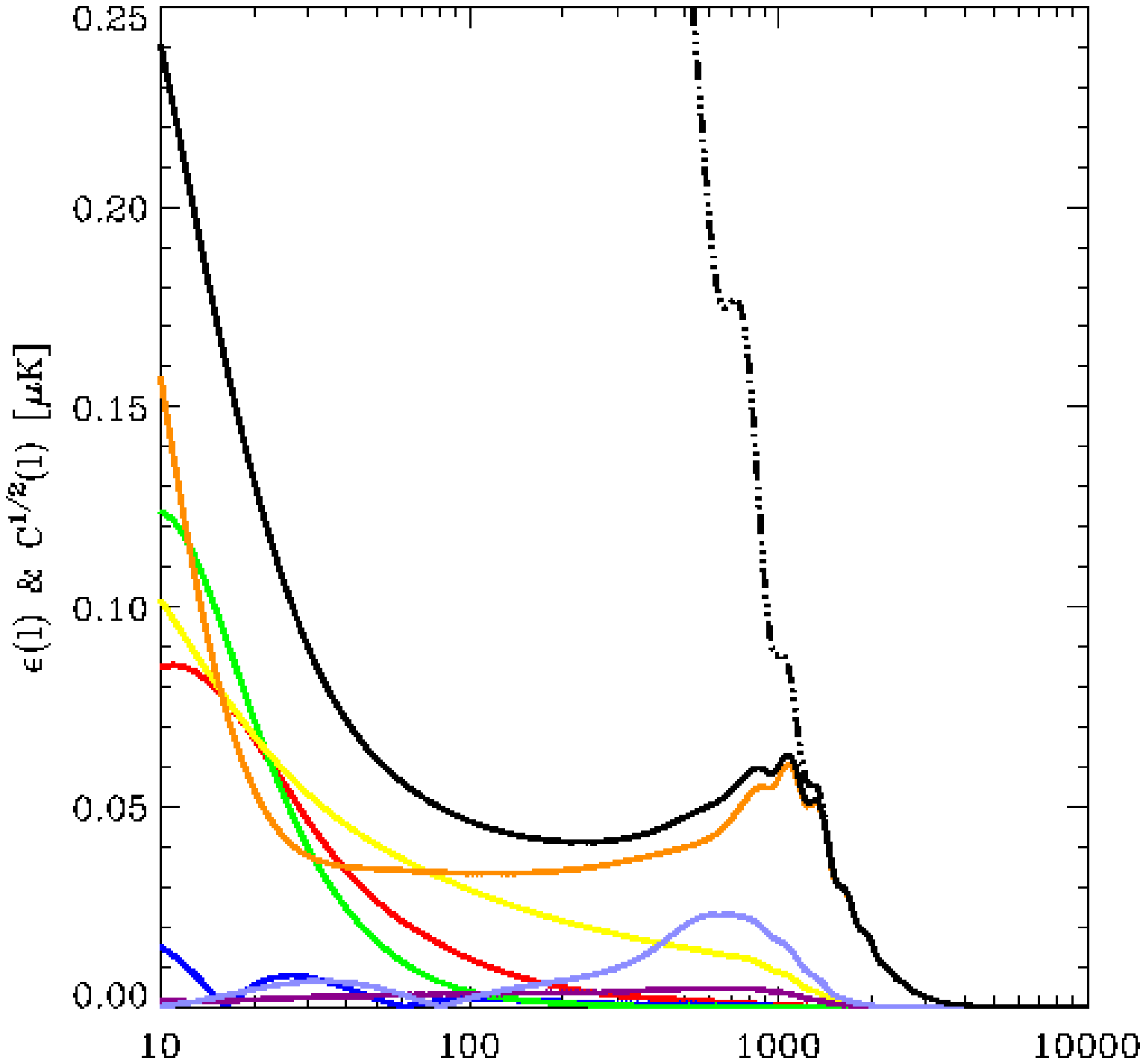,width=0.5\textwidth} } \hbox{
\psfig{figure=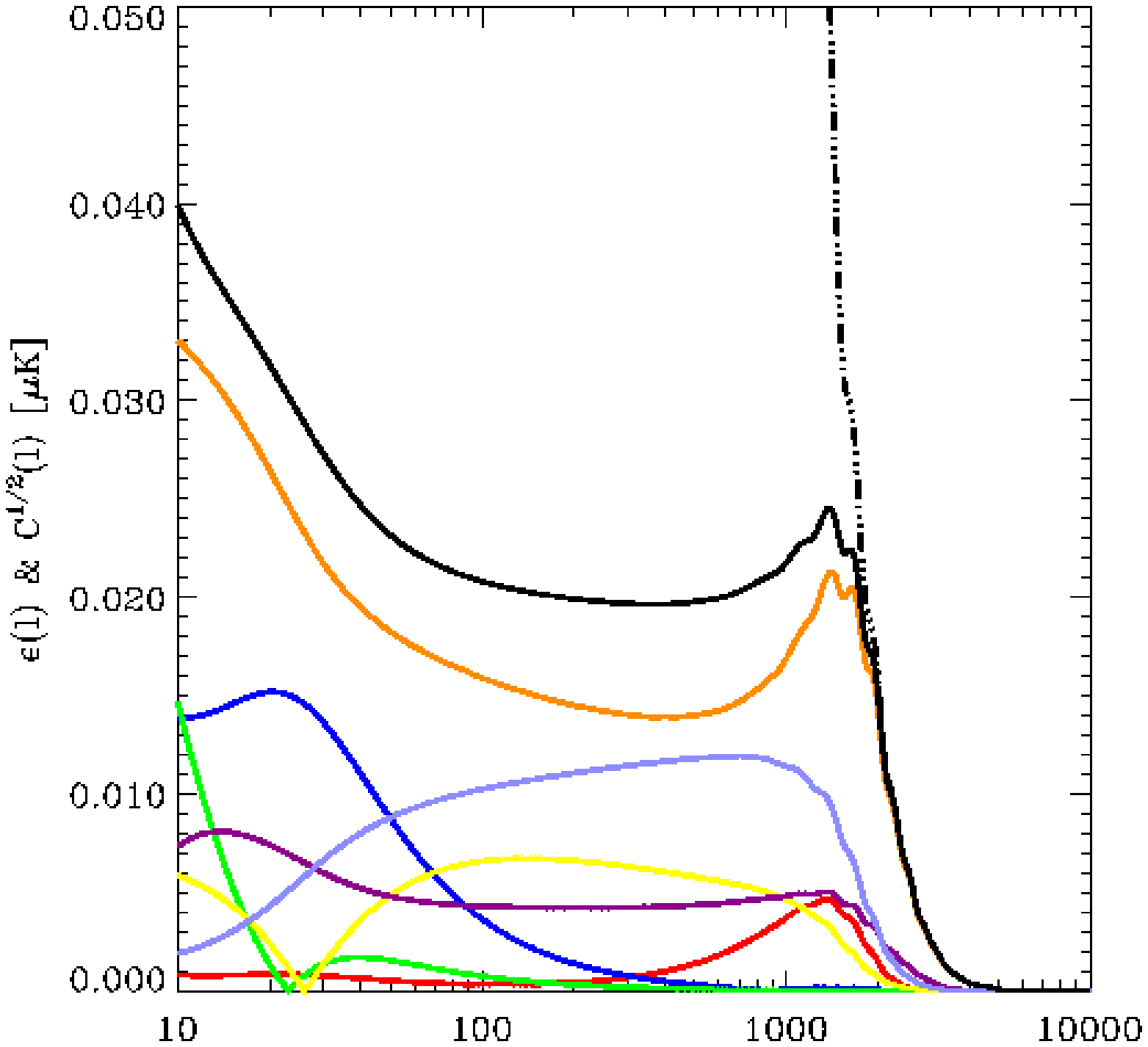,width=0.5\textwidth}
\psfig{figure=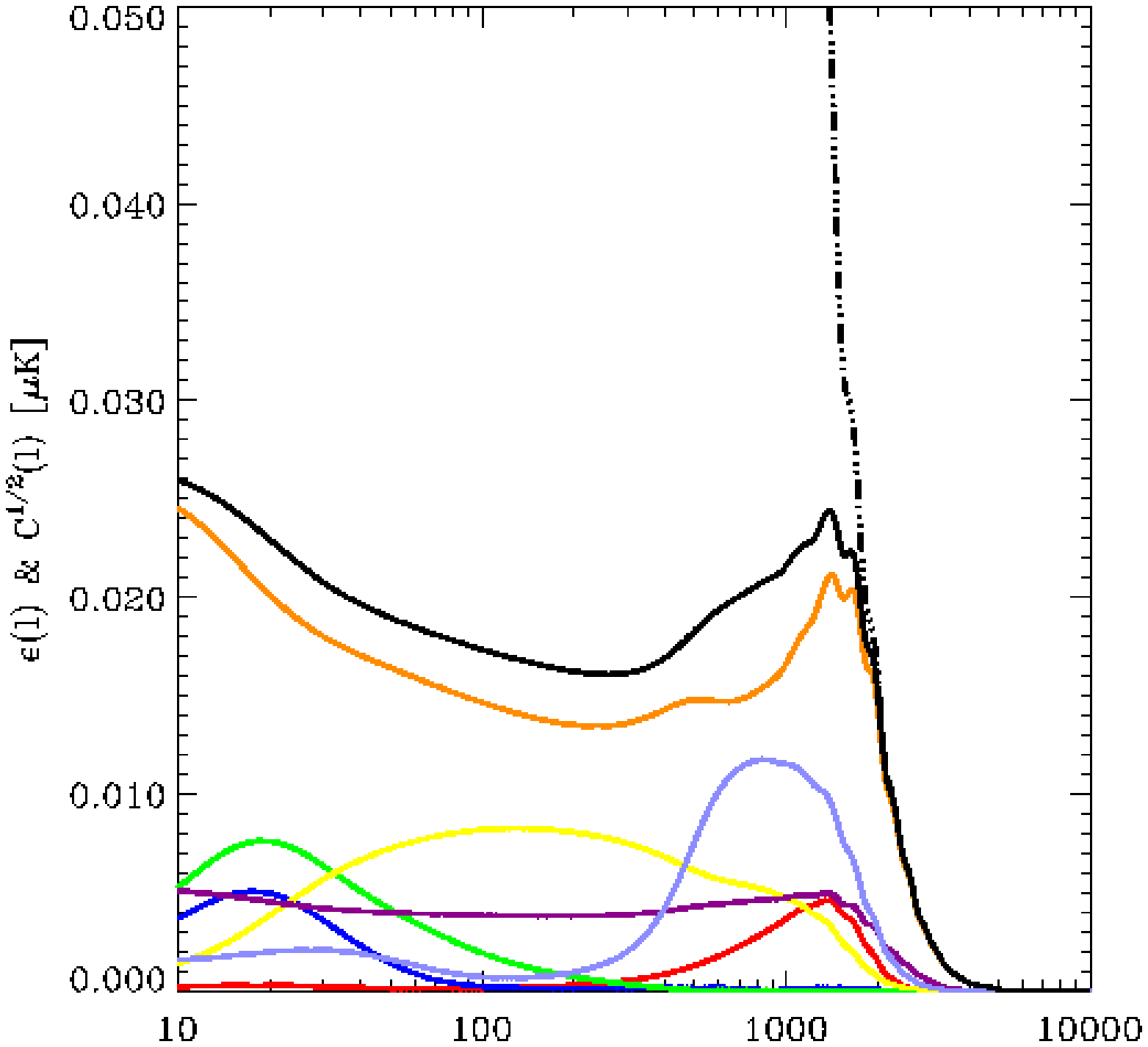,width=0.5\textwidth} } } }
\caption[]{CMB reconstruction error contributed by each component (with the
same line coding than above) and their total in black. The integral of
$\ell(\ell+1)\varepsilon^2/2\pi$ would give the reconstruction error of the
map. The dashed-dotted line shows for comparison $C(\ell)^{1/2}$ for the
CMB. For \planck, the total of the reconstruction error is nearly constant at
all $\ell$ and corresponds to $\varepsilon(\ell) \simlt 0.025$\,\microK, a
factor greater than six below the \maps case.}
\label{fig:errors}
\end{figure}
%++++++++++++++++++++++++++++++++++++++++++++++++++++++++++++++++++++++++++++++

\begin{figure}[htbp]\centering \centerline{ \hbox{
\psfig{figure=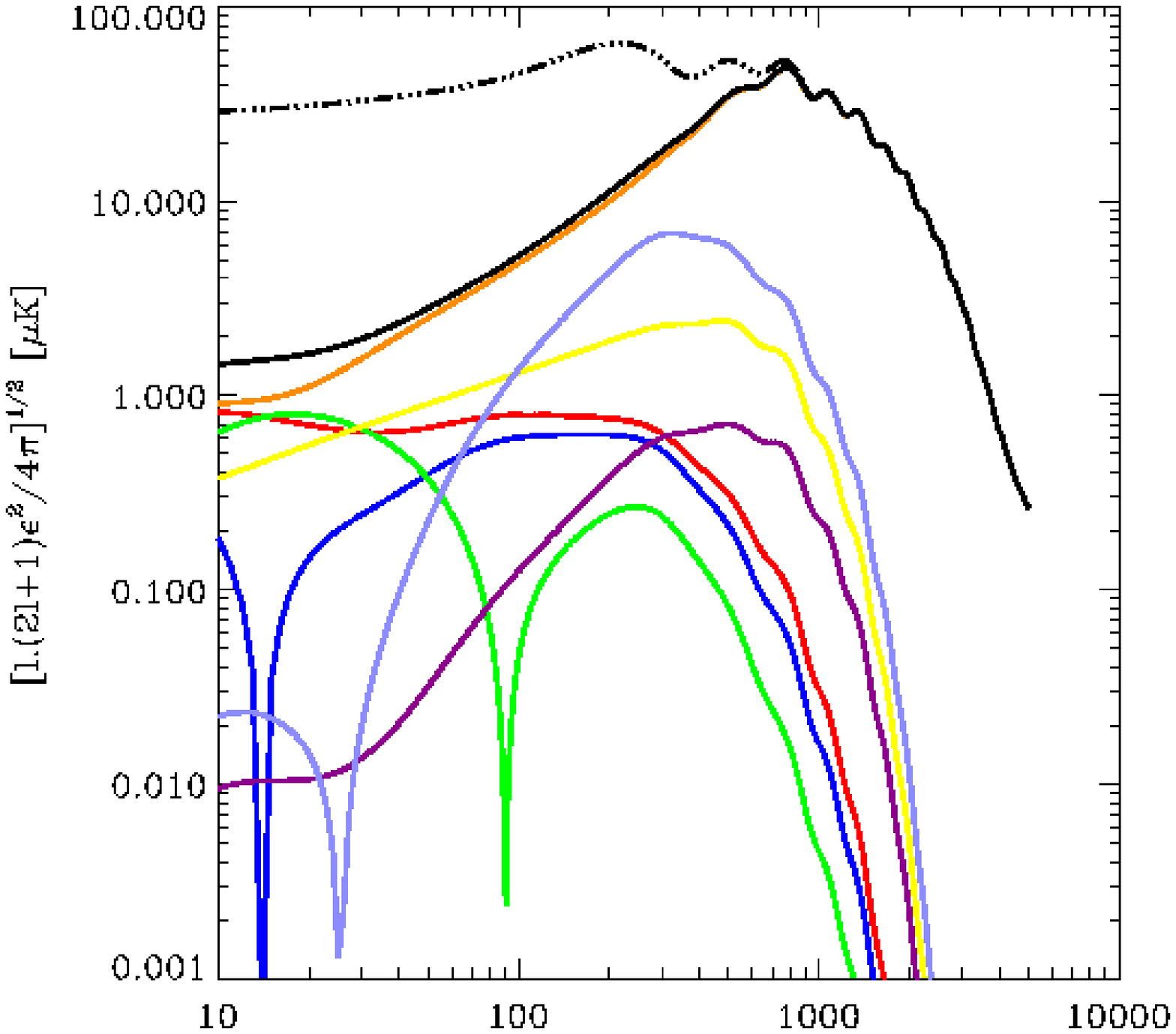,width=0.5\textwidth}
\psfig{figure=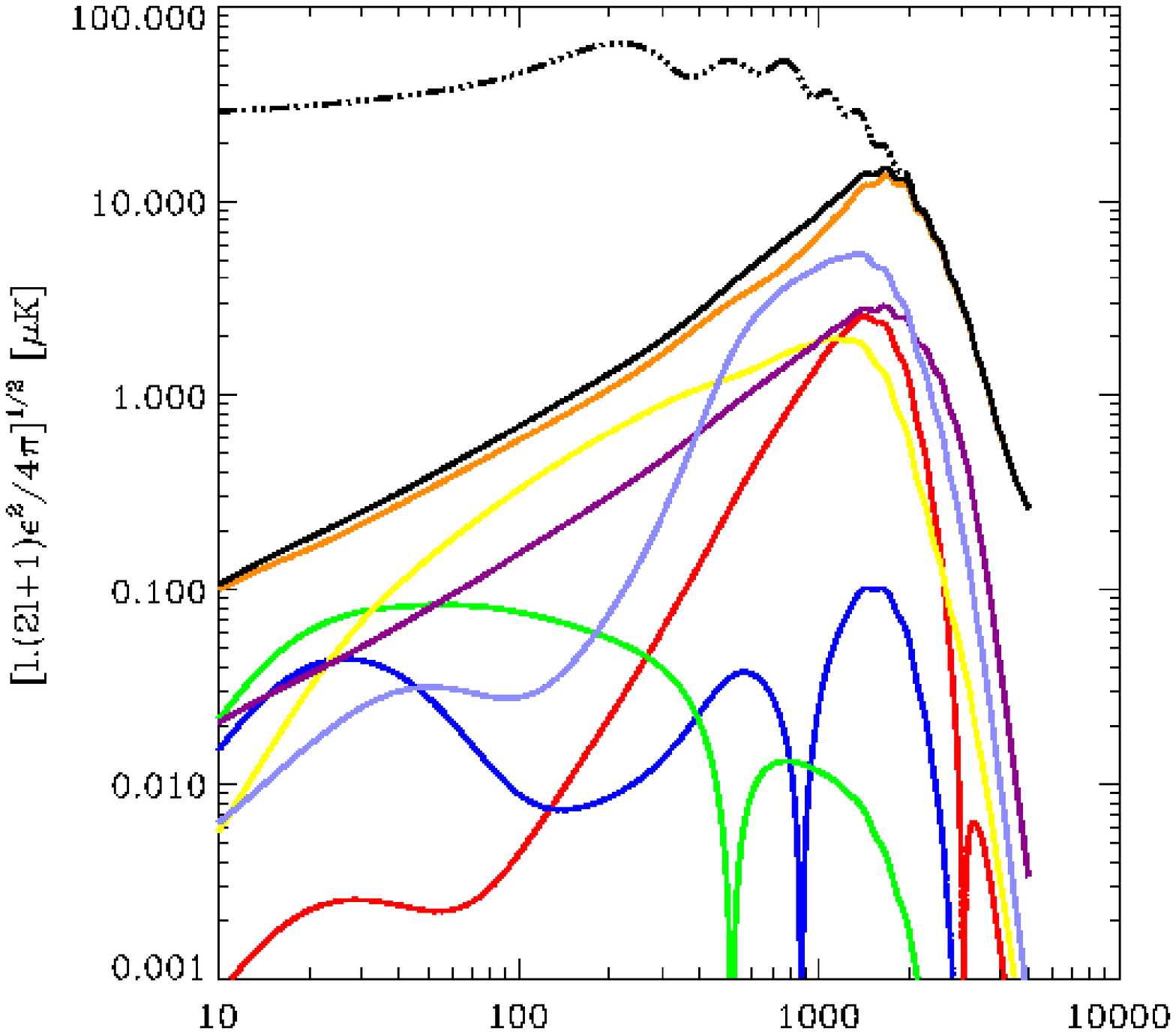,width=0.5\textwidth} } }
\caption[]{CMB reconstruction errors compared with the input spectrum in the
\maps (left) and \plancks case (right).}
\label{fig:errcomp} 
\end{figure}

We now estimate the spectrum of the reconstruction errors in the
map. Figure~\ref{fig:errors} compares these residual errors per individual
$\bl$ mode with the typical amplitude of the true signal. The first thing to
note is that the error spectrum cannot be accurately modelled as a simple
scale-independent white noise (as expected from the $\nu-\ell$ shape of the
Wiener matrix for the CMB) . We also see that the \lfis improves on \map\
mostly by decreasing the noise contribution, but at the largest scales where
the galactic contributions are also lowered. The \lfis taken jointly with the
\hfis helps in reducing the low-$\ell$ galactic emissions, and the power
leakage from the radio-sources background. The contributions per logarithmic
bin of $\ell$ to the variance are compared in figure~\ref{fig:errcomp}. It is
interesting to note that in the \plancks case the largest contribution but
noise comes from SZ clusters in the range $50 \simgt \ell \simgt 500$, to be
then superseded by the leakage from the radio-sources background.

%++++++++++++++++++++++++++++++++++++++++++++++++++++++++++++++++++++++++++++++
% Plots of Errors compared to CMB
%++++++++++++++++++++++++++++++++++++++++++++++++++++++++++++++++++++++++++++++
%\psfig{figure=eps_ratio_bpl.ps,width=0.5\textwidth}

%++++++++++++++++++++++++++++++++++++++++++++++++++++++++++++++++++++++++++++++

\subsection{Foreground and noise induced errors on the CMB power spectrum}
%*************************************************************************

\begin{figure}[htbp] \centering \centerline{ \vbox{
\psfig{file=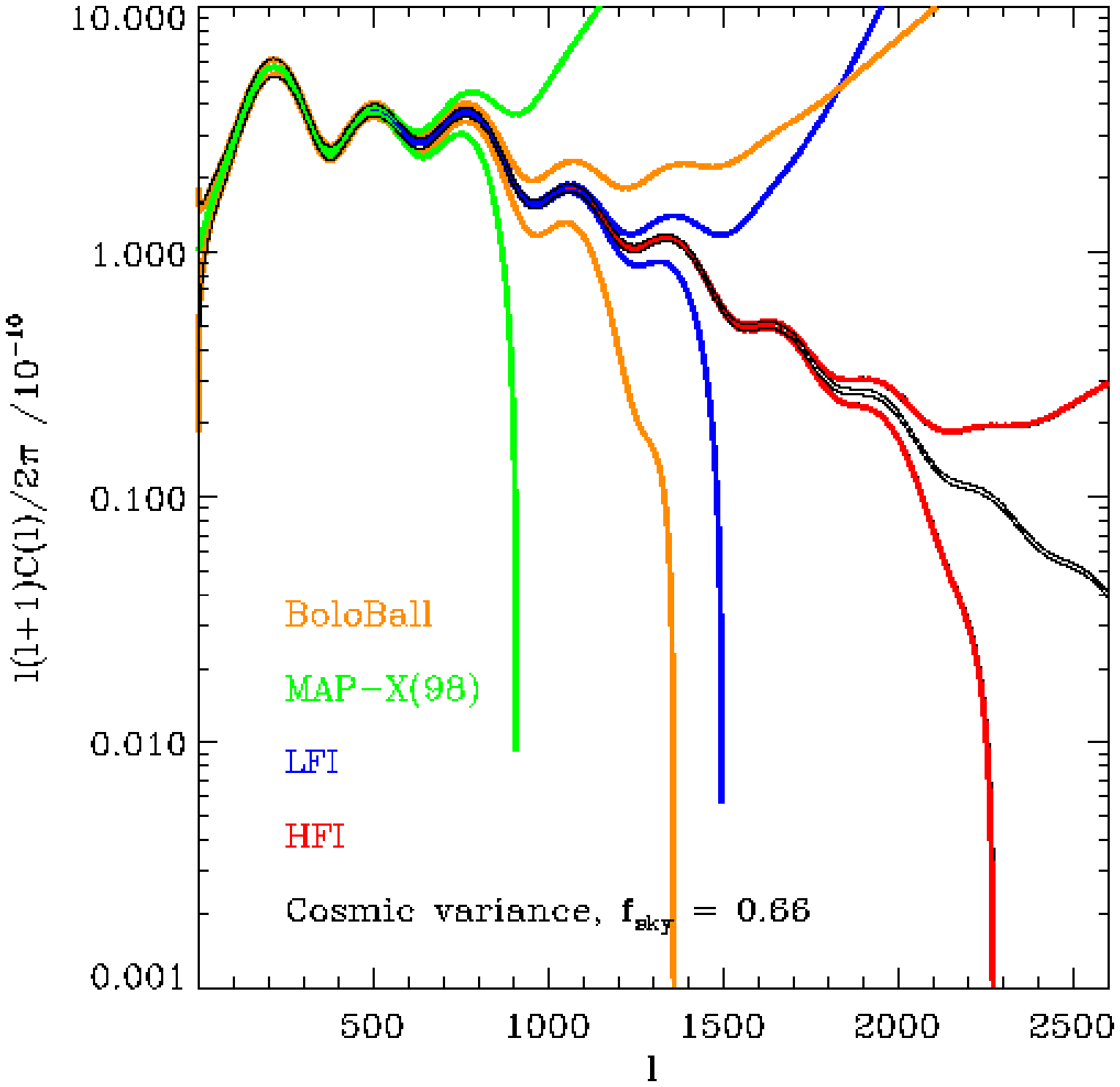,width=0.75\textwidth}
\psfig{file=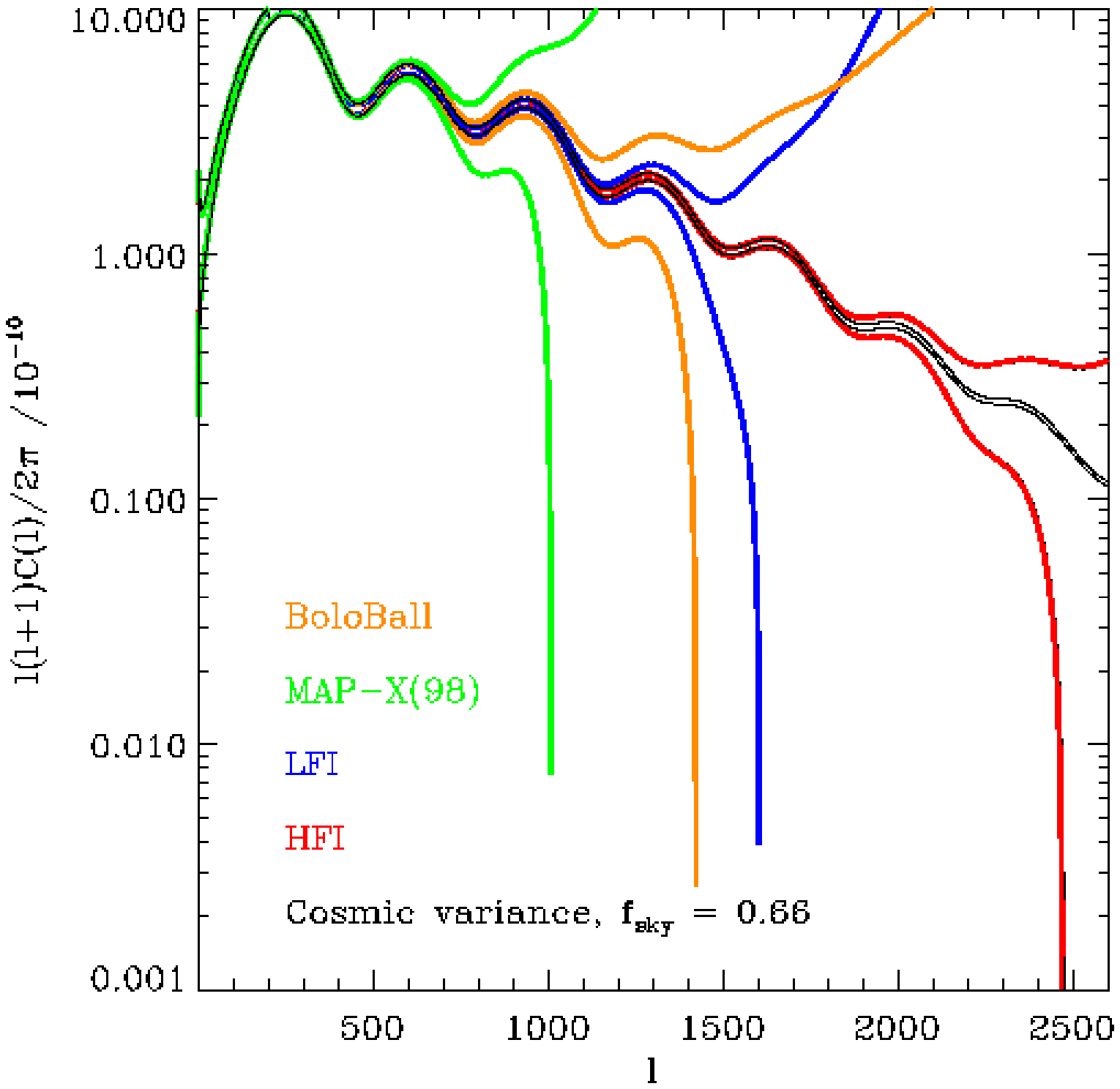,width=0.75\textwidth} } }
\caption[]{Expected errors on the amplitudes of each mode individually (no
band averaging) for different experiments (the full \plancks is
indistinguishable from the \hfis case). The thin central lines gives the
target theory plus or minus the cosmic variance, for a coverage of 2/3 of the
sky. a) Standard CDM b) Lambda CDM (with $\Omega_b=0.05,\ \Omega_{CDM}=0.25,\
\Omega_\Lambda = 0.7,\ h=0.5$).}
\label{fig:qual_comp}
\end{figure}

Finally, we can estimate the uncertainty added by the noise and foreground
removal to the CMB power spectrum. Figure~\ref{fig:qual_comp} shows the
envelope of the 1-$\sigma$ error expected from the the current design of \map
(green), the \lfis (blue), and the \hfis or the full \plancks (red), as well
as a ``boloBall'' experiment meant to show what bolometers on balloon might
achieve soon; the experimental characteristics used in this comparison may be
found in table~\ref{tab:perfs}. Note that there has been no ``band averaging''
in this plot\footnote{The band averaging would reduce the error bars on the
smoothed $C(\ell)$ approximately by the square root of the number of
multipoles in each band, if the modes are indeed independent.}, which means
that there is still cosmological signal to be extracted from the \hfis at
$\ell \sim 2500$. The message from this plot is thus excellent news since it
tells us that even accounting for foregrounds, \plancks will be able to probe
the very weak tail of the power spectrum and allow breaking the near
degeneracy between some of the cosmological parameters.

The second panel of the figure shows the variations induced by changing the
target CMB theory. In this case, a flat universe with a (normalised)
cosmological constant of 0.7 was assumed. What \maps can say on the third peak
would then be much reduced, while on the contrary the \hfis would probe
accurately one more peak. This illustrates the fact that the errors on the
signal will of course depend on the amplitude of the signal; it is a reminder
of the fact that the numbers given so far are meant to be illustrative. As
already stressed, the merit of this approach is to offer a quantitative tool
for comparisons. The impact of changing the foreground model will be addressed
in section \S\ref{sec:varsky}.

%++++++++++++++++++++++++++++++++++++++++++++++++++++++++++++++++++++++++++++++
% Plots C(ell) errors
%++++++++++++++++++++++++++++++++++++++++++++++++++++++++++++++++++++++++++++++
%++++++++++++++++++++++++++++++++++++++++++++++++++++++++++++++++++++++++++++++

\subsection{Wiener versus ``na\"{\i}ve'' error estimates:\label{sec:naive}}
%--------------------------------------------------------------------------

In the previous section we estimated the error bars reachable on the CMB power
spectrum assuming a sky model and a separation method. In effect we found an
equivalent ideal experiment described by an ``on-sky'' noise power spectrum
$[w^{-2} C_N]_{eff} = C_{CMB} [1/Q_{CMB}-1]$. Numerous papers have been
devoted to estimating the possible accuracy of determinations of the
cosmological parameters from CMB experiments. To forecast this accuracy,
various recipes were used to derive an estimate of this effective on-sky noise
spectrum from tables of performance of the experiment such as
table~\ref{tab:perfs}. One option has been to assume that all frequency
channels but one are used to remove the foregrounds contribution from the
``best'' CMB channel, i.e. the one closest to 100 GHz, or the most sensitive
one. A less radical proposition by \citet{1997MNRAS.291L..33B} has been to
assume that all channels but $N_{keep}$ are used to clean the $N_{keep}$
closest to 100 GHz, which are then assumed to contain only the CMB signal and
noise, with an effective ``on-sky'' power spectrum of this noise $[w^{-2}
C_N]_{eff}$ given by
\begin{equation}
        {1 \over [w^{-2} C_N]_{eff} } = \sum_{i=1}^{N_{keep}} {1 \over
        w_i^{-2} C_{N,i} } , \label{eq:naive} .
\end{equation}
This is a generalisation of the formula~\eqref{eq:SigNois} to unequal noise
power spectra, $C_{N,i}$ in different channels. This approach does not tell
how many and which $N_{keep}$ channels can be considered clean once the
$N_c-N_{keep}$ others have been somehow used for foreground removal.

\begin{figure}[htbp] \centering \centerline{ \hbox{ 
\psfig{figure=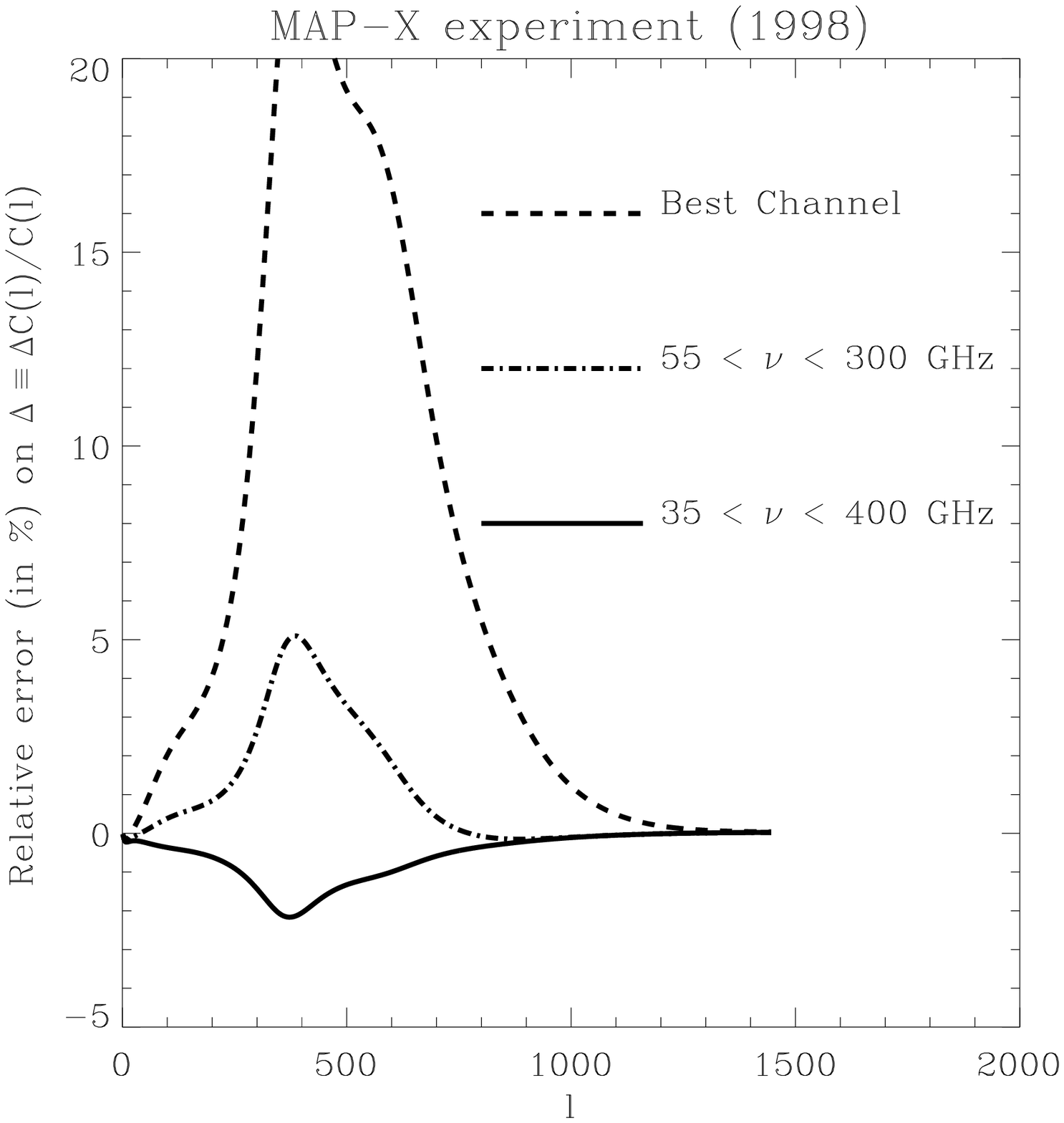,width=0.5\textwidth}
\psfig{figure=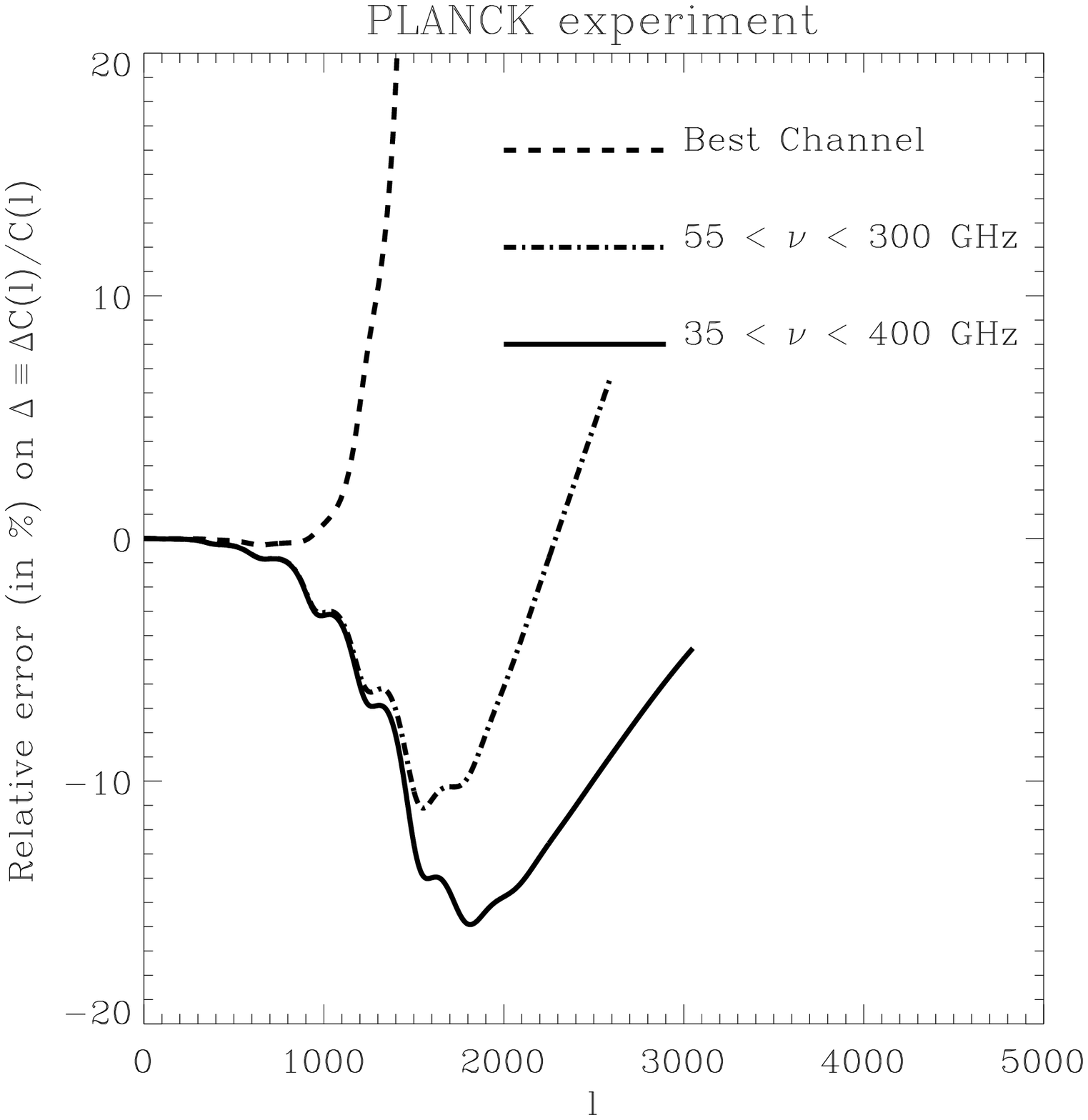,width=0.5\textwidth} } }
\caption[]{Ratio versus scale of na\"{\i}ve error estimates of the CMB power
spectrum to the one derived from the quality factor. This ratio is given both
for \maps (left) and the full \plancks (right).}
\label{fig:naive}
\end{figure}

Figure~\ref{fig:naive} compares the fractional error $\Delta C_{CMB}/C_{CMB}$
expected from Wiener filtering and different prescriptions, namely using the
best channel or all channels in some frequency range combined according to
equation~\eqref{eq:naive}. As expected we see that the error in the
``na\"{\i}ve'' estimates of the uncertainty of the CMB power spectrum goes to
zero at small $\ell$, when cosmic variance dominates over the noise, and at
large $\ell$ when detector noise dominates all foreground contributions. If
only the best channel (dashes) of the experiment is retained (i.e. $\nu = 90,
143$ GHz respectively for \maps \& \planck), the error is of course always
over estimated, i.e. optimal use of all channels by Wiener filtering does much
better than this na\"{\i}ve estimator would suggest.

The error in estimating the noise level is much decreased if the channels
between 55 and 300 GHz (i.e. the 60 \& 90 GHz channels of \maps and the 70,
100, 143 and 217 GHZ channels of \planck) are retained in the
sum~\eqref{eq:naive}. A better estimate for \maps obtains if one also includes
the 40 GHz channel, since one now underestimates the error by only the few per
cent over the full $\ell$ range. For \plancks on the contrary retaining all
channels between 55 and 300 GHz already underestimates the errors by 10 \% at
$\ell \sim 1500$. Assuming that all \plancks channels between 35 and 400\,GHz
can be cleaned with the help of the remaining ones further increases the error
on the CMB uncertainty to more than 15 \% at $\ell \sim 1800$.

More importantly, this error is scale-dependent, \ie it shows that no
``na\"{\i}ve'' prescription gives a fair estimate of the CMB power spectrum
uncertainties at all scales. This is not surprising, given the shapes of the
Wiener matrixes we saw in figure~\ref{fig:wien_mat_cmb}. These shapes indeed
illustrate that an optimal estimate of the CMB component obtains by weighting
differently as a function of scale the informations coming from different
channels, according to their angular resolution, and to the levels of their
detector noise and of their foregrounds contaminations.

The results above show that the relative error between Wiener based and other
error estimates of the CMB spectrum are relatively small, but they are scale
dependent, and the same na\"{\i}ve prescription might either underestimate or
overestimate the errors depending on the experiment considered. This invites
exercising some restraint in using results based on these simple
prescriptions, like the usual comparison of accuracy of cosmological
parameters determinations by different experiments. Of course, our Wiener
based estimates cure only some of the limitations of standard analyses since
many simplifying (and unrealistic) assumptions remain. The main limitations of
course arises from the idealised description of the measurement process
(eq.~\eqref{eq:model_ell}) and the assumed perfect knowledge of the response
matrix, $\bA$, and the correlation matrix, $\bC$.

%++++++++++++++++++++++++++++++++++++++++++++++++++++++++++++++++++++++++++++++
% Plots of wiener vs. naive error estimates
%++++++++++++++++++++++++++++++++++++++++++++++++++++++++++++++++++++++++++++++
%++++++++++++++++++++++++++++++++++++++++++++++++++++++++++++++++++++++++++++++

\subsection{Experiment optimisation and robustness}
%**************************************************

\begin{figure}[htbp] \centering \centerline{ \hbox{  
\psfig{figure=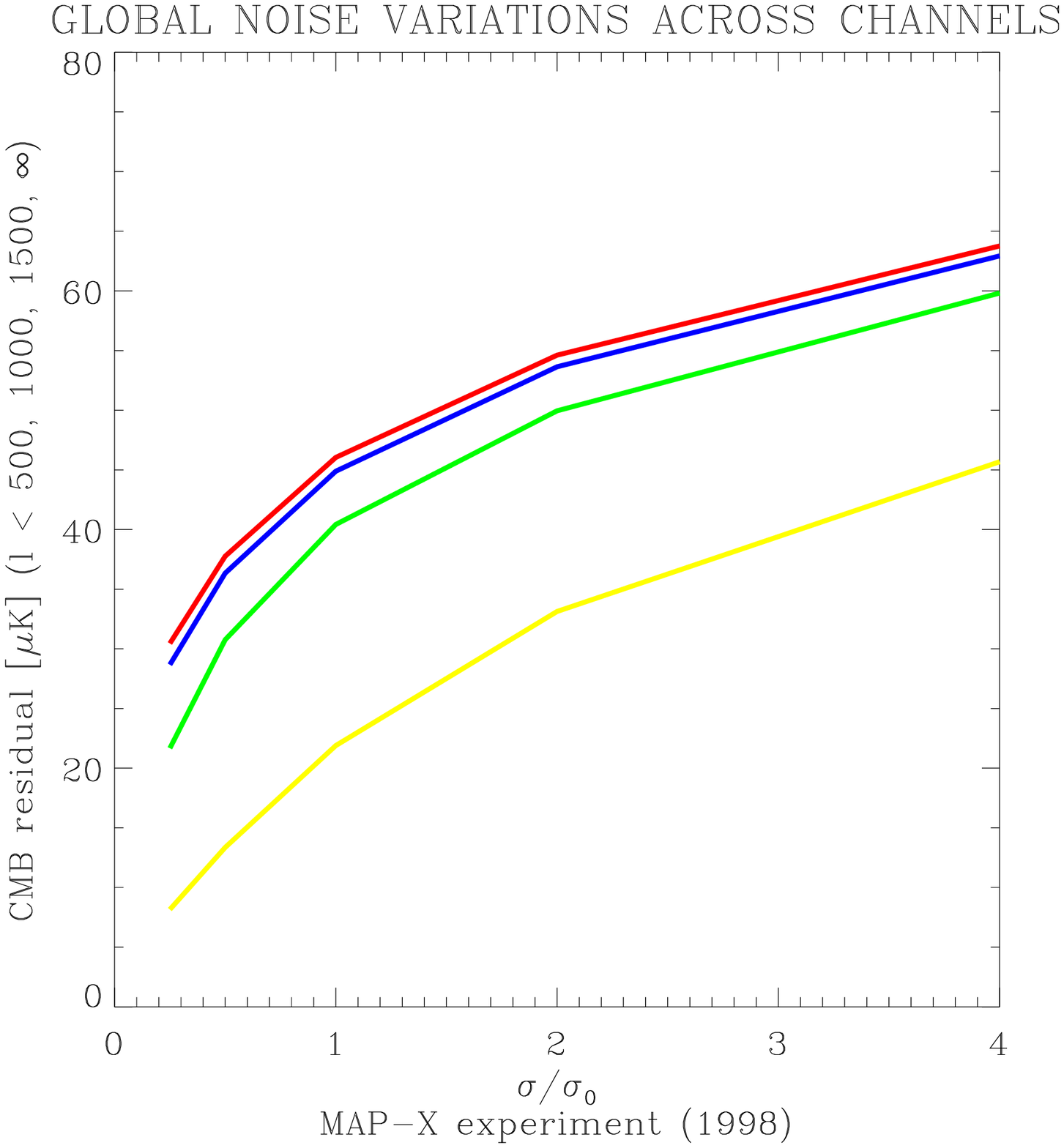,width=0.5\textwidth}
\psfig{figure=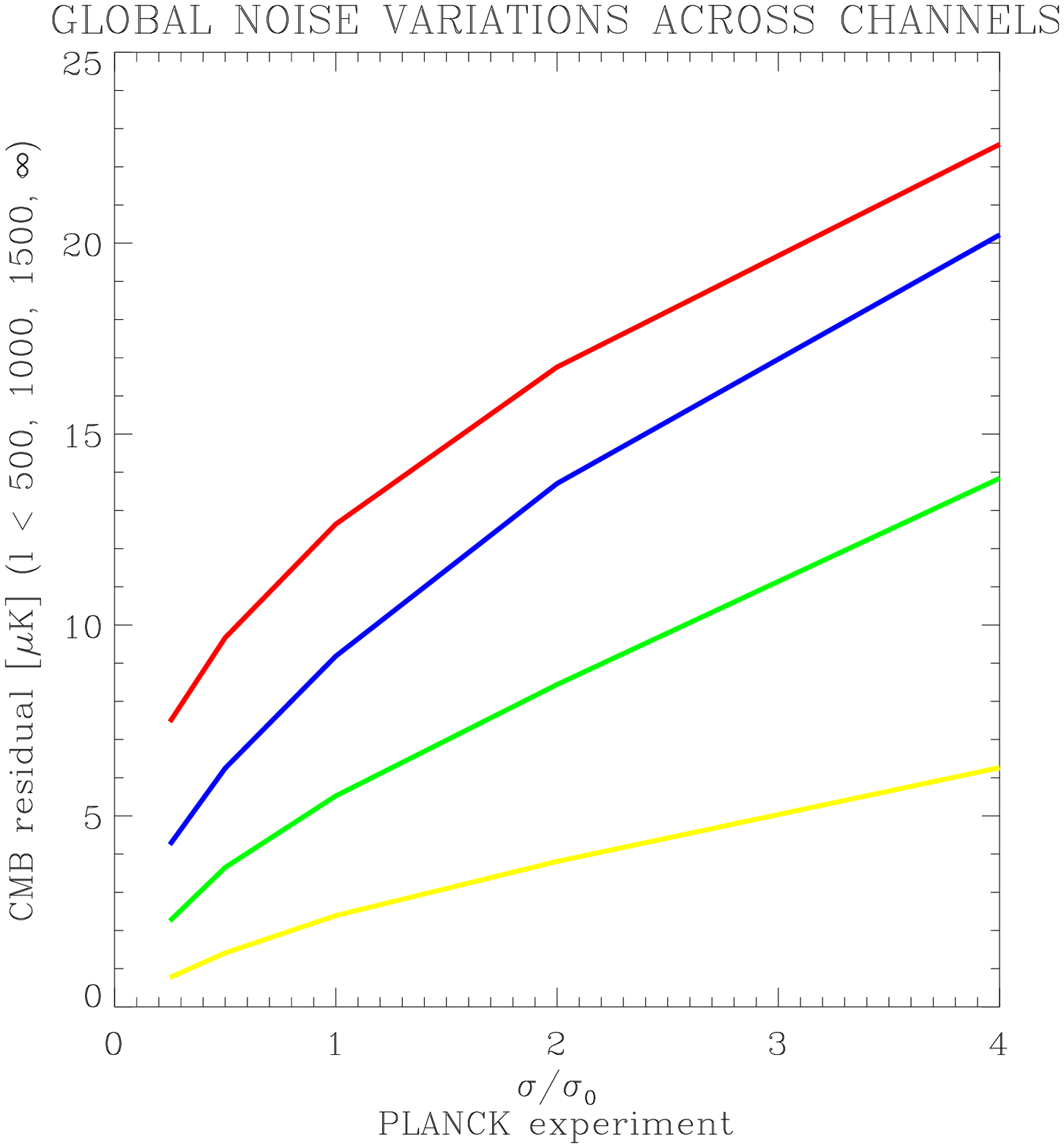,width=0.5\textwidth} } }
\caption[]{Performance variations for global variations across channels of the
detector noise level (\eg with the number of sky coverages done). The yellow,
green, and blue lines correspond to restricting the integration to a maximum
$\ell_{max}$ of respectively $500, 1000, 1500$. The red line corresponds to no
restriction at all.}
\label{fig:cl_noisvar}
\end{figure}

One may wonder whether the CMB residuals will vary linearly with global
variations of the noise figures of an experiment, as could be produced for
instance by variations of the operating temperature, or length of operation of
the experiment (allowing different number of sky coverages), or simply overly
optimistic or pessimistic estimates of the detectors performances.
Figure~\ref{fig:cl_noisvar} shows the variation of the rms of the total
residual contribution from noise and unsubtracted foregrounds to a CMB map as
a function of the noise level compared to the nominal one.  The square of this
rms value is obtained by integrating the residual spectrum of
figure~\ref{fig:errors} till a maximum $\ell_{max}$. Varying $\ell_{max}$
gives a concise description of the contribution of various scales to the map
error.

%++++++++++++++++++++++++++++++++++++++++++++++++++++++++++++++++++++++++++++++
% Plots of reliability & failure
%++++++++++++++++++++++++++++++++++++++++++++++++++++++++++++++++++++++++++++++
%++++++++++++++++++++++++++++++++++++++++++++++++++++++++++++++++++++++++++++++

In the nominal case, $\sigma/\sigma_0 =1$, one sees that for \maps and
\plancks respectively the residuals are 21 and 2.5\microK\ for
``low''-resolution maps retaining only the modes till $\ell_{max} = 500$. For
medium resolution maps with $\ell \leq 1000$, these errors increase to 41 and
5.5\microK, and high-resolution maps with $\ell \leq 1500$ have residuals of
50 and 9\microK\ respectively. If all $\ell$ are retained, the rms of the
residual are then 52 and 13\microK. The difference between \maps and \planck\
in the increase of the residual with the number of modes retained reflects the
difference in angular resolution of the maps produced by Wiener filtering
(figures~\ref{fig:qual_fac} and \ref{fig:qual_beam}).

For ``low'' and medium resolution maps, the residuals for \plancks vary nearly
linearly with the noise level. This is the high-$\ell$ part which is most
affected by variations of the noise level. In the \maps case, the stronger
curvature of the lines indicate that it would be more affected by sub-nominal
performance of the detectors, and conversely that it would benefit more of an
increase of the mission duration; a decrease of the noise level by a factor of
two (\eg 4 years instead of one year of operation) would for instance reduce
the residuals for $\ell < 1000$ from 41 to $\sim$ 30\microK.

%++++++++++++++++++++++++++++++++++++++++++++++++++++++++++++++++++++++++++++++
% Plots of failures
%++++++++++++++++++++++++++++++++++++++++++++++++++++++++++++++++++++++++++++++
\begin{figure}[htbp] \centering \centerline{ \hbox{ 
\psfig{figure=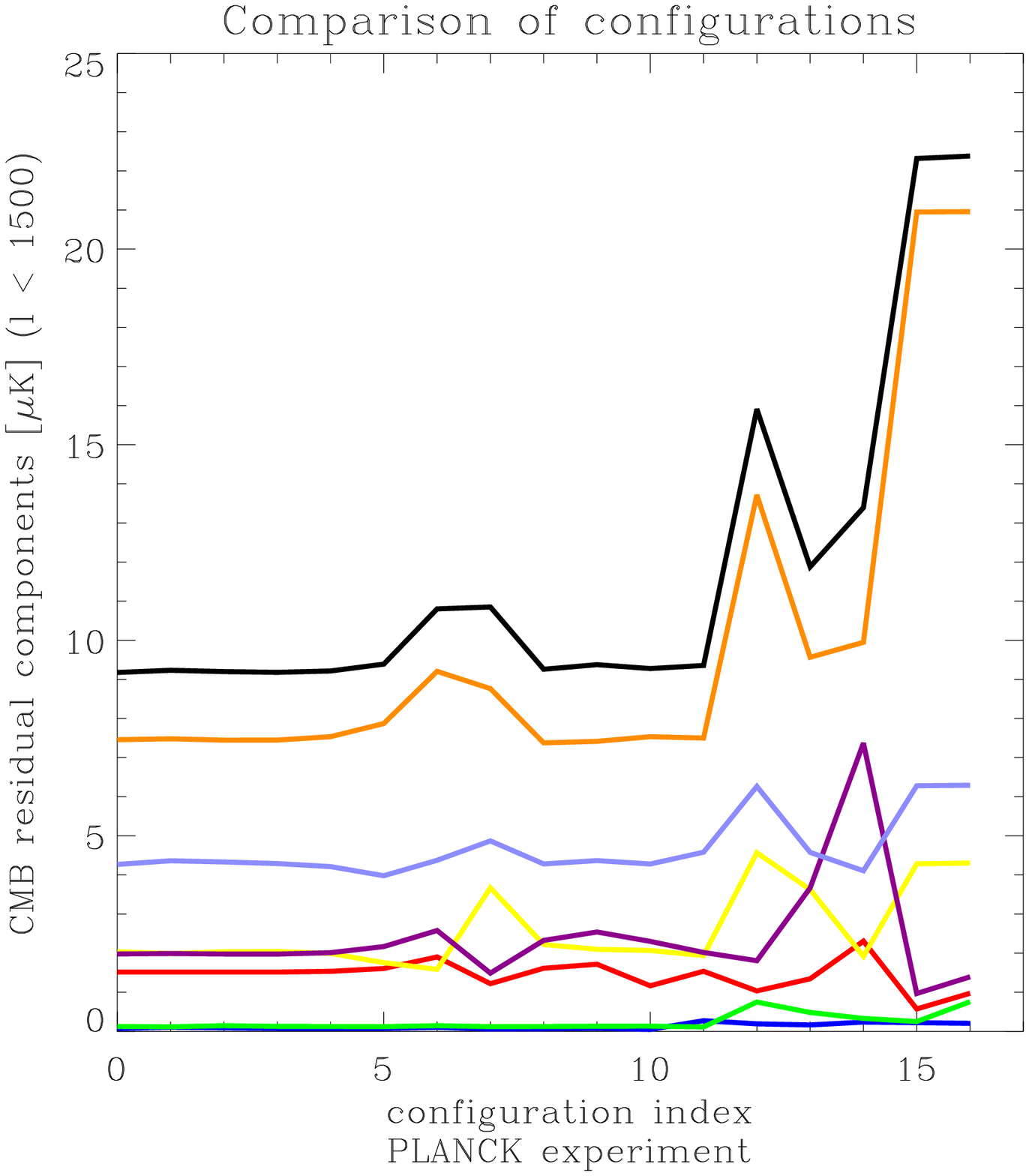,width=0.4\textwidth}
\psfig{figure=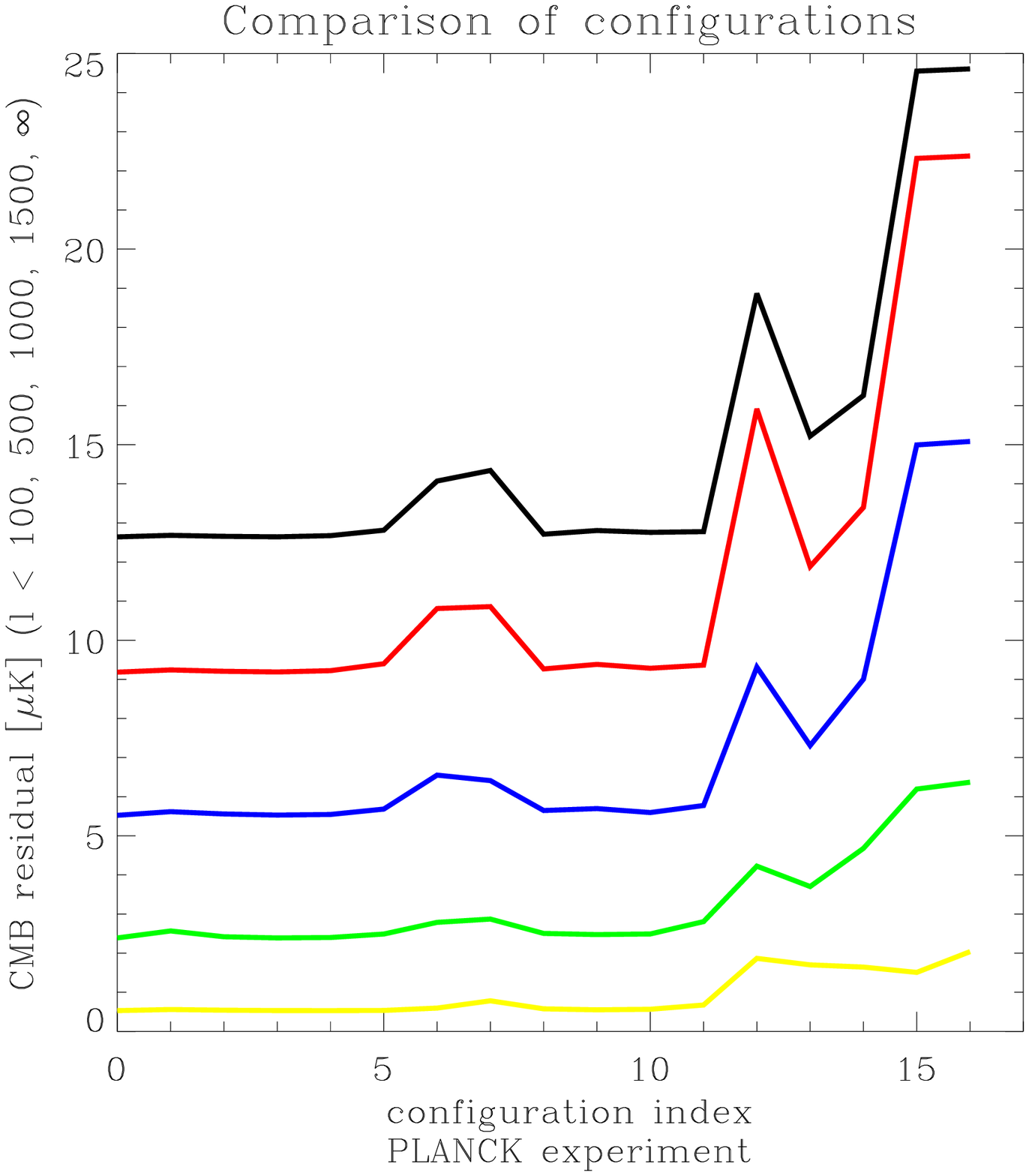,width=0.4\textwidth}
\psfig{figure=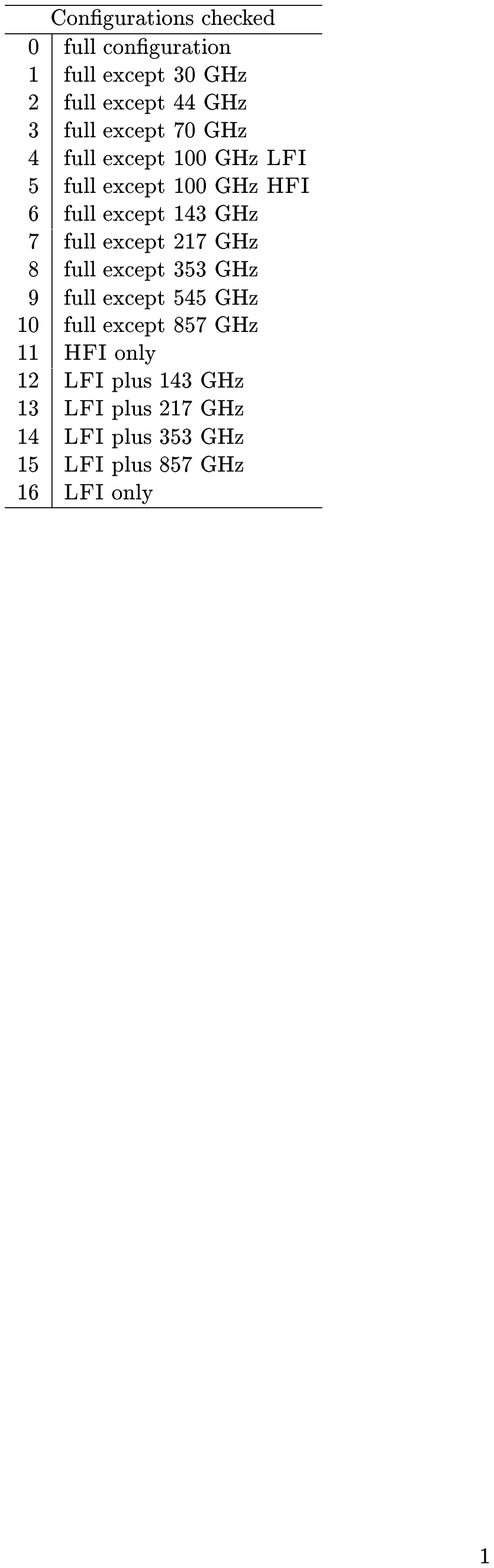,height=0.4\textwidth,bbllx=55pt,bblly=545pt,bburx=210pt,bbury=795pt}
} }
\caption[]{a) CMB residuals (in a map with all modes till $\ell_{max} =1500$)
retained) in various configurations for \planck. The black curve displays the
total, and the other lines detail the various contributions from noise (orange
curve) and foregrounds with the same colour coding than figures
\ref{fig:contribs} or \ref{fig:rmsmodel} (\ie red, blue, and green curves
correspond to the Galactic residuals, yellow is for the thermal SZ, and the
light blue and purple curves are for the unresolved sources backgrounds). b)
Total residual in different configurations, for different $\ell_{max}$ (as in
figure~\ref{fig:cl_noisvar}, the yellow, green, and blue lines correspond to
restricting the integration to a maximum $\ell_{max}$ of respectively $500,
1000, 1500$, while the red line corresponds to no restriction at all).  c)
Table of configuration indexes in the adjacent figures.}
\label{fig:nois_configs}
\end{figure}
%++++++++++++++++++++++++++++++++++++++++++++++++++++++++++++++++++++++++++++++

The quality factor approach may also be used to test whether the failure of a
single channel of an experiment would be critical for the outcome of the
experiment, \ie whether it is robust. Here we estimate the variation of the
CMB residuals in different ``crippled'' configurations for
\planck. Figure~\ref{fig:nois_configs} shows that i) the residuals are
dominated by noise ii) there is nearly no impact at all if a single \planck\
channel is lost except for the 143 \& 217\,GHz channels iii) loosing the \hfi\
would more than double the residuals, but the \lfis completed by the \hfis 217
or 353\,GHz channel would already do rather well.

\subsection{Uncertainties of the sky model \label{sec:varsky} }
%**************************************************************

%++++++++++++++++++++++++++++++++++++++++++++++++++++++++++++++++++++++++++++++
% Power spectrum (P) and quality factor (Q) for sky model variations
%++++++++++++++++++++++++++++++++++++++++++++++++++++++++++++++++++++++++++++++
\begin{figure}[htbp] \centering \centerline{ \vspace{-12pt} \vbox{ \hbox{  
\psfig{figure=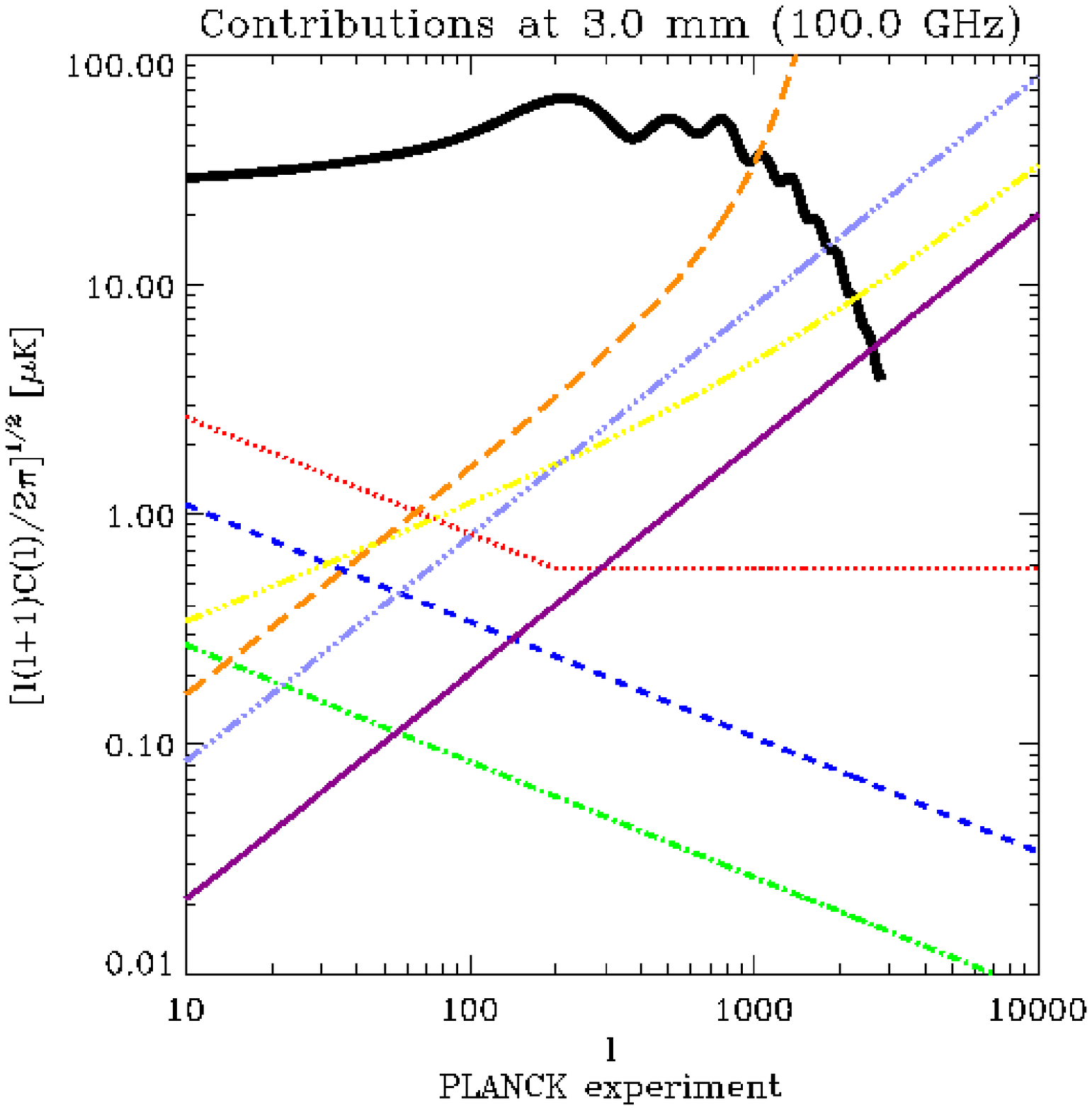 , width=0.33\textwidth} \psfig{figure=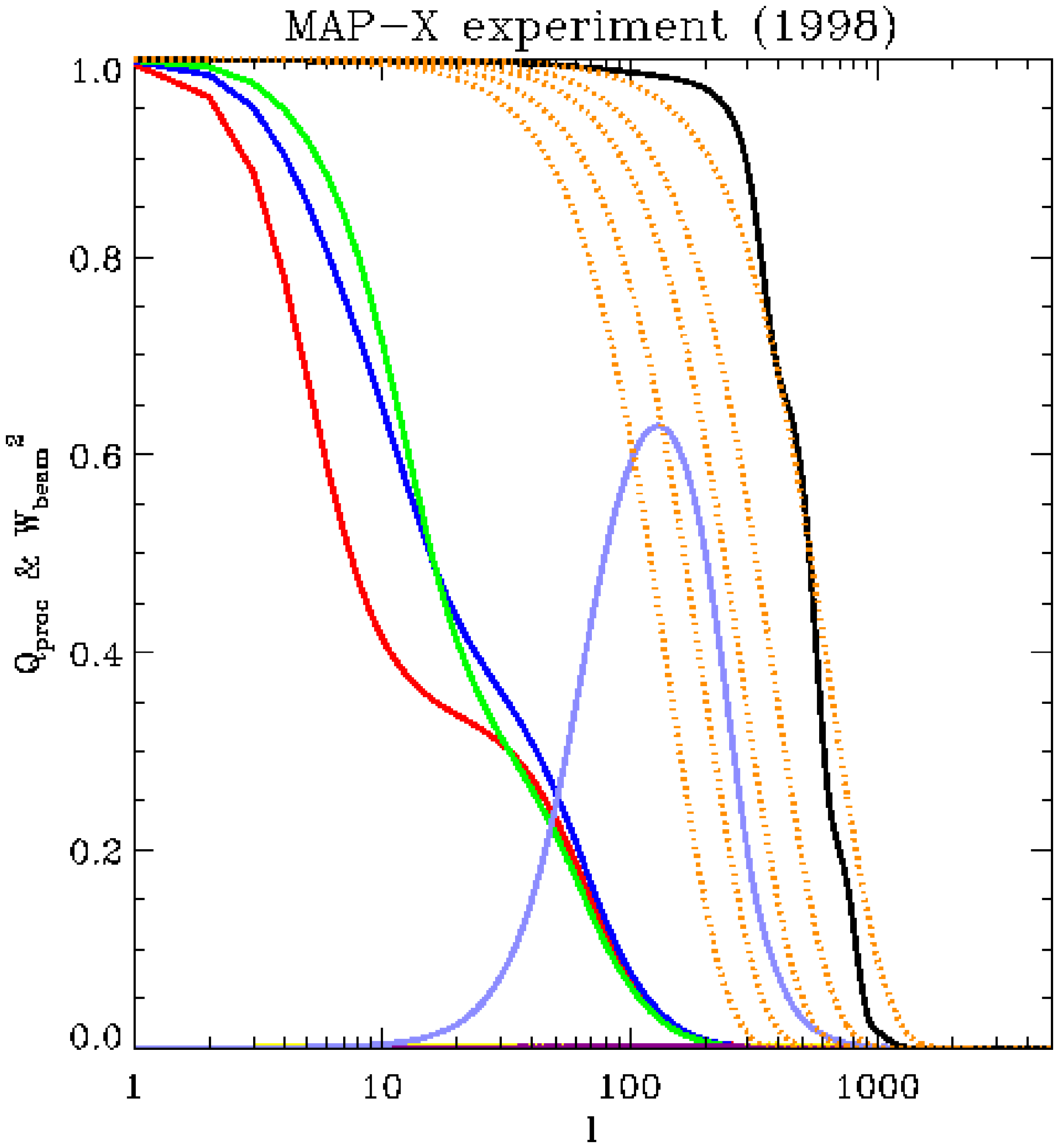,
width=0.33\textwidth} \psfig{figure=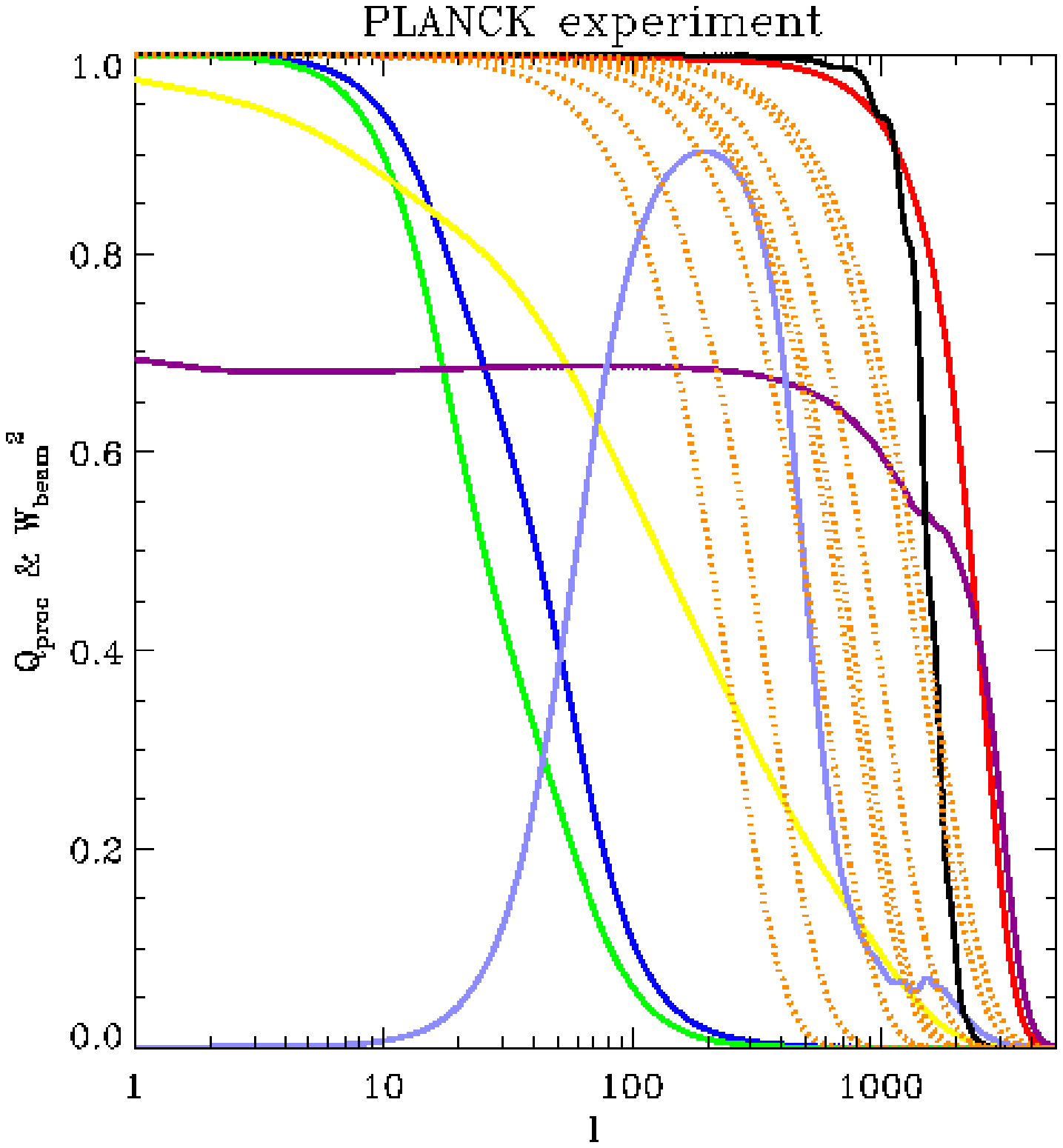, width=0.33\textwidth} } \hbox{
\psfig{figure=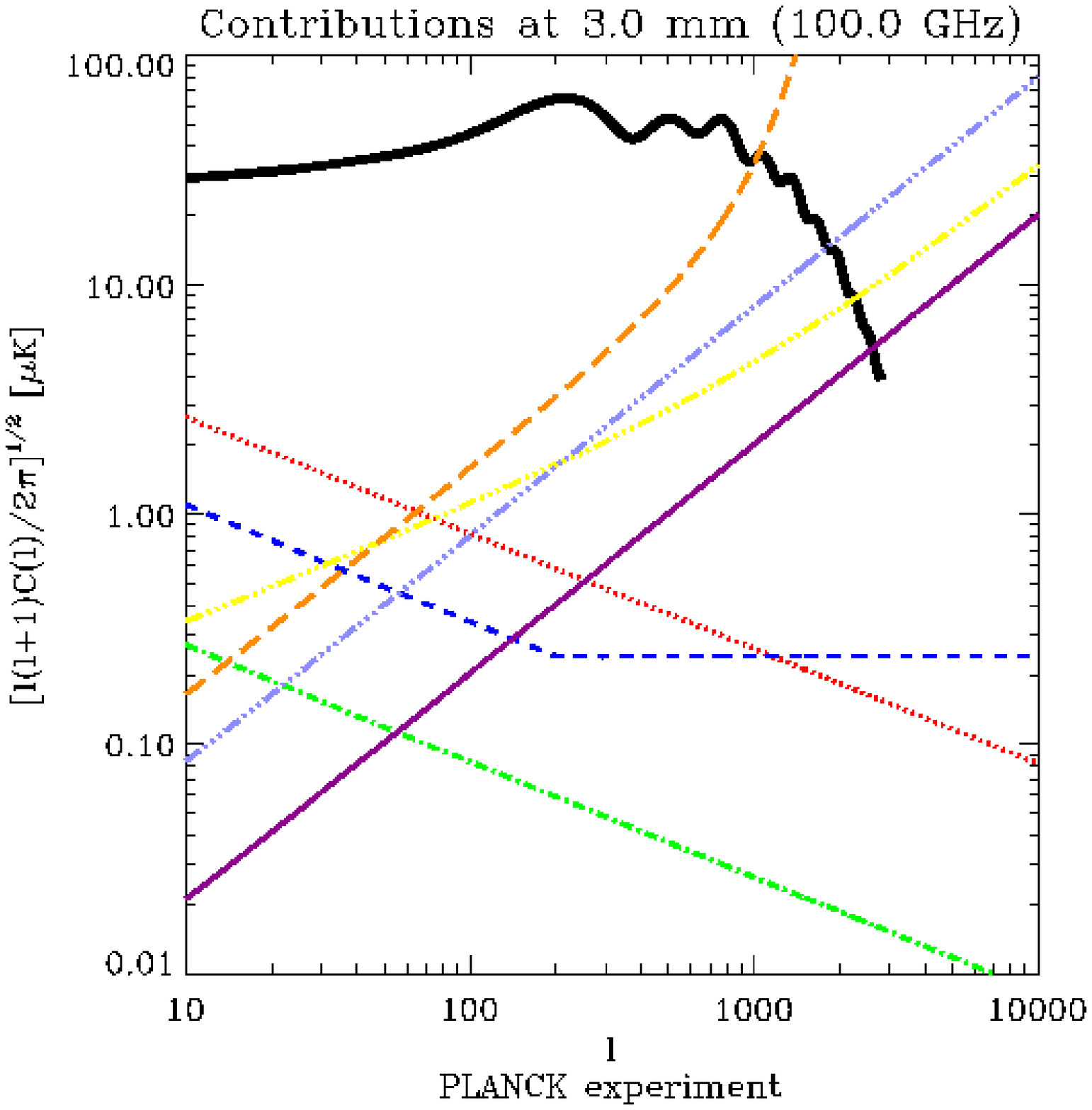 ,width=0.33\textwidth} \psfig{figure=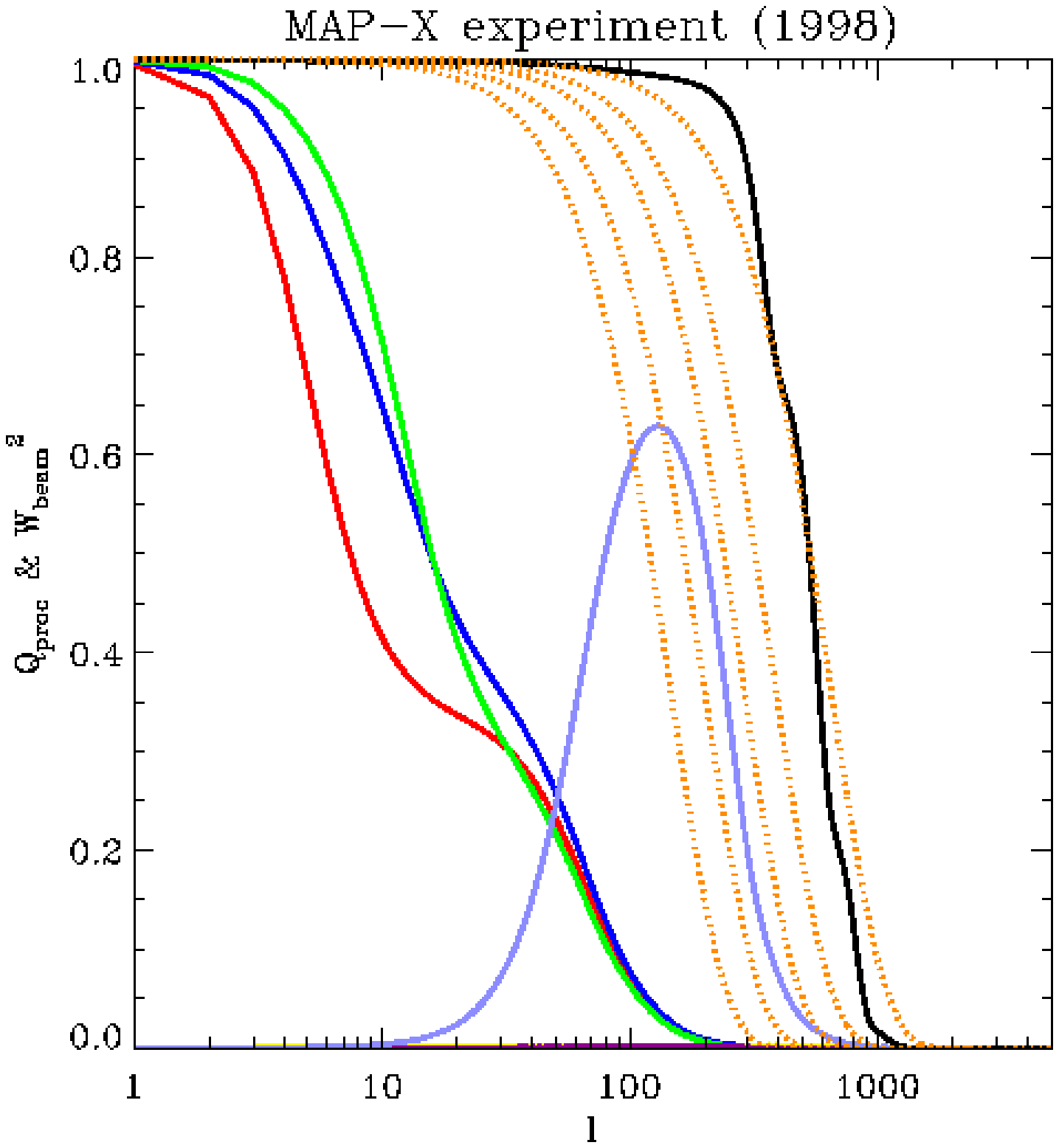,
width=0.33\textwidth} \psfig{figure=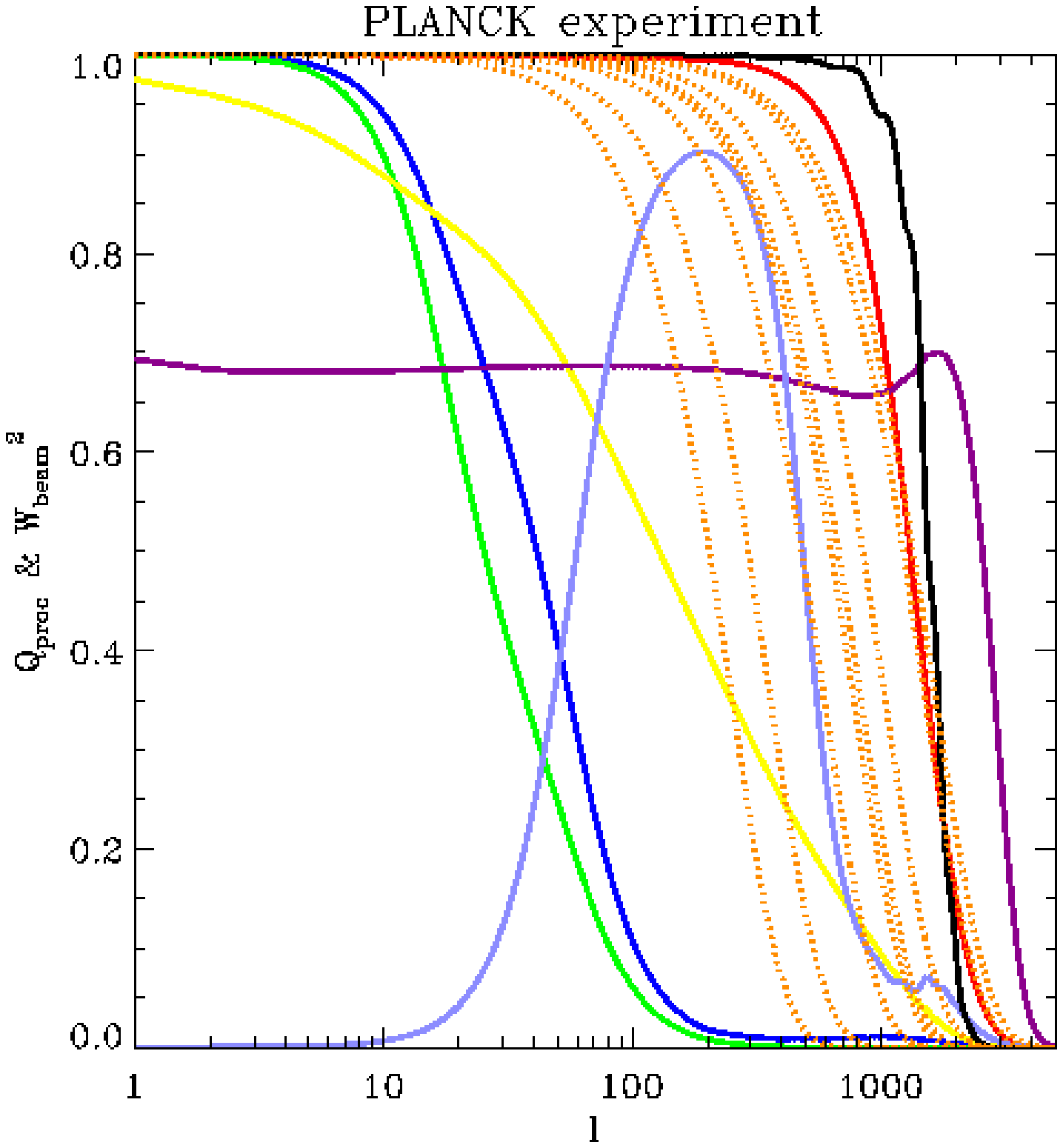, width=0.33\textwidth} } \hbox{
\psfig{figure=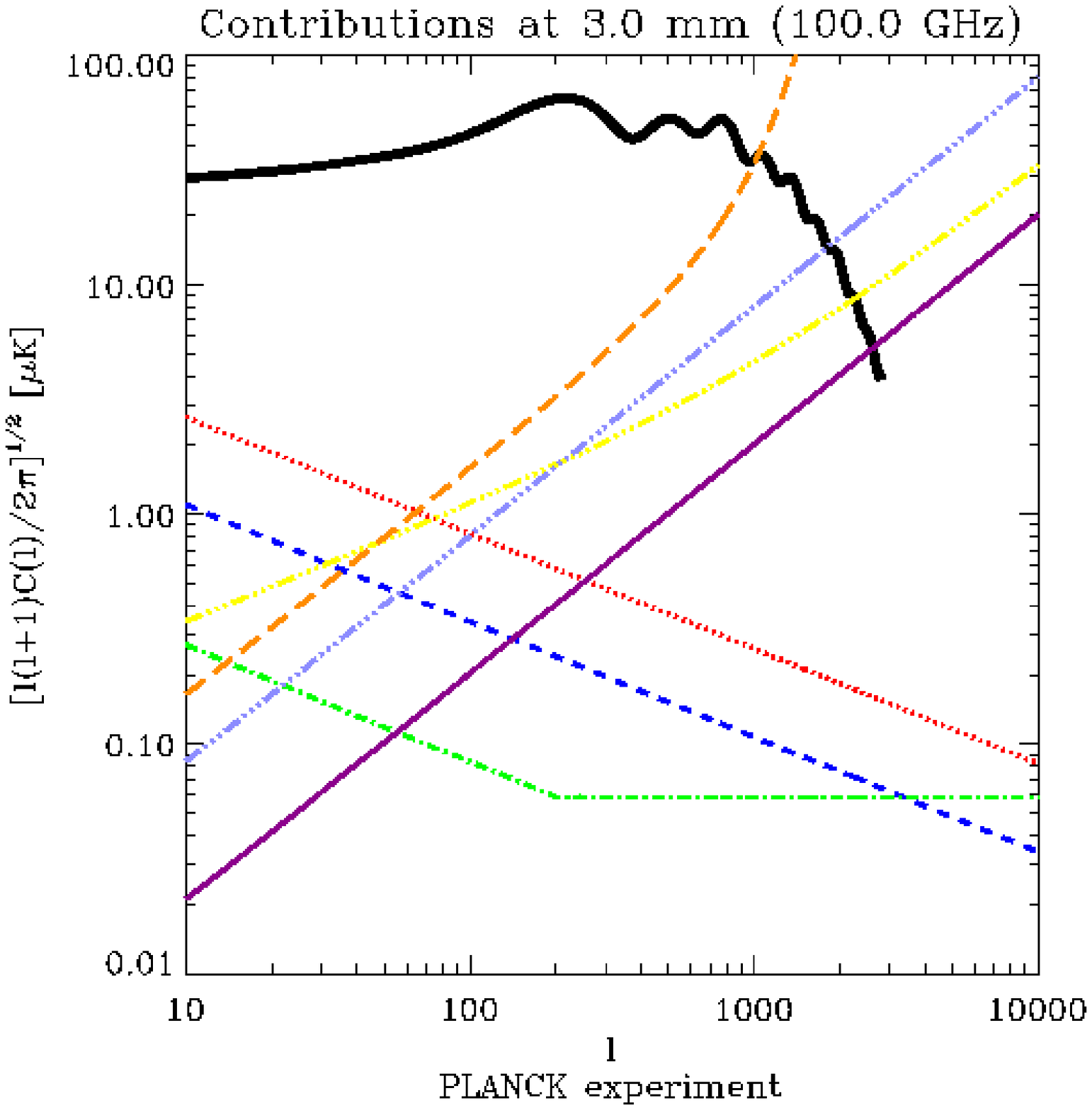 ,width=0.33\textwidth} \psfig{figure=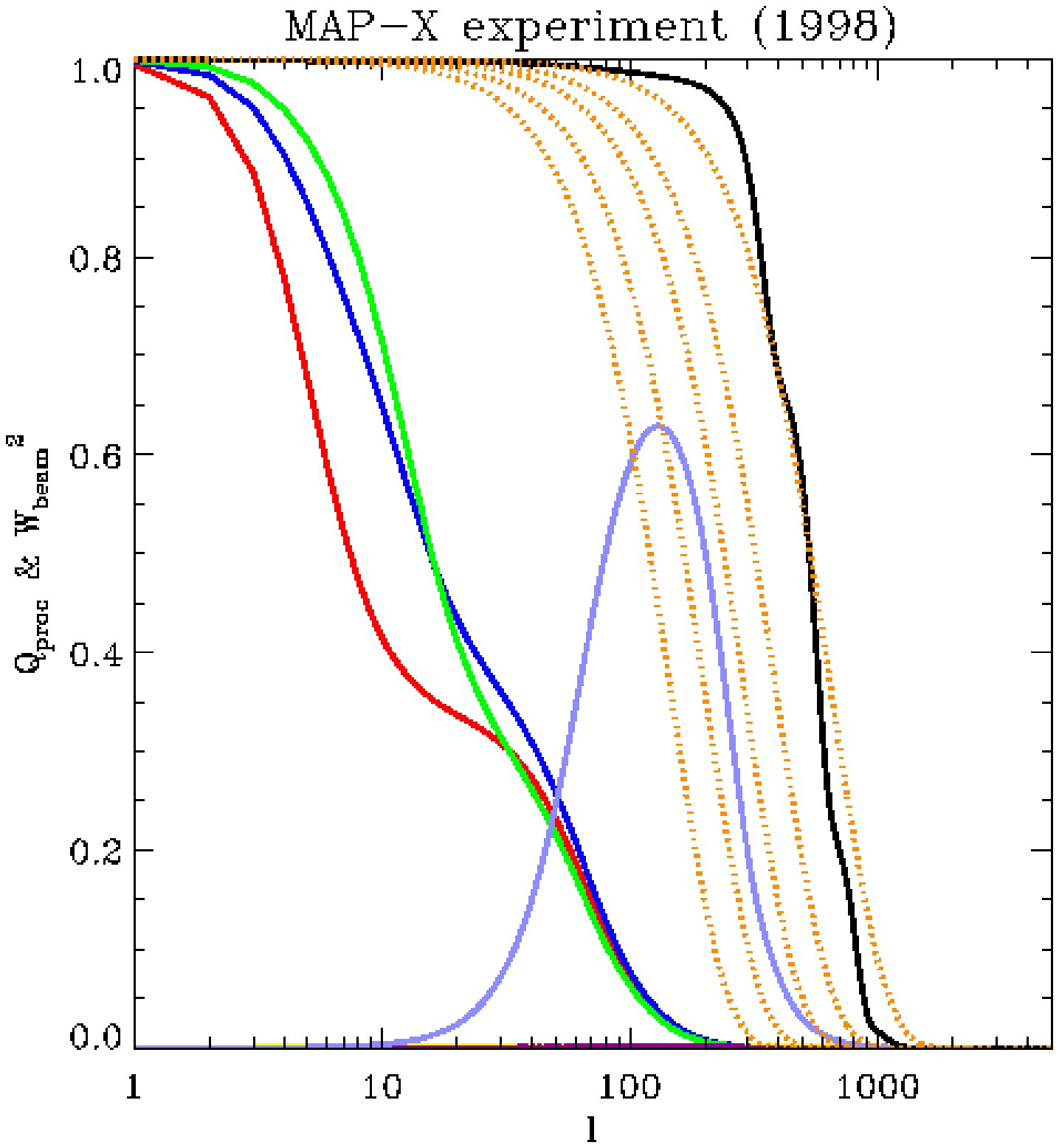,
width=0.33\textwidth} \psfig{figure=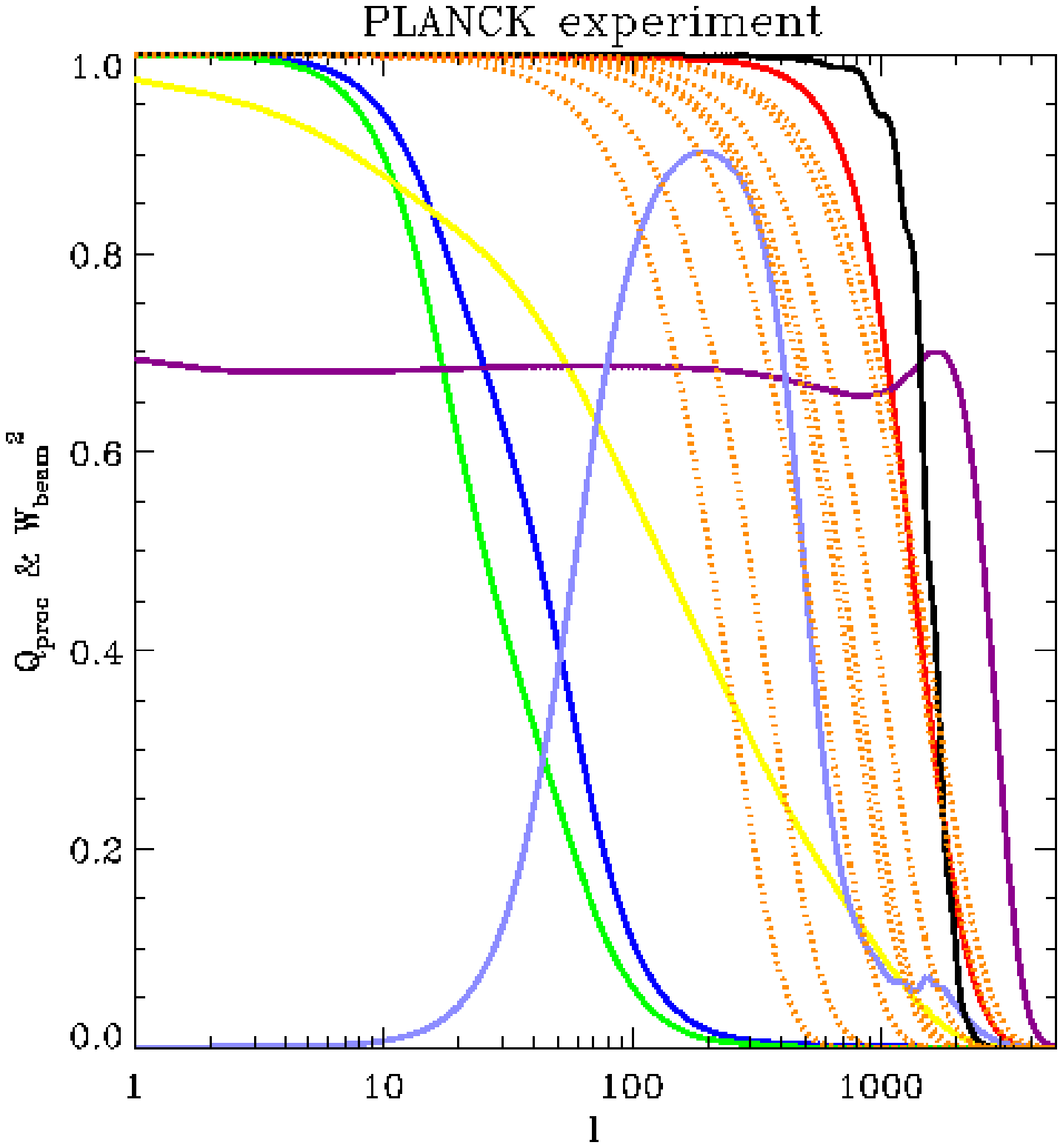, width=0.33\textwidth} } \hbox{
\psfig{figure=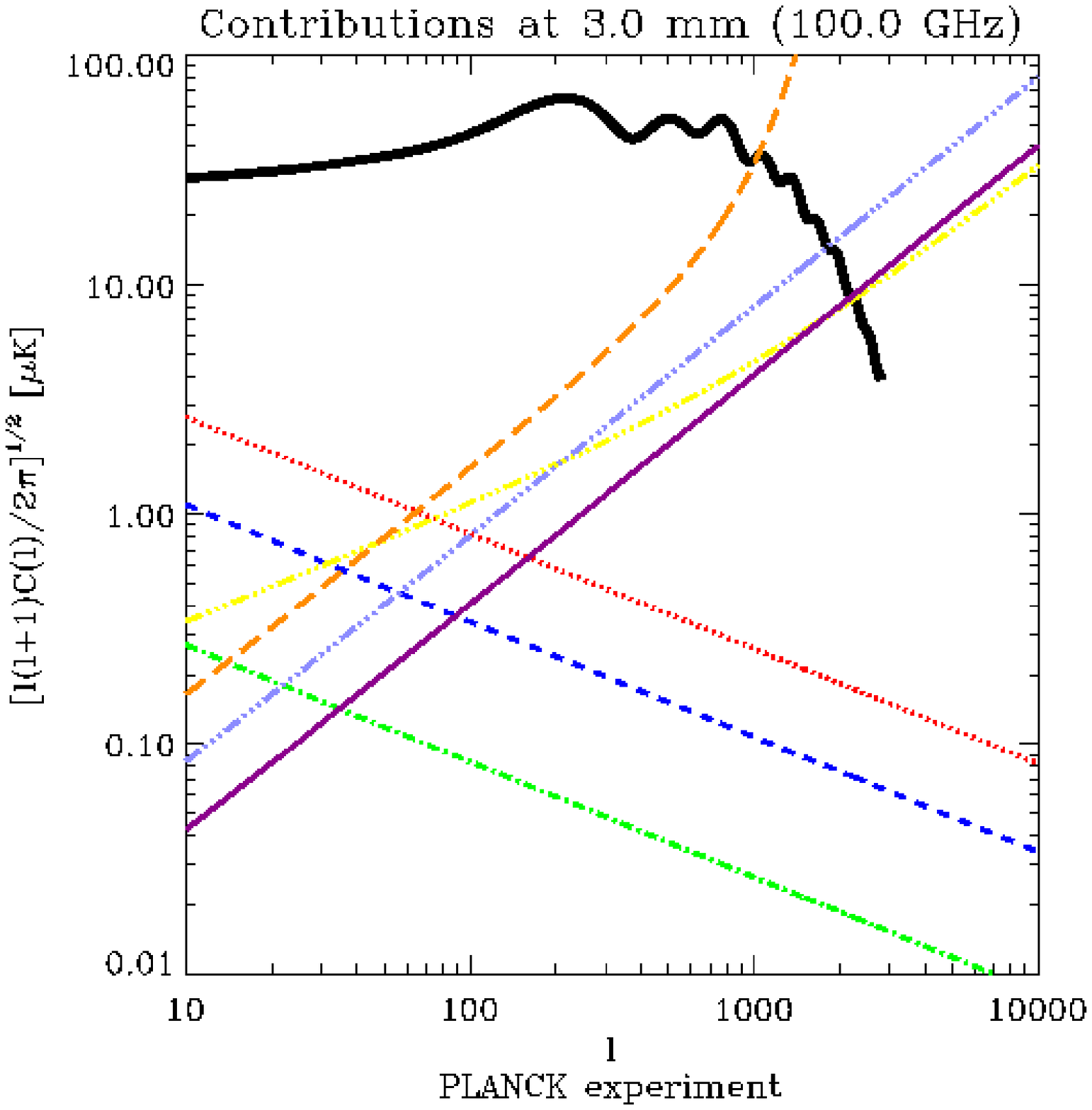 ,width=0.33\textwidth} \psfig{figure=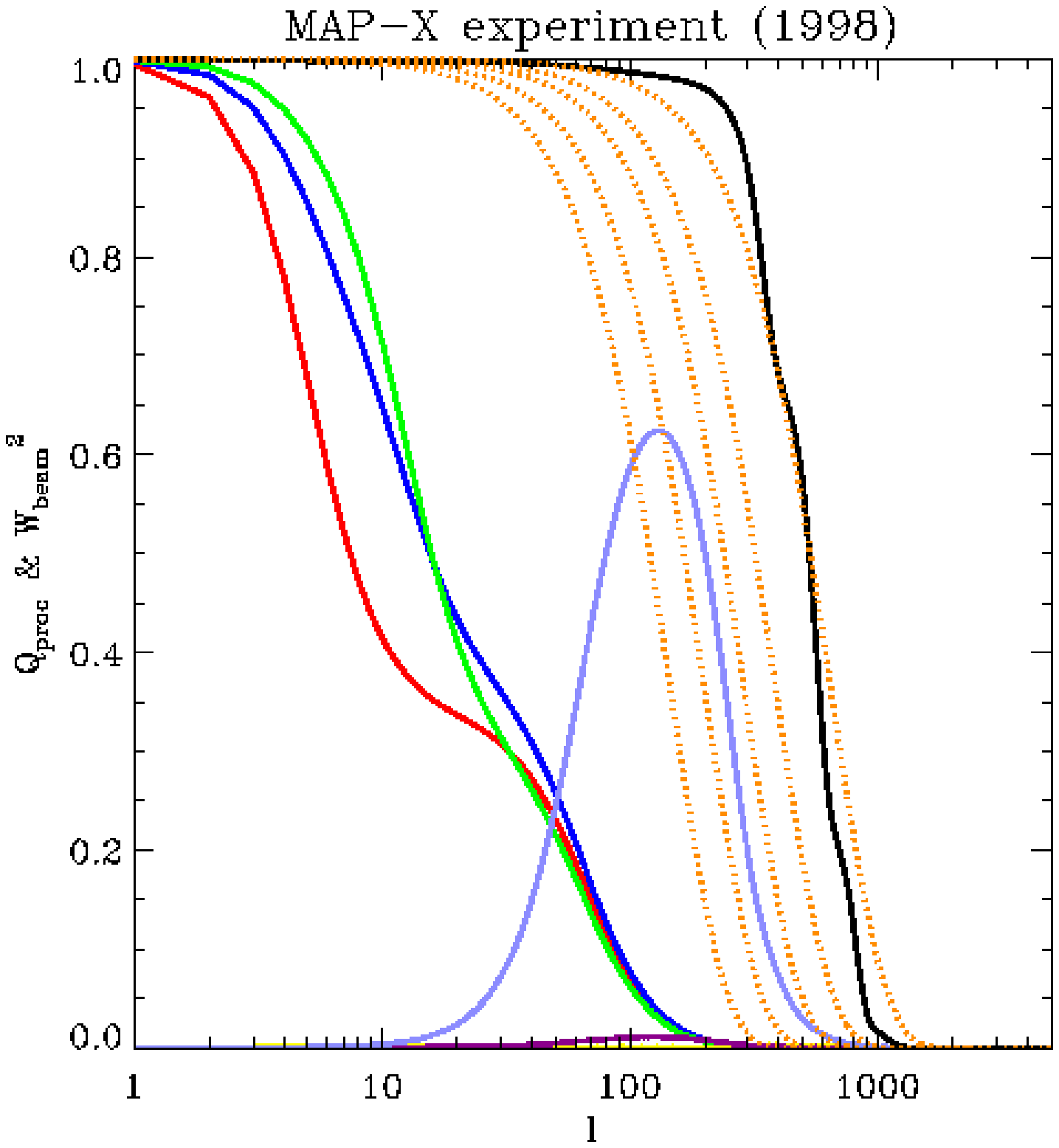,
width=0.33\textwidth} \psfig{figure=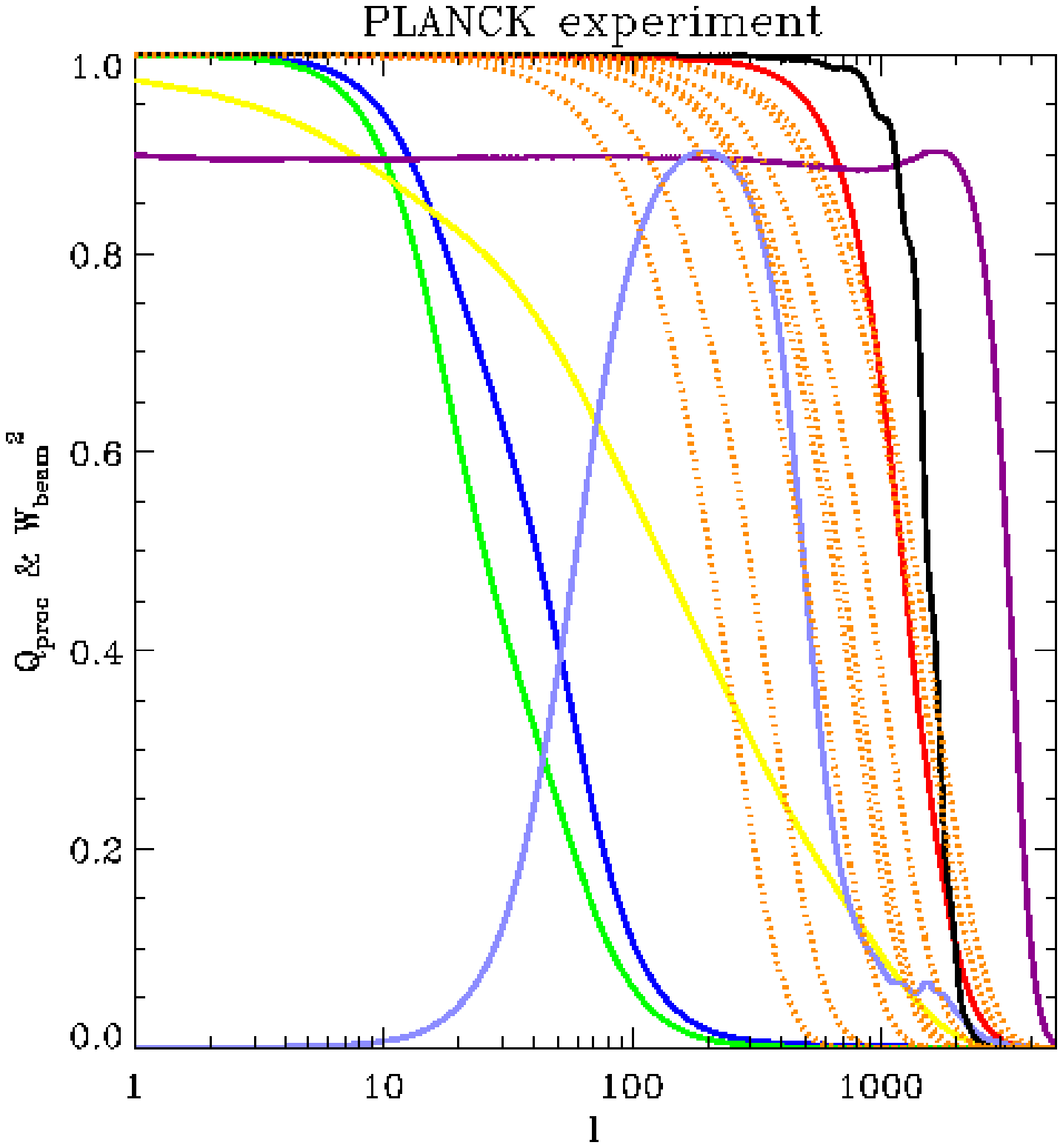, width=0.33\textwidth} } \hbox{
\psfig{figure=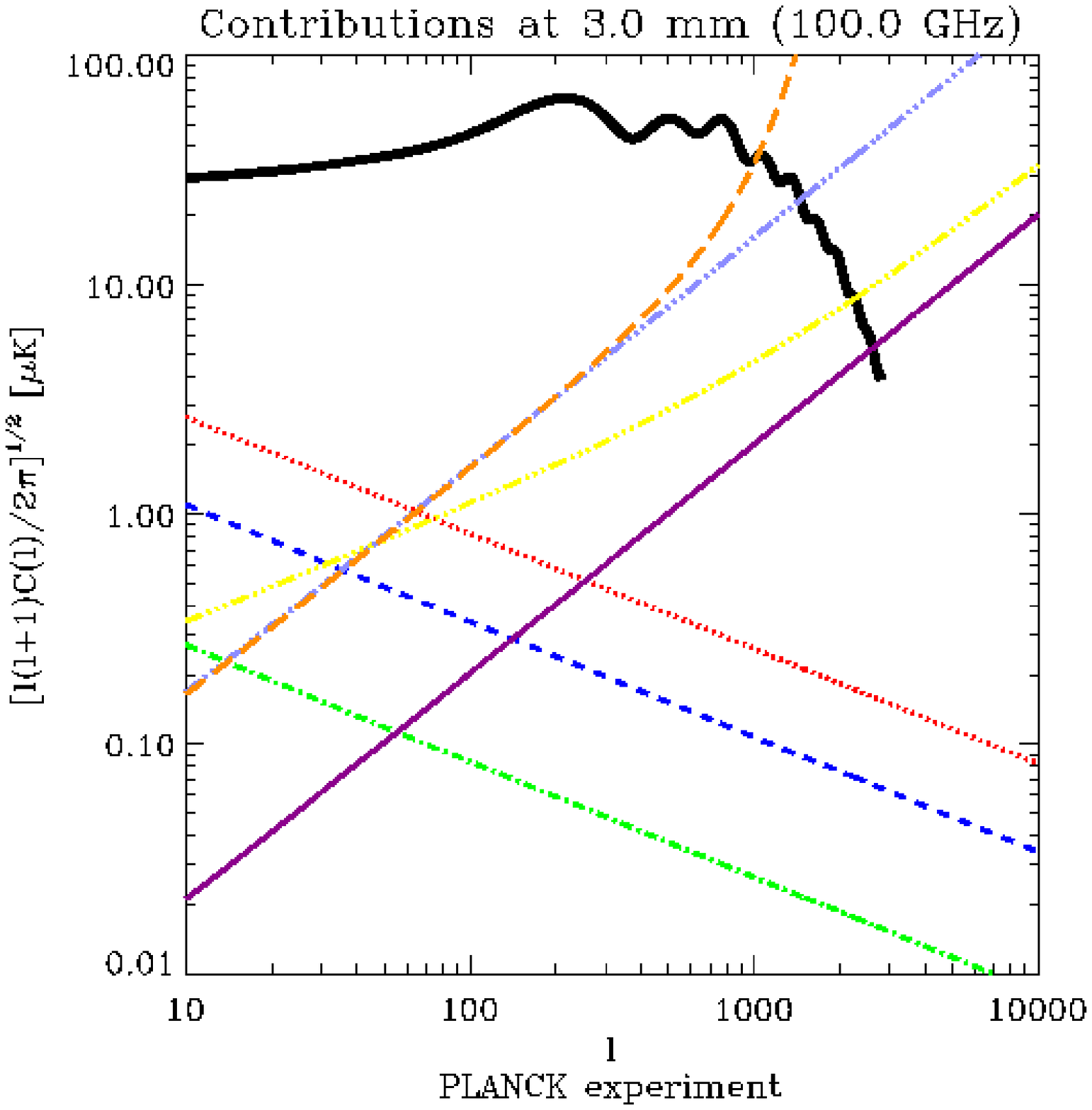 ,width=0.33\textwidth} \psfig{figure=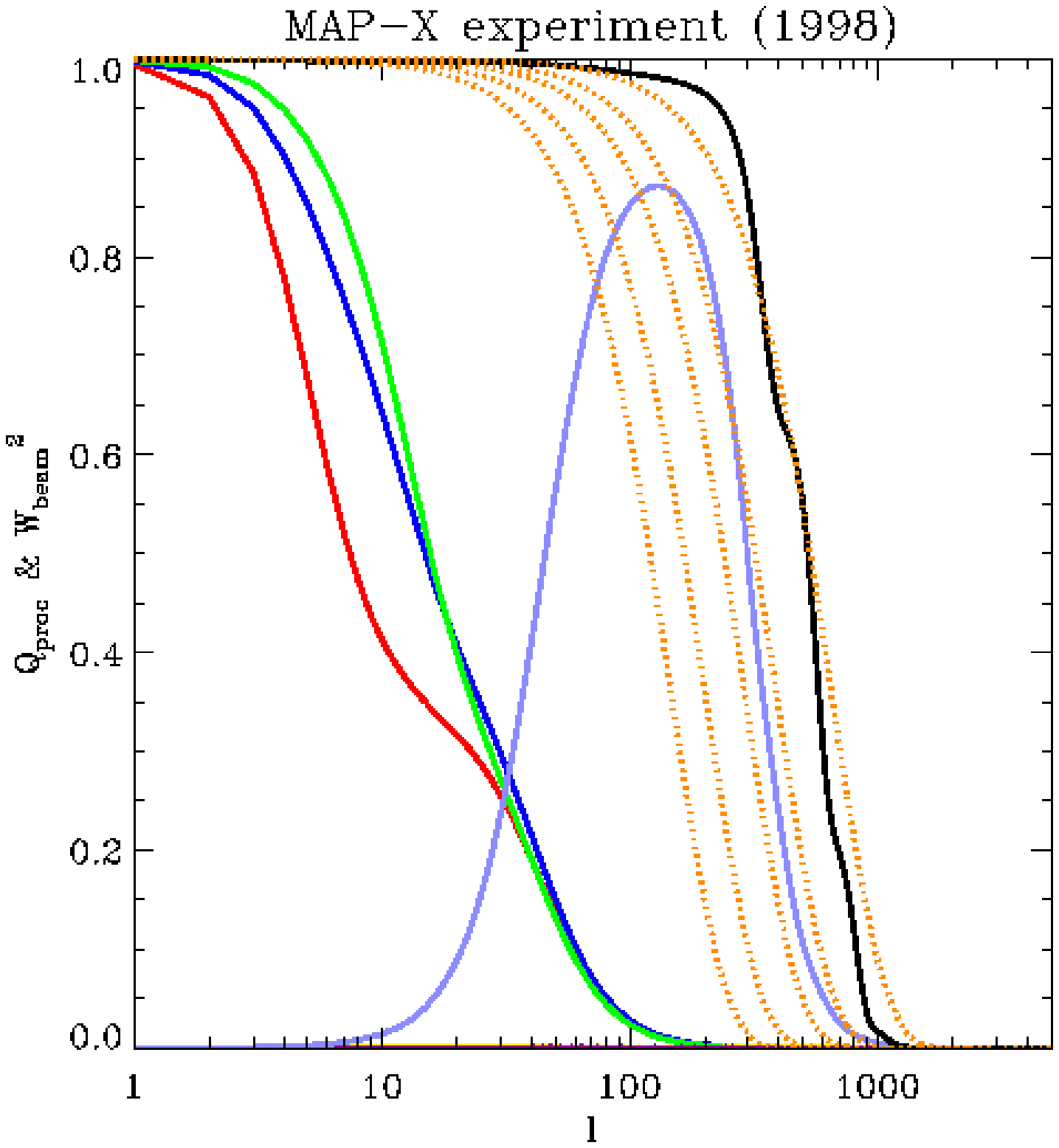,
width=0.33\textwidth} \psfig{figure=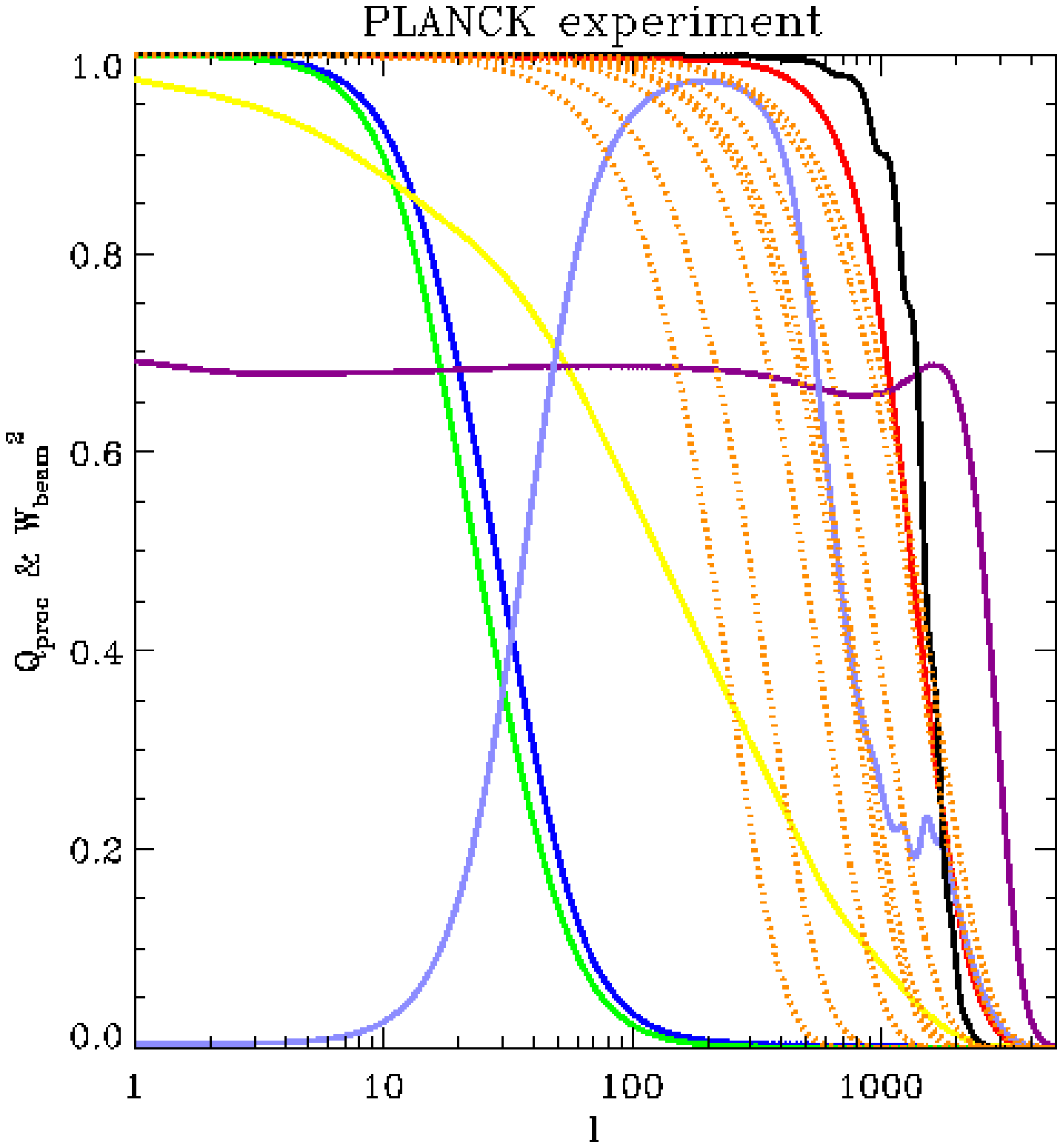, width=0.33\textwidth} } } }
\caption[]{Power spectra at 100\,GHz (including \hfi\ noise, left column) and
corresponding quality factors for \maps and \plancks (respectively middle and
right column) for various variations of the sky model. The colour coding of
the various foregrounds is the same than in similar previous figures. In the
top row, the spectrum of \HI-correlated component flattens to $\ell{-2}$ at
$\ell > 200$, while this flattening is rather for the \HI-uncorrelated
component (second row) and the synchrotron emission (third row). The last two
rows corresponds to higher levels of the backgrounds from unresolved infrared
sources (fourth row) or radio-sources. }
%Top to bottom: C, U, S, I, R...
\label{fig:varsky_PQ}
\end{figure}
%++++++++++++++++++++++++++++++++++++++++++++++++++++++++++++++++++++++++++++++

Figure~\ref{fig:qual_comp} illustrated the fact that the deduced 1-$\sigma$
errors on the CMB spectrum depend on the amplitude of that signal by comparing
the results for a standard and a ``Lambda-CDM'' model. Given the observed
variations one may worry that all the results obtained so far might critically
depend on the details of the sky model assumed. This is potentially
frightening since clearly the situation regarding foregrounds emission might
be quite more complicated than assumed here (e.g. some free-free may locally
be emitted at higher temperature, the backgrounds from unresolved sources
might be higher, etc).

We have thus considered variations on the basic sky model used so far. For the
three galactic components, we considered cases when each power spectrum might
level off and become $\propto \ell^{-2}$ at $\ell > 200$, resulting in an
increase of power by a factor of 10 at $\ell =2000$ as compared to our basic
assumption of a constant slope of -3 at all scales. We also considered that
the level of the free-free like emission uncorrelated with the HI-emission
might be four times lower. We also increased independently the normalisation
of each unresolved foreground component by a factor of four. We allowed each
of these 6 variation to occur in combination with all the others, thereby
creating 64 different foreground models.

Figure~\ref{fig:varsky_PQ} shows the resulting 100\,GHz power spectra in 5 of
these 64 configurations, as well as the corresponding quality factors for
\maps and \planck. As anticipated, the effective transmission of a foreground
component is improved when and where the amplitude of that component is
increased. But the transmission of the other components does not appear
affected. This is not really surprising either, since Wiener filtering is a
signal to (signal+noise) weighting, where (signal+ noise) obtains for the
contribution from detectors noise and of of all the astrophysical
components. The lack of sensitivity of the transmission of all components but
the one varied simply reflects the fact that the dominant terms are the CMB
and detector noise, which are precisely those which were kept untouched here.

%++++++++++++++++++++++++++++++++++++++++++++++++++++++++++++++++++++++++++++++
% Summary of variations of errors on C(l)
%++++++++++++++++++++++++++++++++++++++++++++++++++++++++++++++++++++++++++++++
\begin{figure}[htbp] \centering \centerline{ \hbox{  
\psfig{figure=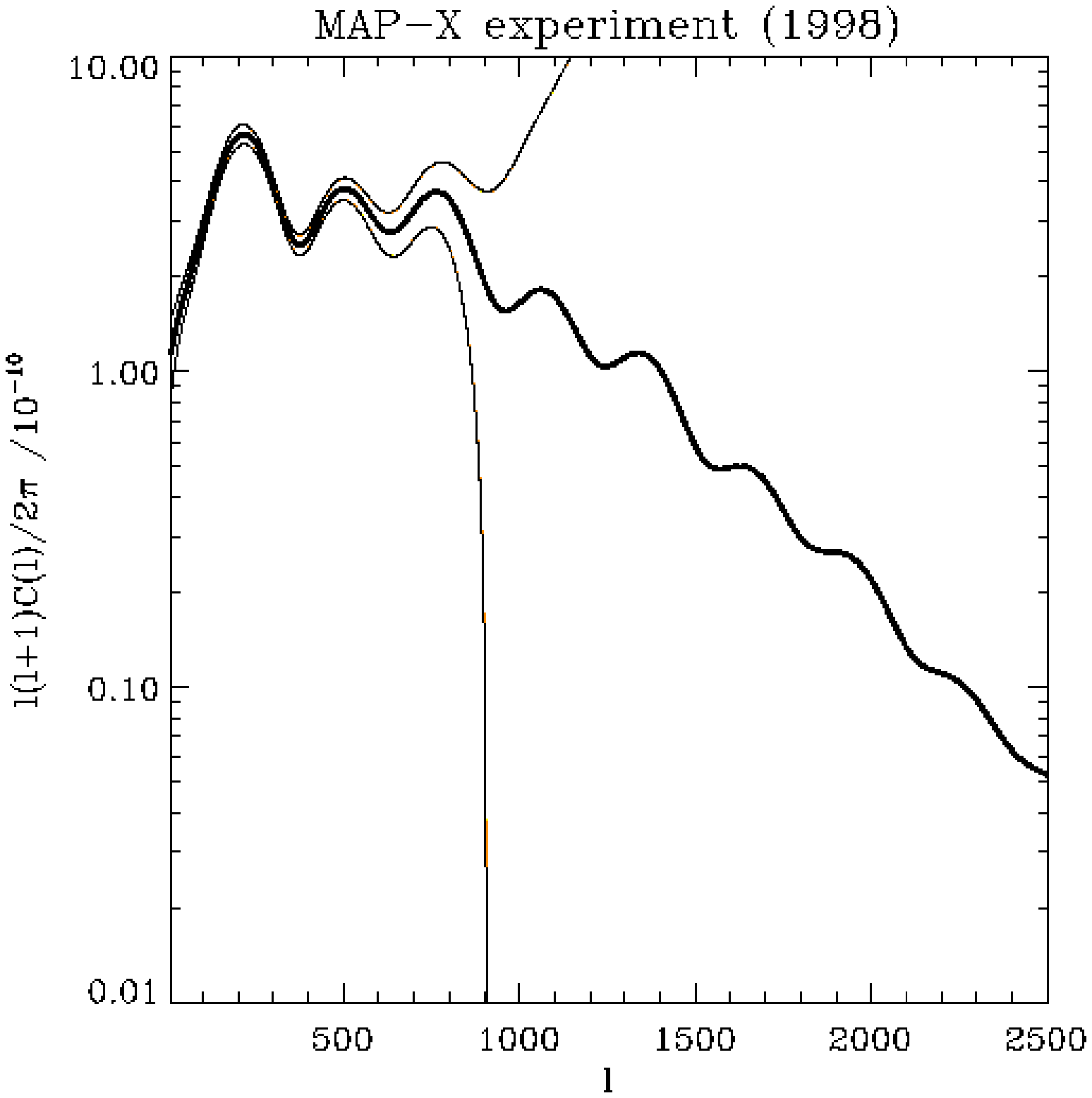, width=0.5\textwidth} \psfig{figure=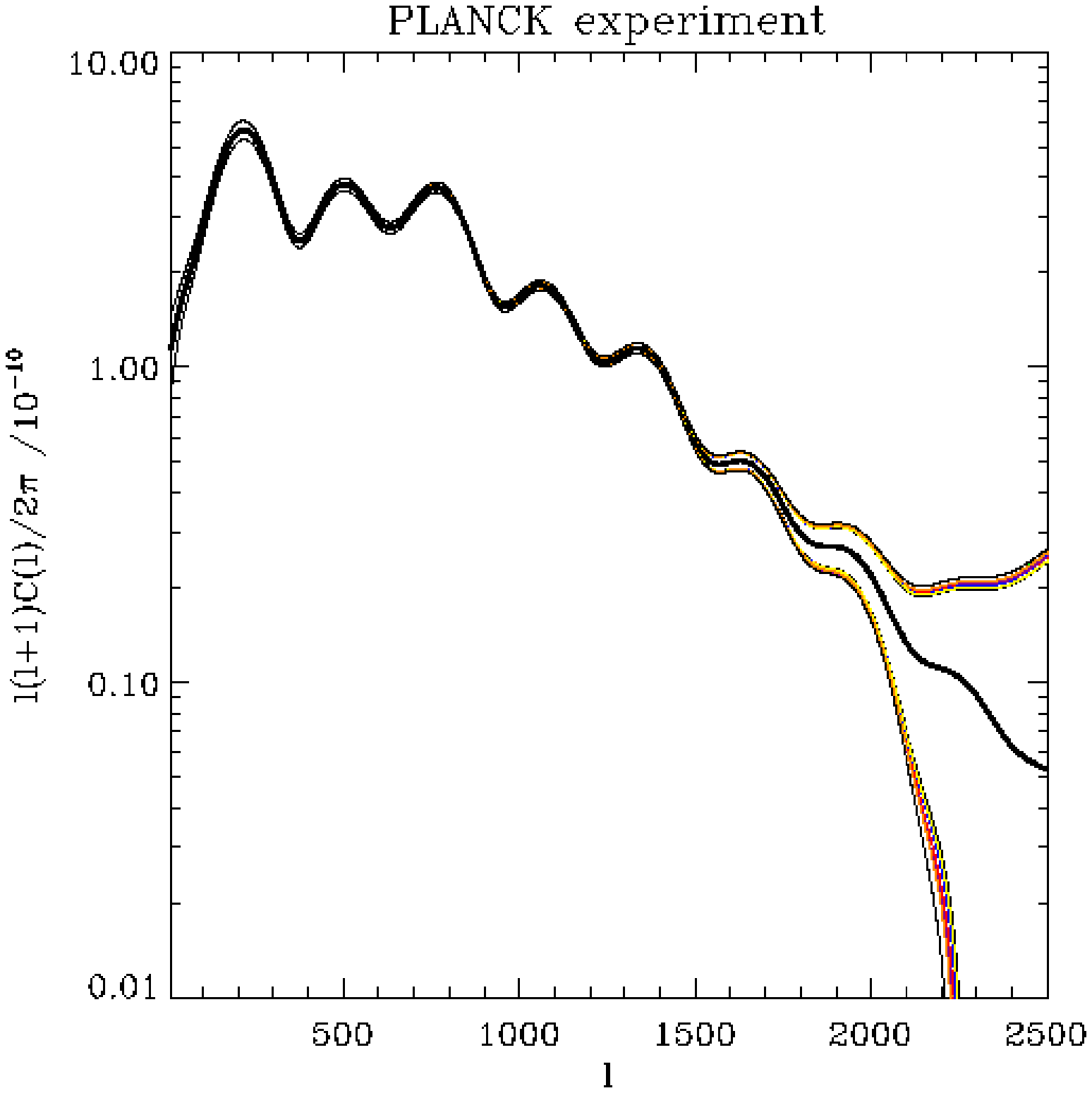,
width=0.5\textwidth} } }
\caption[]{Effect of variations in the sky model (see text) on the deduced
$1-\sigma$ error area for the CMB power spectrum. Both in the \maps (left) and
\plancks (right) case, the 64 variations investigated barely affect the result
because the dominant contributions to the signal in the channels of most
weight for the CMB determination are the CMB itself and the detector noise.}
\label{fig:varsky_C}
\end{figure}
%++++++++++++++++++++++++++++++++++++++++++++++++++++++++++++++++++++++++++++++

Figure~\ref{fig:varsky_C} displays the extent of $1-\sigma$ error for the CMB
power spectrum in the case of \maps and \plancks for the 64 sky models
explored. Nearly all curves are superimposed, which demonstrates that the
deduced errors are rather robust to variations in the details of the sky
model. This is simply because the dominant contributions to the signal in the
channel of most weight for the CMB determination remained the CMB itself and
the detector noise. Thus if the main features of the sky model proposed here
are correct, it is unlikely that most results obtained in this paper will be
much modified.

%******************************************************************************
\section{Conclusions}
%******************************************************************************

In order to assess the expected level of foreground contamination to be
expected in CMB experiments we have developed a simple sky model based on
current knowledge. In this model, the microwave emission of the sky is the
superposition of the CMB, three galactic emission mechanism, in addition to
two unresolved backgrounds from sources, one associated with radio-sources and
the uncorrelated other one arising from infrared sources. Further fluctuations
are imprinted by the Sunyaev-Zeldovich effect of clusters. While not known
precisely, we believe that the non-CMB components are reasonably well
constrained by current data supplemented with modest theoretical
extrapolations.

We then analysed aspects of the data analysis of multi-frequency
multi-resolution observations. Wiener filtering is shown to allow building
optimal linear estimates of the spatial templates of the physical processes,
as defined by their spectral behaviour.  We computed specific examples of
Wiener filters for our sky model and several experimental set-up. The shape of
the CMB Wiener filter in the $\nu-\ell$ space allows to build intuition on how
the information from different frequency channels is combined at different
angular scales to minimise the residuals in a CMB map. We further showed as a
function of scale the fraction of residuals which arise from detector noise
and from partly unsubtracted foregrounds. Integrating these residuals show for
instance that the rms reconstruction error from modes $\ell < 1000$ should be
$\sim 40$ \microK\ for \maps and smaller than 6\microK\ for \planck.

We also introduced a simple measure of the effective resolution of
multi-frequency multi-resolution experiments in the presence of
foregrounds. It's harmonic transform is the effective point spread function by
which the underlying maps of the component emissions would have been convolved
once the analysis is complete. This effective transmission or window function
is directly related to the expected uncertainty on the power spectra
determinations once foregrounds are accounted for (under the assumption of
Gaussian foregrounds though). Typically, \maps will probe well the CMB power
spectrum till $\ell \sim 900\ (1000)$ and \plancks till $\ell \sim 2300\
(2500)$ for a standard (or a Lambda) CDM model. We find that simpler estimates
make a scale-dependent error in estimating the spectrum uncertainty and that
depending on the experiment they might yield an over- or an under-estimates.

As further examples of the usefulness of this simple analytical framework, we
computed how the rms reconstruction error is affected by global noise level
variations in an experiment, as induced for instance by a variation in the
mapping duration of the experiment. \maps is relatively sensitive to such
variations, while the variation is closer to linear for \planck. We also
analysed the robustness of \plancks to various failures leading to the loss of
some channels. The most important channels are the 143 and 217\,GHz, but even
their total loss would result in less than a 10 \% increase in the residual
rms. Loosing all \hfis channels but the 217\,GHz one would only yield a 20 \%
increase, but it's loss would then nearly double the residual rms!  Clearly,
optimisation trade-off should not result in decreasing the sensitivity of the
217\,GHz and neighbouring channels.

Of course, assessing the effect of foregrounds relies to some extent on the
realism of the foregrounds model. We considered 64 sky models differing by
what appears to be the least well established features. We found that the
quality of the CMB recovery should be basically unaffected, mainly because the
dominant high-$\ell$ contributions in the channels of most weight for the CMB
determination are the CMB itself and detector noise. This excellent news
should nevertheless be taken with a grain of salt. Indeed, it was always
assumed a nearly optimal filter could be built, which entails some knowledge
of the foregrounds (\ie their very existence, their power spectra, and their
frequency dependence). Paper II will address some of these points by means of
direct numerical simulations. We can already note though that a strong
foreground at least in some channels should be no problem since the needed
information will be easily obtained from the data itself. Forgetting a very
weak foreground in all channels should not lead to great errors. The difficult
case will of course be to make sure that no unsuspected foreground is sizeable
mostly in the primary CMB channels.

But much new data is becoming available to improve on our current knowledge of
the microwave sky, either from the analysis of data from the ISO satellite, of
ongoing $H_\alpha$ and radio surveys, and of many ground and balloon
experiment. And of course the future generations of CMB satellite themselves
were designed to much increase this knowledge by mapping most of the sky in a
large frequency range and with much improved noise levels as compared to
COBE. The analysis above thus makes us rather confident that foregrounds
should not prevent the future CMB satellites from finally unveiling the CMB
power spectrum with high accuracy.

%******************************************************************************
%Acknowledgements
%******************************************************************************

\begin{ack}

We are grateful to many colleagues, and in particular N. Aghanim,
F. Boulan\-ger, D. Bond, B. Guiderdoni, S. Hancock, E. Hivon, S. Prunet,
J.-L. Puget for useful informations on foregrounds, stimulating discussions,
and access to yet unpublished results.

\end{ack}

%******************************************************************************
% Bibliography
%******************************************************************************

%\bibliography{Bibliography_CMB} % contains the bibtex codes

%******************************************************************************
\end{document}